\documentclass{article}
\usepackage[utf8]{inputenc}
\usepackage{secdot, amsfonts, amsmath, amssymb, bm, breqn, mathtools, graphicx, derivative, caption, bbm, dsfont, float, subcaption, tikz, relsize, amsthm, fullpage, array, authblk}
\usepackage{natbib}

\usepackage{algorithm}
\usepackage{algpseudocode}
\usepackage[export]{adjustbox}
\usepackage{multirow}
\usepackage{color}
\usetikzlibrary{automata, fit, shapes.geometric} 
\usetikzlibrary{bayesnet}
\usetikzlibrary{arrows}
\tikzset{
  latentnode/.style={draw, minimum width=10mm, shape=circle, scale = 1, ultra thick, black, transform shape, minimum size = 2cm},
  dagconn/.style={arrows=->, black, thick},
  plate/.style={draw, shape=rectangle, rounded corners=0.5ex, thick,
    minimum width=3.1cm, text width=3.1cm, align=right, inner sep=10pt, inner ysep=10pt,label={[xshift=-14pt,yshift=14pt]south east:#1}}
}

\usepackage[title]{appendix}
\makeatletter
\def\Vdots{\vbox{\baselineskip12\p@ \lineskiplimit\z@
  \kern1\p@\hbox{.}\hbox{.}\hbox{.}\hbox{.}\hbox{.}\hbox{.}\hbox{.}\hbox{.}\hbox{.}}}
\makeatother

\allowdisplaybreaks % Breaks align environment
\usepackage[colorlinks=false, linktocpage=true]{hyperref}
\hypersetup{
  colorlinks,
  citecolor=blue,
  linkcolor=blue,
  urlcolor=blue}

\usepackage[capitalise]{cleveref} % cleveref to be loaded after hyperref
\usetikzlibrary{automata, fit, shapes.geometric}
\crefrangelabelformat{equation}{(#3#1#4--#5#2#6)} % Macro for referencing range of equations
\usepackage[export]{adjustbox} 
\usepackage[margin=1in]{geometry}
\newcommand{\beginsupplement}{%
        \setcounter{section}{0}
        \renewcommand{\thesection}{\Alph{section}} 
        \setcounter{lemma}{0}
        \setcounter{equation}{0}
        \setcounter{table}{0}
        \renewcommand{\thetable}{S\arabic{table}}%
        \setcounter{figure}{0}
        \renewcommand{\thefigure}{S\arabic{figure}}%
     }

\usepackage{setspace}
\begin{document}

\title{\bf Global-Local Dirichlet Processes for Identifying Pan-Cancer Subpopulations Using Both Shared and Cancer-Specific Data}
\date{}

\author[1]{\small Arhit Chakrabarti}
\author[2]{\small Yang Ni}
\author[3]{\small Debdeep Pati}
\author[1]{\small Bani K. Mallick}

\affil[1]{\footnotesize \textit{Department of Statistics, Texas A\&M University}}
\affil[2]{\footnotesize \textit{Department of Statistics and Data Sciences, University of Texas at Austin}}
\affil[3]{\footnotesize \textit{Department of Statistics, University of Wisconsin-Madison}}

\maketitle

\begin{abstract}
\sloppy We consider the problem of clustering grouped data for which the observations may include group-specific variables in addition to the variables that are shared across groups. This type of data is common in cancer genomics where the molecular information is usually accompanied by cancer-specific clinical information. Existing grouped clustering methods only consider the shared variables, thereby ignoring valuable information from the cancer-specific variables. To allow for these cancer-specific variables to aid in the clustering, we propose a novel Bayesian nonparametric approach, termed global-local (GLocal) Dirichlet process, that models the ``global-local'' structure of the observations across groups. We characterize the GLocal Dirichlet process using the stick-breaking representation and the representation as a limit of a finite mixture model, which leads to an efficient posterior inference algorithm. We illustrate our model with extensive simulations and a real pan-gastrointestinal cancer dataset. The cancer-specific clinical variables included carcinoembryonic antigen level, patients' body mass index, and the number of cigarettes smoked per day. These important clinical variables refine the clusters of gene expression data and allow us to identify finer sub-clusters, which is not possible in their absence. This refinement aids in the better understanding of tumor progression and heterogeneity. Moreover, our proposed method is applicable beyond the field of cancer genomics to a general grouped clustering framework in the presence of group-specific idiosyncratic variables.
\end{abstract}
\sloppy \noindent%

\vfill

\newpage
\section{Introduction}
\subsection{Molecular- and Clinical-Based Pan-Cancer Classification}
In the current clinical practice, the classification of cancer greatly depends on the tumor site of origin. However, many recent studies (e.g., \citealt{hoadley2014multiplatform, hoadley2018cell, sanchezvega2018oncogenic}) suggest that tumors from different sites of origin may have significant clinical and molecular similarities. Classifying tumors beyond the site of origin using clinical and molecular information, therefore, can improve our understanding of both within-tumor and between-tumor heterogeneity and potentially repurpose existing cancer treatments from one tumor site to another \citep{Schein2021, Rodrigues2022}.
Large-scale cancer genomics studies such as The Cancer Genome Atlas (TCGA) have generated molecular and clinical profiles for many human cancers, making a systematic molecular- and clinical-based pan-cancer classification possible.
In such studies, molecular information is often shared across different tumors. For example, mRNA gene expression data on a common gene set are readily available for different tumors from the TCGA database. 
However, clinical variables may not be shared across cancers. For instance,  prostate-specific antigen is only recorded for prostate cancer patients. The cancer-specific clinical variables may provide invaluable insights into the study of gene expressions, either directly or indirectly. Discarding such readily available cancer-specific clinical variables might result in the loss of information that would greatly refine pan-cancer classification. Moreover, it is of scientific interest to investigate if patients with different clinical characteristics show differential gene expression patterns.
Thus, while it is desirable to utilize both molecular and clinical information to identify pan-cancer subpopulations, their varying availability across cancers makes it a challenging statistical problem. This paper proposes a novel method of incorporating cancer-specific clinical information with molecular data for a coherent and systematic classification of pan-cancer subpopulations.
\subsection{Motivating Application: Pan-Gastrointestinal Cancer}\label{sec:mapgc}
Gastrointestinal (GI) cancer is a group of cancers that develop along the GI tract. The GI tract starts from the food pipe carrying food from the mouth to the stomach, also known as the esophagus or gullet, and ends at the anus. 
Classified according to their primary site of origin \citep{valladares2011prognostic, zheng2017prognostic}, esophageal, stomach, colon, and rectal cancers are the four most common cancers of the GI tract.
Esophageal cancer is a cancer that develops in one of the layers of the food pipe. The malignant tumors of this cancer often originate near its junction with the stomach and may even spread to the stomach. On the other hand, stomach cancer originates in the cells lining the stomach.  Esophageal and stomach cancer together constitute cancers of the upper GI tract. In contrast, colon cancer and rectal cancer \citep{ColonCancer, RectalCancer} are cancers of the lower GI tract. They have many overlapping features and are often jointly termed colorectal cancer \citep{CRC}. 
In this paper, we are interested in jointly studying the tumor heterogeneity both within and across these four GI cancers, which can potentially shed light on individualized cancer prognosis, treatment, and management. The publicly available TCGA datasets consist of the (log-transformed) gene expression measurements for a common set of 60,483 genes in patients with the corresponding tumors. In particular, data for each cancer is a matrix with rows corresponding to genes and columns representing the respective cancer patients. In our later analysis, we considered the gene expression data from 92, 407, 173, and 120 patients for esophageal, stomach, colon, and rectal cancer, respectively. Following common practice, we performed uniform manifold approximation and projection (UMAP, \citealp{McInnes2018}) on the combined gene expression data from the four cancers to reduce the dimension of the data on a common manifold.  \par
The TCGA data on GI cancers include additional clinical information, which often includes prognostic markers that provide valuable insights into disease progression and tumor heterogeneity. Recent studies have shown that several common cancers including colon cancer have been linked to obesity \citep{Cancer_obesity}. \citealp{BMI_colon} believes that BMI measurement is important to understand the obesity-related risk of developing colon cancer. Moreover, for colorectal cancer, the carcinoembryonic antigen (CEA) is an important prognostic marker for monitoring tumor progression \citep{CEA_Rectal, CEA_CRC}. This naturally raises some pertinent queries: can variations in CEA levels serve as indicators of tumor subpopulations within colorectal cancer? Do colon cancer patients with high obesity risk or BMI show differential gene expression pattern in comparison to patients having lower risk? CEA, however, does not provide meaningful insights into the other GI cancer progression and is hence not collected, making it unique to colorectal cancer. Additionally, smoking has been identified as a major risk factor for esophageal cancer \citep{fan2008alcohol} and therefore is collected as a clinical variable. We will include these cancer-specific clinical variables, i.e., the number of cigarettes smoked per day for esophageal cancer, pre-operative and pre-treatment CEA for both colon and rectal cancers, and BMI measurements specific to colon cancer patients in our analysis to identify pan-GI cancer subpopulations, as they, together with genomic information, can provide invaluable insights into the understanding of tumor heterogeneity.   \par
\begin{figure}[!ht]
\centering
\begin{subfigure}{0.45\textwidth}
  \centering
    \includegraphics[width= 1\linewidth]{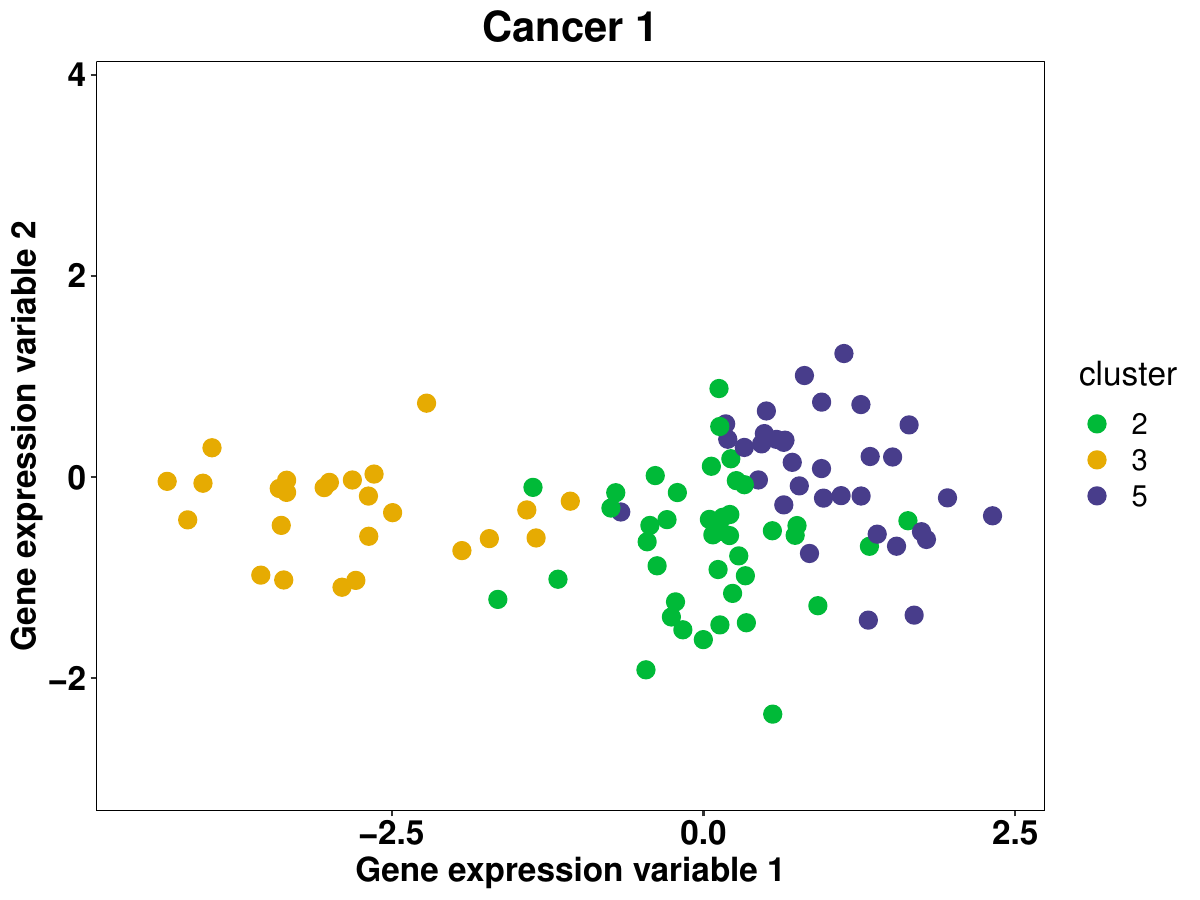}
    \caption{}
    \label{fig:Cancer1_global}
\end{subfigure}
\begin{subfigure}{0.45\textwidth}
    \centering
    \includegraphics[width= 1\linewidth]{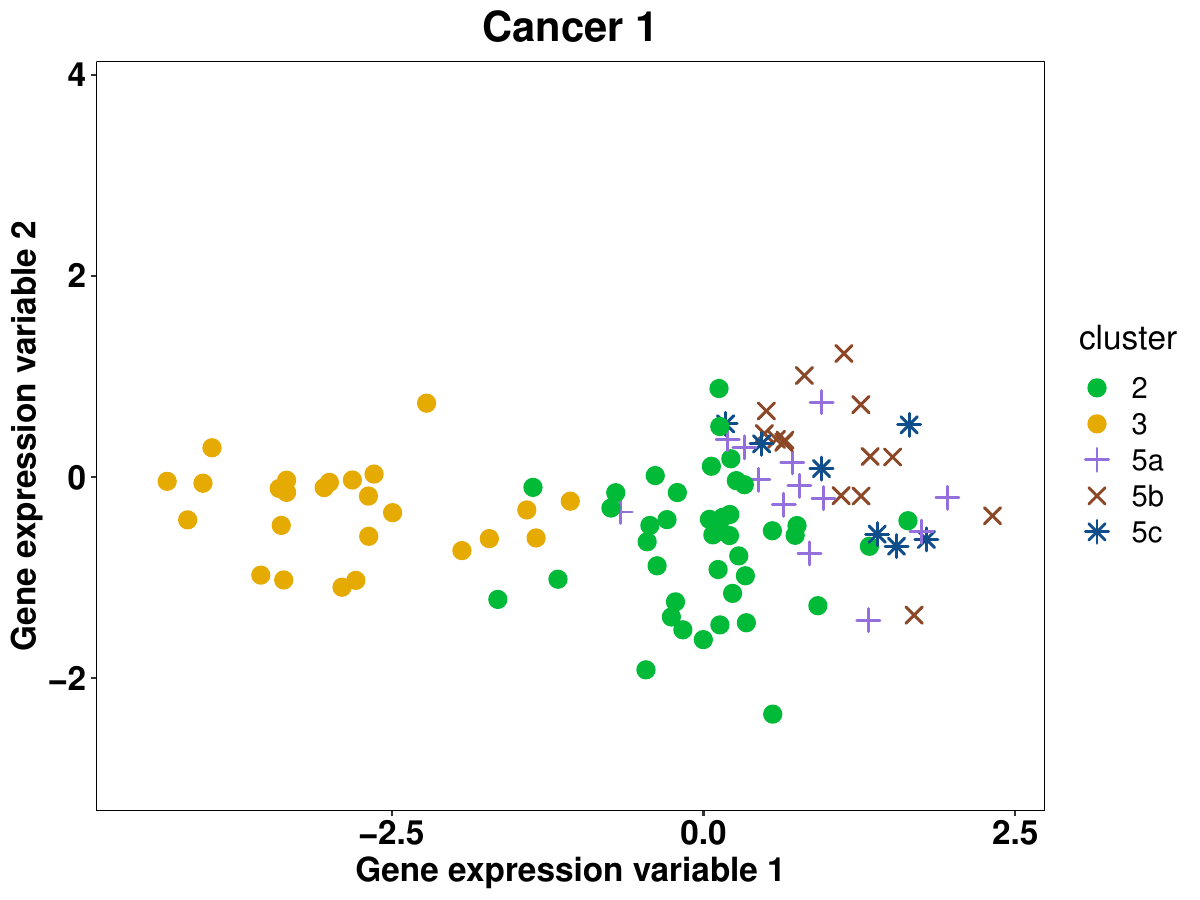}
    \caption{}
    \label{fig:Cancer1_local}
\end{subfigure}
\par
\begin{subfigure}{0.45\textwidth}
    \centering
    \includegraphics[width= 1\linewidth]{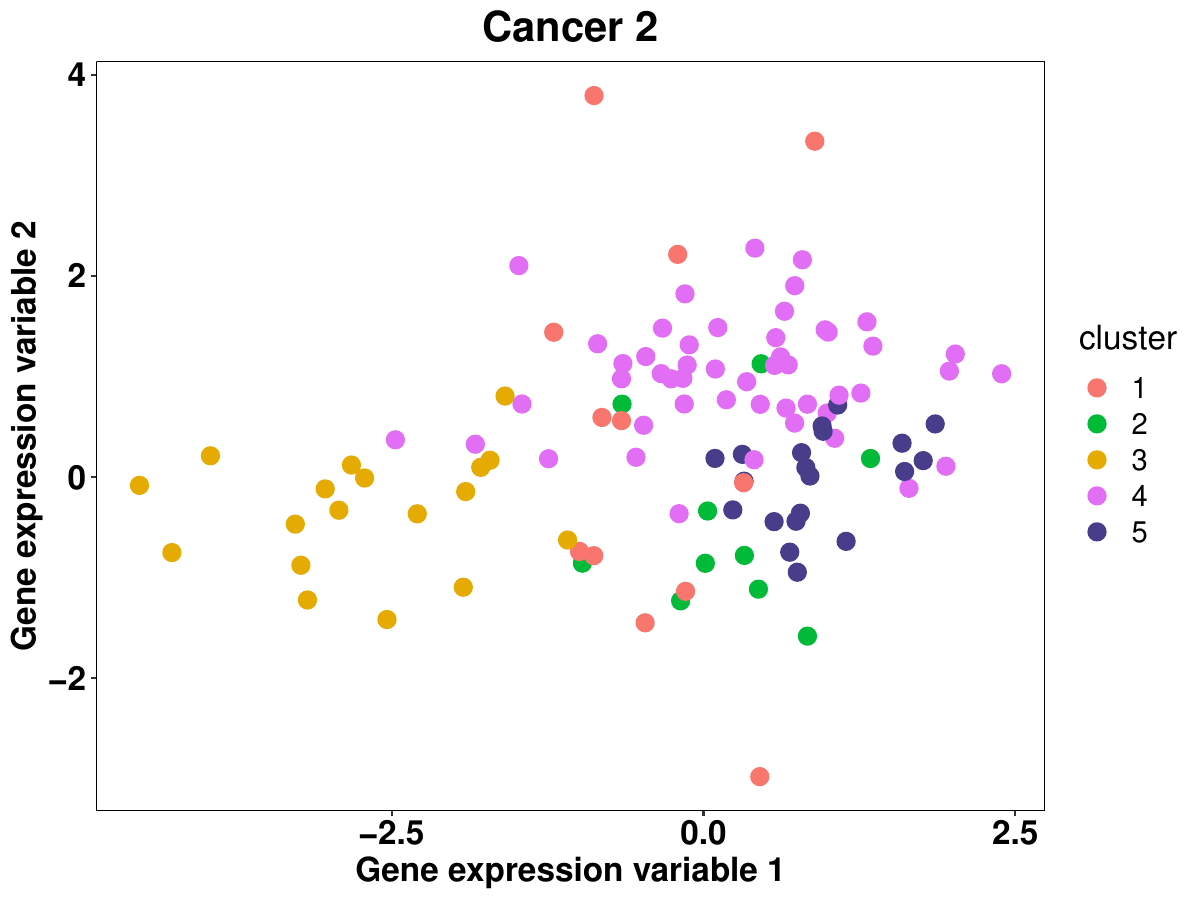}
    \caption{}
    \label{fig:Cancer2_global}
\end{subfigure}
\begin{subfigure}{0.45\textwidth}
    \centering
    \includegraphics[width= 1\linewidth]{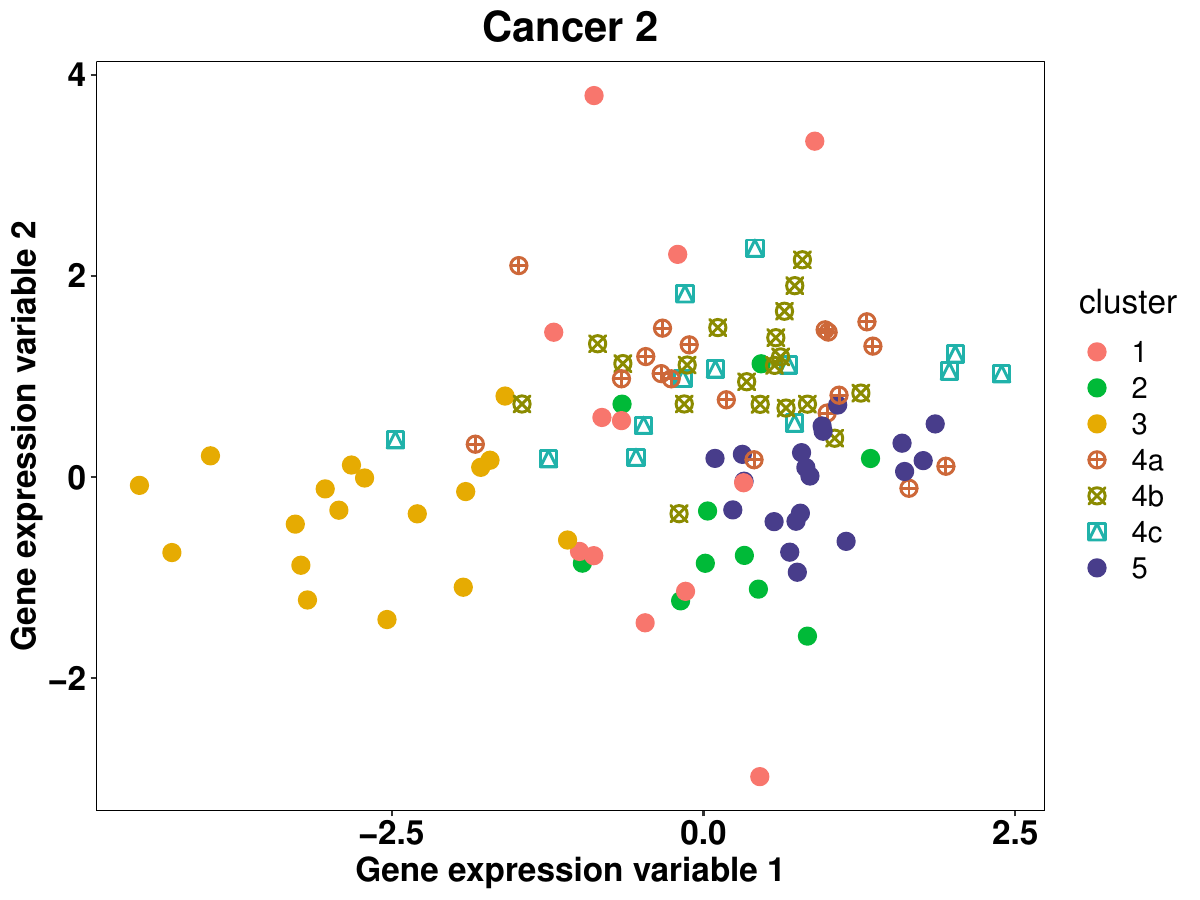}
    \caption{}
    \label{fig:Cancer2_local}
\end{subfigure}
\caption{Illustrative simulated example. Panels (a) and (c) denote the clustered gene expression data for cancer populations 1 and 2. Panels (b) and (d) show the finer sub-clustering of gene expression data induced by the cancer-specific biomarkers.}
\label{fig:Cancer_simulation_motivation2}
\end{figure}

\begin{figure}[htp]
\centering
\begin{subfigure}{0.45\textwidth}
  \centering
    \includegraphics[width= 1\linewidth]{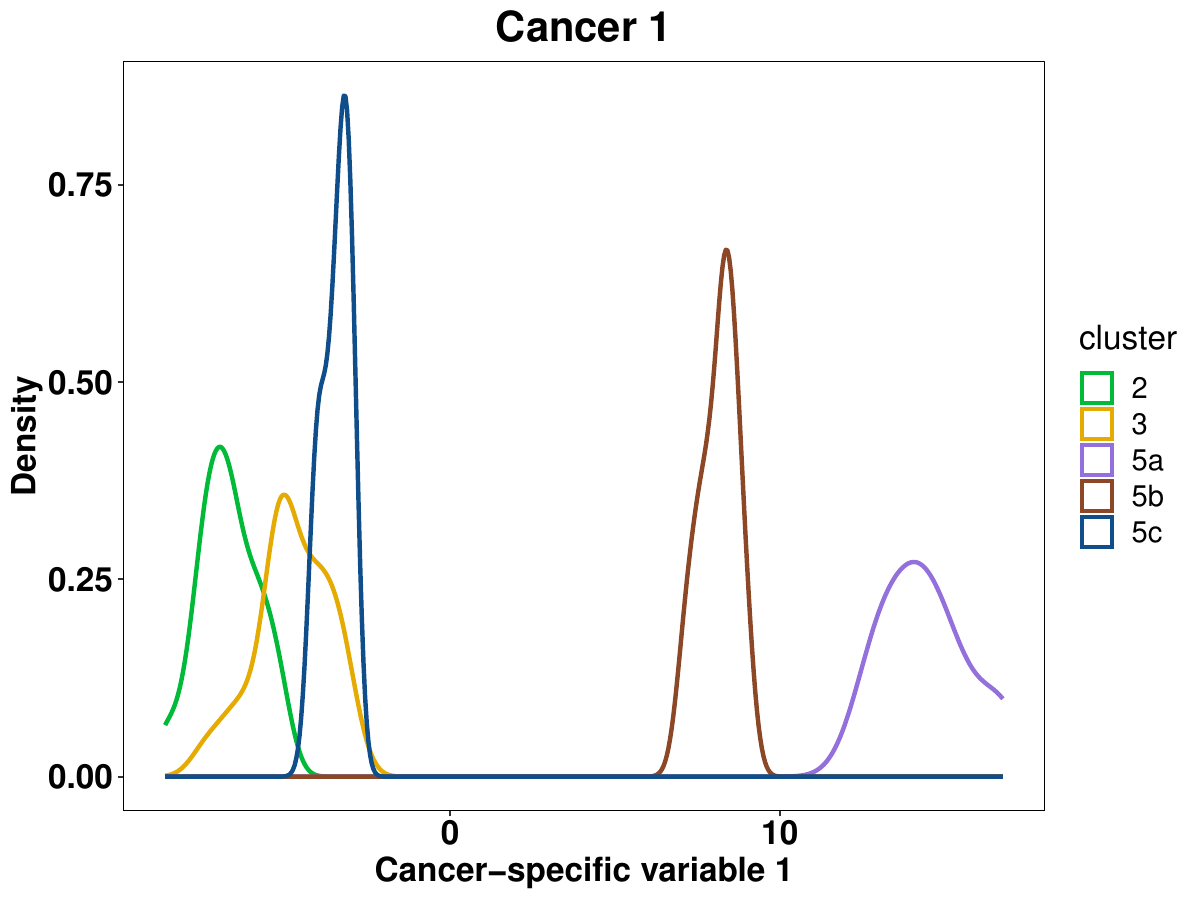}
    \caption{}
    \label{fig:Cancer1_localvariable}
\end{subfigure}
\begin{subfigure}{0.45\textwidth}
    \centering
    \includegraphics[width= 1\linewidth]{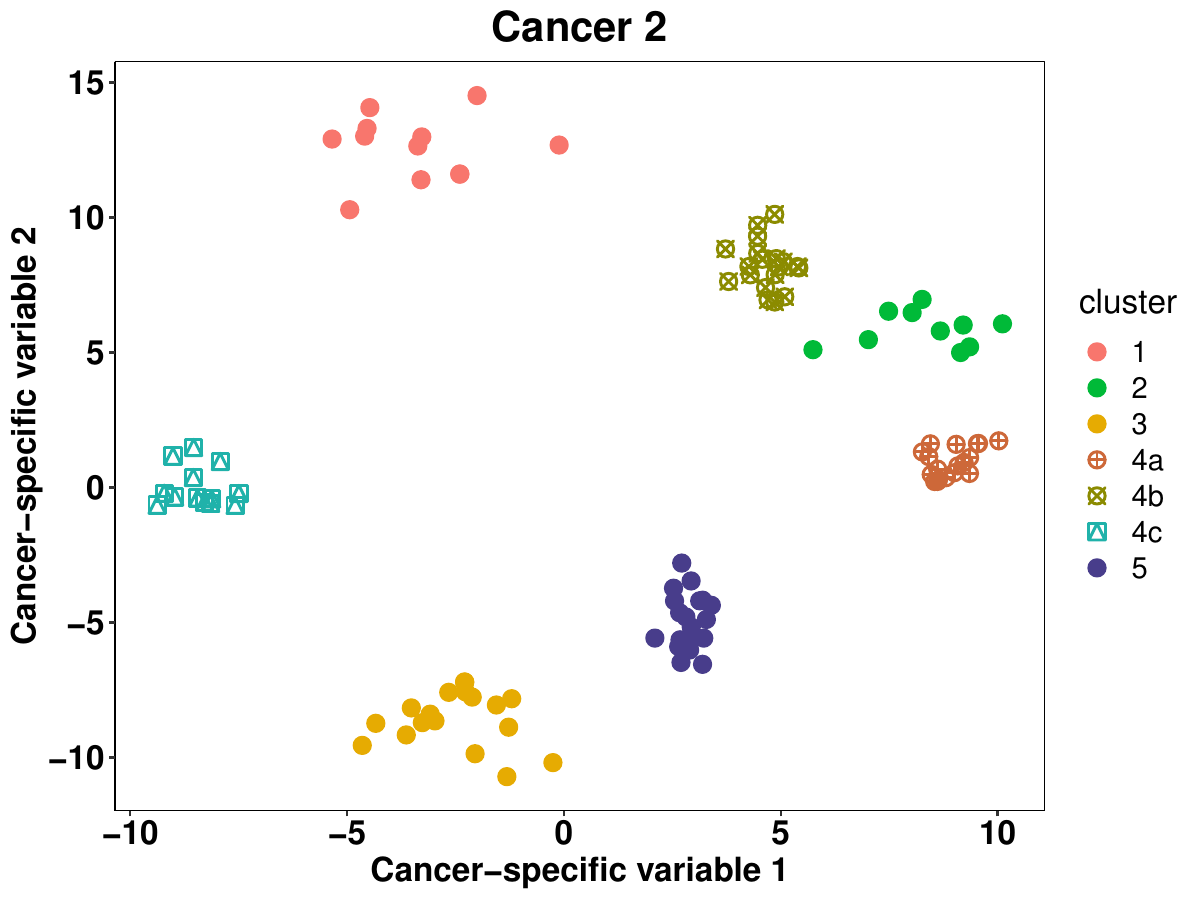}
    \caption{}
    \label{fig:Cancer2_localvariable}
\end{subfigure}
\caption{Illustrative simulated example. Panels (a) and (b) show the density and scatterplot of the cancer-specific clinical variables.}
\label{fig:Cancer_simulation_motivation}
\end{figure}

 
Clustering cancer genomic data is a common and powerful approach to identify distinct molecular subtypes within a specific cancer type. The goal is to uncover underlying biological differences that may have implications for cancer diagnosis, prognosis, and treatment response.  In this paper, we consider clustering of the gene expression of the pan-GI cancers, incorporating the cancer-specific clinical information. As a toy example, we simulated a two-dimensional gene expression dataset with two groups/cancers (e.g., colon and rectal cancers) accompanied by 1 or 2 cancer-specific clinical variables for each cancer. Figure \ref{fig:Cancer_simulation_motivation2} shows the gene expression data where we labeled the observations by two levels of clustering -- global (Figure \ref{fig:Cancer1_global} and Figure \ref{fig:Cancer2_global}) and local (Figure \ref{fig:Cancer1_local} and Figure \ref{fig:Cancer2_local}). The global-level clusters may be shared across cancers whereas the local-level clusters are unique to each cancer.  
The two cancers share the global clusters 2, 3, and 5, while the global clusters 1 and 4 are unique to the cancer population 2. The local variable in cancer population 1 refines the global cluster 5 into three finer local clusters 5a, 5b, and 5c (Figure \ref{fig:Cancer1_local}). This refinement of the global cluster may be associated with the levels of the cancer-specific local variable (say, a prognostic biomarker for cancer 1). Similarly, the local variables in cancer population 2 refine the global cluster 4 into three finer local clusters 4a, 4b, and 4c (Figure \ref{fig:Cancer2_local}). The density plots and scatterplots of the cancer-specific clinical variables are shown in Figure \ref{fig:Cancer_simulation_motivation}. The separation of the clinical variable(s) explains the refinement of cancer subpopulations, which would not be possible in the absence of cancer-specific variables. Furthermore, the cancer-specific variables also help form the global clusters. For instance, in this simulated example, the local variables are better separated than the global variables (Figure \ref{fig:Cancer2_localvariable}), which would assist in the detection of the highly overlapped global clusters (e.g., clusters 4 and 5 in Figure \ref{fig:Cancer2_global}). 

\subsection{Literature Review on Clustering}
Clustering methods have been used for gene expression data such as hierarchical clustering \citep{ SEAL2005143, do2008clustering, hossen2015methods},  k-means clustering \citep{handhayani2015intelligent,  jothi2019dk}, WGCNA \citep{langfelder, tian2020identification, hou2021k}, self organizing map \citep{nikkila2002analysis, brameier2007co}, and consensus clustering \citep{monti2003consensus, galdi2015consensus}; see \citealp{kerr2008techniques} and \citealp{oyelade2016clustering} for a comprehensive review and \citealp{cluster_gene} for a comparative study of many clustering algorithms. Some of these methods have been used in numerous cancer studies to identify genes associated with tumor development and their progression \citep{ma2009identification, kim2018}. Furthermore, clustering of gene expression data have been used in cancer-subtype detection and prediction \citep{saha2013gene, nidheesh2017enhanced}. Cancer subtype detection or the grouping of patients according to the subtype of their disease is a critical step in the development of novel targets for cancer therapy \citep{yu2017clustering, gao2022}. However, a clustering algorithm typically requires the user to pre-specify  the number of clusters or chooses it based on some \textit{ad hoc} criteria.
In this paper, we consider Bayesian nonparametric clustering methods because we do not know the number of cancer subpopulations \textit{a priori}, and Bayesian nonparametric methods provide an elegant way to automatically infer it from the data \citep{Mallick}. 
The celebrated \emph{Dirichlet process} (DP, \citealp{ferguson}) is at the core of numerous model-based Bayesian nonparametric clustering methods \citep{antonaik,escobar&west,mallickwalker, maceacher_muller,hjort2010bayesian, muller2015bayesian}. The DP, $\mbox{DP}(\alpha_0, G_0)$, is a probability measure on random probability measures, where $\alpha_0 > 0$ is the concentration parameter and $G_0$ is a base probability measure. The random probability measure drawn from the DP is almost surely discrete and, therefore, is useful for clustering when used as a mixing distribution in a mixture model.
One of the advantages of DP mixture models (\citealp{lo1984}; \citealp{escobar&west};  \citealp{maceacher_muller}) is its ability to perform clustering without having to fix the number of clusters \emph{a priori}.  Note that \citealp{miller_harrisonDP,miller_harrisonPY, yang2020posterior} showed that DP mixture models are not consistent for the number of clusters. However, such inconsistency can be avoided by simply imposing a hyperprior on the concentration parameter \citep{ClusteringconsistencywithDPM}.
\par
When considering grouped data (e.g., the groups are the tumor tissues of origin in our application), naively, one could apply either a separate DP mixture model to each group on one extreme or a single DP mixture model ignoring the groups on the other extreme. However, it is often desirable to identify group-specific clusters while allowing the groups to be linked so that clusters are comparable across groups. While there are numerous non-Bayesian algorithms for clustering, to the best of our knowledge, there are no such frequentist algorithms for jointly clustering grouped data that allow cluster information to be shared across groups, while clustering group-specific observations. The two popular Bayesian approaches to this problem include the hierarchical Dirichlet process (HDP, \citealp{hdp}) and the nested Dirichlet process (nested DP, \citealp{nestedDP}).
Let $G_j$ denote the group-specific random probability measure for $j=1,\dots,J$. HDP assumes that conditional on $G_0$, each $G_j$ is independently and identically distributed as $\mbox{DP}(\alpha_0, G_0)$ where $\alpha_0$ is the shared concentration parameter and $G_0$ is the shared base probability measure for all groups. They further assume that $G_0$ follows another DP, $G_0\sim \mbox{DP}(\gamma, H)$. Since $G_0$ is almost surely discrete, the group-specific probability measure $G_j$ shares the same set of atoms. The corresponding HDP mixture model is thus capable of identifying group-specific clusters indicated by the occupied atoms while sharing cluster information across groups. Contrarily, the nested DP assumes that $G_j$ follows a DP-distributed random probability measure with another DP as the base measure, i.e., conditionally, $G_j\sim Q$ and $Q\sim \mbox{DP}(\alpha_0,\mbox{DP}(\gamma,H))$. The nested DP clusters groups as well as observations within each group cluster. It restricts the distribution of observations within each group to be either identical or completely unrelated across groups. Additionally, the nested DP is known to suffer from a degeneracy property \citep{latent_nDP} -- two distributions sharing even one atom in their support are automatically assigned to the same cluster.

Both the HDP and the nested DP fall under the general framework of dependent DP \citep{DDP_MacEachern1, DDP_MacEachern2}. See \citealp{DDP_review_paper} for a recent review of different dependent DPs. Several recent works \citep{beraha2021semi, balocchi2022clusteringarealunitsmultiple, bi2023class, hidden_HDP} have been proposed to take advantage of the cluster-sharing feature of the HDP and the group-clustering feature of the nested DP. In contrast to methods relying on the HDP or its variants, some other works rely on models with additive structure or common atoms \citep{latent_nDP, denti2023common, BNPnestedDP, dangelodenti2024}. \citealp{CAM_regression} considers a Bayesian nonparametric common atoms regression model to generate synthetic controls in clinical trials.  Their underlying goal is to introduce matched clusters of patients in a treatment-only trial dataset with historical control trials or real world datasets to provide a reliable comparison of the treatment effect. Furthermore, the authors extend their method to include covariates of different data-types and missing values by accommodating variable dimensional covariates. In particular, they consider the scenario wherein observations $i$ and $i'$ within a group may be of different dimension due to missing values for some variables. However, all existing methods assume that the observations across the groups are measured on the same set of variables (with possible missing values for some variables within a group), which is not the case in our application where some clinical variables are unique to a specific cancer. 
To the best of our knowledge, there is no existing Bayesian or non-Bayesian method that can accommodate this idiosyncratic data structure in the clustering of grouped data. 
\subsection{Our Major Contributions}
We introduce a novel Bayesian nonparametric approach for clustering grouped data by incorporating both the shared (e.g., gene expression from different tumors) and the group-specific variables (e.g., cancer-specific biomarkers). Specifically, let $\bm{x}_{ji}$ denote the observation $i$ from group/cancer $j$. We assume that the observations are \emph{partially exchangeable} \citep{definnetti}, entailing that observations are exchangeable within each group but not across the groups.
Suppose that the observations are partitioned into $\bm{x}_{ji} = \left(\bm{x}_{ji}^L, \bm{x}_{ji}^G\right)$, where $\bm{x}_{ji}^G$ denotes the set of variables shared across the groups (e.g., age, sex, and gene expression) and $\bm{x}_{ji}^L$ denotes the set of group-specific variables idiosyncratic to group $j$ (e.g., prostate-specific antigen, which is a biomarker specific to prostate cancer). Note that HDP assumes that the observations are measured on exactly the same set of variables across the groups, i.e., $\bm{x}_{ji} = \bm{x}_{ji}^G$. We refer to $\bm{x}_{ji}^G$ as \emph{global} variables and $\bm{x}_{ji}^L$ as \emph{local} variables. 
We will cluster these grouped observations by a new Bayesian nonparametric approach that incorporates the ``global-local'' structure of the observations.\par 
To model both global and local variables, we let the group-specific random measure $G_j$ be supported on a space that consists of a common subspace shared across groups and an idiosyncratic subspace specific to each group. 
More precisely, we assume that conditionally on $\alpha$ and $V$, $G_j$ is independently (but not identically) distributed as $\mbox{DP}(\alpha, U_j\otimes V)$ where $U_j$ is an idiosyncratic base measure, $V$ is a common base measure, and $\otimes$ is the measure product. To allow the clustering information to be shared across groups, we assume that the common base measure is also conditionally DP distributed $V \sim \mbox{DP}(\gamma,H)$, conditional on the concentration parameter $\gamma$. 
We refer to this new model as the global-local (GLocal) DP. By the construction of the proposed GLocal DP, there is a positive probability that $G_j$ has shared atoms in the common subspace across the groups, thereby allowing the ``global'' clustering information to be shared. Moreover, the idiosyncratic base measure $U_j$ modifies the global clusters and refines them into smaller ``local'' clusters through the local variables. GLocal DP includes HDP as a special case in the absence of local variables for all the groups. Unlike HDP, the groups are not exchangeable in GLocal DP because it has group-specific base measure. Even though HDP can be generalized to avoid the exchangeability of the groups by introducing group-specific hyperparameters, i.e., $G_j  \mid G_0 \sim \mbox{DP}(\alpha_j,G_0)$. Nonetheless, even in this case, $G_j$'s share the same support and, thus, HDP cannot be used to cluster observations with different variables across groups; enabling the clustering of such data is the main novel contribution of GLocal DP.

We will characterize the proposed GLocal DP by the stick-breaking representation and the infinite limit of finite mixture model representation. These representations pave the way to develop a simple and efficient posterior sampler for the GLocal DP. We provide extensive simulations to demonstrate our method. Furthermore, we analyze a real pan-cancer dataset using the GLocal DP. In particular, we cluster pan-GI cancer gene expression data, incorporating cancer-specific biomarkers. Our goal was to cluster the ``global'' variables (UMAP embeddings of gene expression data), while allowing the cancer-specific clinical variables to aid in clustering. In summary, our main contribution is three-fold: 
\begin{enumerate}
    \item We propose a general Bayesian nonparametric approach, GLocal DP, to incorporate group-specific local variables for clustering of grouped data. 
    \item We provide two characterizations of GLocal DP, each providing a different perspective and paving the way for an efficient algorithm for posterior inference.
    \item In the pan-GI cancer application, we identified shared subpopulations between the two upper-GI cancers, esophagus and stomach, and between the two lower-GI cancers, colon and rectum, but no shared subpopulations across upper- and lower-GI cancers.
    Clinical variables further refine the subpopulations and aid in the understanding of tumor progression and heterogeneity, which would not be captured by existing methods. In particular, the local variables help in the classification of survival patterns of cancer patients with the levels of associated risk factors, concurrent with existing scientific knowledge. Moreover, our analysis exclusively shows a disparate differentially expressed gene set characterizing the subpopulations, which would not have been possible using existing grouped-clustering methods that only identify the shared clusters. The upregulation of marker genes in tumor subpopulations and its corresponding effect on the prognostic clinical biomarkers were identified, which is further corroborated by existing literature. Furthermore, the application of the GLocal DP is not exclusively limited to the field of cancer genomics. The proposed method can be used for a general grouped clustering framework, wherein the available data consists of important group-specific variables apart from the shared variables.
\end{enumerate}
The rest of the paper is organized as follows. Section \ref{sec:prelim} provides a brief overview of some preliminaries needed for the remainder of the paper. Section \ref{sec:glocal DP} introduces the proposed GLocal DP and the corresponding mixture model. We present two representations of the proposed GLocal DP in Sections \ref{subsec:stick_breaking_repr} and \ref{subsec:fmm_repr}. Section \ref{sec:posterior inference} outlines the proposed Markov chain Monte Carlo (MCMC) algorithm for posterior inference. Section \ref{sec:Real_data_analysis} presents a real data analysis using the proposed method on pan-cancer genomics data. In Section \ref{sec:simulations}, we provide simulation studies. The paper concludes with a brief conclusion in Section \ref{sec:discussion}.  The code used for analysis, encompassing simulations and real data, as well as the datasets themselves, are available in the GitHub repository: \href{https://github.com/Arhit-Chakrabarti/GLocalDP}{https://github.com/Arhit-Chakrabarti/GLocalDP}.

\section{Preliminaries}\label{sec:prelim}
\subsection{Infinite mixture model}
We present a brief overview of infinite mixture models for a single population, the DP mixture model, and for multiple exchangeable populations, the HDP mixture model. 
\subsubsection{Dirichlet process mixture model}
For a single population, let $x_i$ denote the $i$th realization of a random variable $X$. Consider the following mixture model,
\begin{equation}
\label{DP mixture model}
\begin{aligned}
    \theta_i \mid G   & \overset{iid}{\sim} G,\\
    x_i \mid \theta_i & \overset{ind}{\sim} F(\theta_i),
\end{aligned}
\end{equation}
\noindent where $F(\theta_i)$ denotes the distribution of $x_i$ parameterized by $\theta_i$. The parameters $\theta_i$'s are conditionally independent given the prior distribution $G$. In a DP mixture model, $G$ is assigned a DP prior, $G \sim \mbox{DP}(\alpha_0, G_0)$ with concentration $\alpha_0$ and base probability measure $G_0$.\par
\citealp{Sethuraman} presented the \emph{stick-breaking representation} of the DP based on independent sequences of i.i.d. random variables $(\pi_k')_{k=1}^{\infty}$ and $(\phi_k)_{k=1}^{\infty}$,  which is given by,
\begin{align}
\label{DP stick breaking 1}
    \pi_k' & \overset{iid}{\sim} Beta(1, \alpha_0), & & \phi_k \overset{iid}{\sim} G_0,\\
    \label{DP stick breaking 2}
    \pi_k & = \pi_k'\prod_{l=1}^{k-1}(1-\pi_l'), & & G = \sum_{k=1}^{\infty}\pi_k\delta_{\phi_k},
\end{align}
where $\delta_{\phi}$ is a point mass at $\phi$ and $\phi_k$'s are called the \emph{atoms} of $G$. The sequence of random weights $\boldsymbol{\pi}=(\pi_k)_{k=1}^{\infty}$ constructed from \eqref{DP stick breaking 1} and \eqref{DP stick breaking 2} satisfies $\sum_{k=1}^{\infty}\pi_k = 1$ with probability one. The random probability measure on the set of integers is denoted by $\boldsymbol{\pi} \sim \text{GEM}(\alpha_0)$ for convenience where GEM stands for Griffiths, Engen and McCloskey \citep{pitman_GEM}. It is clear from \eqref{DP mixture model} and \eqref{DP stick breaking 2} that $\theta_i$ takes the value $\phi_k$ with probability $\pi_k$. Let $z_i$ be a categorical variable such that $z_i=k$ if $\theta_i=\phi_k$. An equivalent representation of a Dirichlet process mixture is given by, 
\begin{equation}
    \begin{aligned}
        \boldsymbol{\pi} & \sim \text{GEM}(\alpha_0), &  z_i \mid \boldsymbol{\pi} & \overset{iid}{\sim} \boldsymbol{\pi},\\
        \phi_k  & \overset{iid}{\sim} G_0,  & x_i\mid z_i, (\phi_k)_{k=1}^{\infty} & \overset{ind}{\sim} F(\phi_{z_i}).
    \end{aligned}
\end{equation}
\subsubsection{Hierarchical Dirichlet process mixture model}
Suppose observations are now organized into multiple groups. Let $x_{ji}$ denote the observation $i$ from group $j$. 
Let $F(\theta_{ji})$ denote the distribution of $x_{ji}$ parameterized by $\theta_{ji}$, and let $G_j$ denote a prior distribution for $\theta_{ji}$. The group-specific mixture model is given by, 
\begin{equation*}
\label{hdp mixture model}
    \begin{aligned}
        \theta_{ji}\mid G_j &\overset{ind}{\sim} G_j,\\
        x_{ji}\mid \theta_{ji} &\overset{ind}{\sim} F(\theta_{ji}).
    \end{aligned}
\end{equation*}
As with the DP mixture model, when the random measures $G_j$'s are assigned an HDP prior,
\begin{equation}
\label{hdp eq1}
\begin{aligned}
    G_0  & \sim \mbox{DP}(\gamma, H),\\
    G_j \mid  G_0 & \sim \mbox{DP}(\alpha_0, G_0),
\end{aligned}
\end{equation}
the corresponding mixture model is referred to as the HDP mixture model. The global random probability measure $G_0$ is distributed as a DP with concentration parameter $\gamma$ and base probability measure $H$. The group-specific random measures $G_j$'s are conditionally independent given $G_0$ and hence are exchangeable. They are distributed as DP with the base measure $G_0$ and some concentration parameter $\alpha_0$. 
Because DP-distributed $G_0$ is almost surely discrete, the atoms of $G_j$’s are necessarily shared across groups. This leads to a positive probability of shared clusters across different groups.
\section{GLocal Dirichlet Process}\label{sec:glocal DP}
When data contain varying sets of variables across groups, the HDP prior \eqref{hdp eq1} is not appropriate (e.g., $G_j$ does not have the correct support). Our solution to the problem of clustering such grouped data with varying variable sets is to specify a joint distribution of $G_j$'s that takes into account both the local and global variables via the novel GLocal DP.
\par Recall that $\bm{x}_{ji} = \left(\bm{x}_{ji}^L, \bm{x}_{ji}^G\right)$ denotes the $i$th observation from the group $j$. We assume that each observation is drawn independently from a mixture model with $\bm{\theta}_{ji}$ denoting the factor (parameter) specifying the mixture component associated with the observation $\bm{x}_{ji}$. Similar to the observations, the factor $\bm{\theta}_{ji}$ can be partitioned into local and global factors, $\bm{\theta}_{ji} = \left(\bm{\theta}_{ji}^L, \bm{\theta}_{ji}^G\right)$. By later construction, there is a positive prior probability that the global factors are equal across groups  (e.g., $\bm{\theta}_{ji}^G=\bm{\theta}_{j'i'}^G$), thereby inducing the sharing of global clusters. Furthermore, the local factors ($\bm{\theta}_{ji}^L$) can modify the global clusters and may refine them into smaller local clusters.

Let $F(\bm{x}_{ji}\mid \bm{\theta}_{ji})$ denote the distribution of the observation $\bm{x}_{ji}$, conditional on the factor $\bm{\theta}_{ji}$. For simplicity, we assume that the distribution can be factorized as, 
 \begin{equation}
     \label{eq:global_local_dist_assump}
     F(\bm{x}_{ji}\mid \bm{\theta}_{ji}) = F_1(\bm{x}_{ji}^L\mid \bm{\theta}_{ji}^L) \, F_2(\bm{x}_{ji}^G\mid \bm{\theta}_{ji}^G),
 \end{equation}
 where $F_1(\bm{x}_{ji}^L\mid \bm{\theta}_{ji}^L)$ denotes the conditional distribution of the local variables $\bm{x}_{ji}^L$, conditioned on the local factors $\bm{\theta}_{ji}^L$, and $F_2(\bm{x}_{ji}^G\mid \bm{\theta}_{ji}^G)$ denotes the conditional distribution of the global variables $\bm{x}_{ji}^G$, given the global factors $\bm{\theta}_{ji}^G$. In other words, $\bm{x}_{ji}^G$ and $\bm{x}_{ji}^L$ are conditionally independent. But note that marginally they are not independent. If additional dependency is desired between $\bm{x}_{ji}^G$ and $\bm{x}_{ji}^L$, one can replace $F_1(\bm{x}_{ji}^L\mid \bm{\theta}_{ji}^L)$ in \eqref{eq:global_local_dist_assump} by $F_1(\bm{x}_{ji}^L\mid \bm{x}_{ji}^G,\,  \bm{\theta}_{ji}^L)$ but we do not pursue this direction in this paper.
 Let $G_j$ denote the group-specific prior distribution for the factors $\bm{\theta}_{ji}$. We assume that the factors are conditionally independent given $G_j$, leading to the following probability model,
 \begin{equation}
 \label{eq:global_local_DP_model}
 \begin{aligned}
     \bm{\theta}_{ji} = \left(\bm{\theta}_{ji}^L, \bm{\theta}_{ji}^G\right) \mid G_j & \sim G_j 
 \end{aligned}
 \end{equation}
Let $\left(\Theta_j, \mathcal{A}_j\right)$ denote the measurable space corresponding to the local factors of group $j$ and $\left(\Omega, \mathcal{B}\right)$ denote the measurable space corresponding to the shared global factors across the groups. The proposed GLocal DP defines a set of random probability measures $G_j$, one for each group, on the product space $\left(\Theta_j \times \Omega, \mathcal{A}_j \otimes \mathcal{B}\right)$,
  \begin{equation}
 \label{eq:global_local_DP_def1}
    G_j \mid \alpha, V \sim \mbox{DP}(\alpha, U_j\otimes V),
 \end{equation}
 where $\alpha$ denotes the positive concentration parameter. The base measure $U_j\otimes V$, defined on the same product space $\left(\Theta_j \times \Omega, \mathcal{A}_j \otimes \mathcal{B}\right)$, is a random product probability measure of the local measure $U_j$ and the global measure $V$, where $U_j$ is defined on $\left(\Theta_j, \mathcal{A}_j\right)$ and $V$ is defined on $\left(\Omega, \mathcal{B}\right)$. To allow for the sharing of global factors across the groups, we further assume,
   \begin{equation}
 \label{eq:global_local_DP_def2}
    V \mid \gamma \sim \mbox{DP}(\gamma, H),
 \end{equation}
 where $\gamma$ and $H$ are the concentration parameter and base probability measure, respectively. Equations \eqref{eq:global_local_dist_assump} and \eqref{eq:global_local_DP_model} along with the prior specifications given in \eqref{eq:global_local_DP_def1} and \eqref{eq:global_local_DP_def2} complete the specification of the proposed GLocal DP mixture model. We note that GLocal DP reduces to HDP in the absence of group-specific local variables (hence local factors) for all the groups. But when the local variables are present, they play a significant role in the clustering of grouped data. Apart from defining group-specific local clusters, the local variables can also affect the clustering of global variables across populations, which will be explained at the end of Section \ref{subsec:stick_breaking_repr}. This makes our method different from HDP even on the global level. Furthermore, following \cite{ClusteringconsistencywithDPM}, we assume non-informative gamma priors on the concentration parameters. 
 
  In the next two subsections, we provide the stick-breaking representation and the infinite limit of finite mixture model representation of the proposed GLocal DP, which form the building blocks for an efficient posterior inference procedure.
 \subsection{The stick-breaking representation}\label{subsec:stick_breaking_repr}
Since the global measure $V$ is distributed as a DP, it can be expressed using a stick-breaking representation \citep{Sethuraman},
 \begin{equation}
     \label{eq:stick-breaking_global}
     V = \sum_{k = 1}^{\infty} \beta_k\delta_{\phi_k},
 \end{equation}
 where $\bm{\beta}=(\beta_k)_{k=1}^{\infty} \mid \gamma \sim \text{GEM}(\gamma)$ and $\phi_k \overset{iid}{\sim} H$ independent of $\bm{\beta}$. Furthermore, as each $G_j$ is distributed as a DP, a similar stick-breaking representation gives,
  \begin{equation}
     \label{eq:stick-breaking_local}
     G_j = \sum_{t = 1}^{\infty}  {\pi}_{jt}\delta_{\psi_{jt}},
 \end{equation}
  where $ {\bm{\pi}}_j=( {\pi}_{jt})_{t=1}^{\infty} \mid \alpha  \sim \text{GEM}(\alpha)$ and $\psi_{jt} \mid V \overset{ind}{\sim} U_j\otimes V$ independent of $ {\bm{\pi}}_j$. Since each factor $\bm{\theta}_{ji}$ is distributed according to $G_j$, it takes on the value $\psi_{jt}= \left(\psi_{jt}^L, \psi_{jt}^G\right)$ with probability $ {\pi}_{jt}$, %The group-specific atoms are $\psi_{jt} = \left(\psi_{jt}^L, \psi_{jt}^G\right)$, 
  where
  $\psi_{jt}^L\overset{iid}{\sim} U_j$ and $\psi_{jt}^G \mid V \overset{iid}{\sim} V$. Because $V$ has support at the points $\bm{\phi} = (\phi_k)_{k=1}^{\infty}$, the marginal distribution of each $G_j$ with $\psi_{jt}^L$ marginalized out also has support at these points through $\psi_{jt}^G$. In other words, the atoms $\left(\psi_{jt}^G\right)_{t=1}^{\infty}$ are necessarily the same as $(\phi_k)_{k=1}^{\infty}$. In fact, $\psi_{jt}^G$ takes on the value $\phi_k$ with probability $\beta_k$.  
  This sharing of global factors across the groups ensures the sharing of clustering of the global variables. To make the clustering aspect of our model  explicit, we introduce the latent variables $t_{ji}$ and $k_{jt}$, where
  \begin{align}
     t_{ji}\mid  {\bm{\pi}}_j &\overset{ind}{\sim}  {\bm{\pi}}_j,\\
     k_{jt}\mid \bm{\beta} &\overset{ind}{\sim} \bm{\beta},
 \end{align}
 such that, conditional on the latent indicators $t_{ji}$ and $(k_{jt})_{t=1}^{\infty}$, we have $\bm{x}_{ji}\sim F_1(\bm{x}_{ji}^L \mid \psi_{jt_{ji}}^L)F_2(\bm{x}_{ji}^G \mid \phi_{k_{jt_{ji}}})$. We refer to the latent indicator $k_{jt_{ji}}$ as the \emph{global-level} cluster label as it indicates the shared clustering across groups. For instance, if $k_{jt_{ji}}=k_{j't_{j'i'}}$, then the $i$th observation from group $j$ and the $i'$th observation from group $j'$ belong to the same global cluster. Likewise, we refer to the latent variable $t_{ji}$ as the \emph{local-level} cluster label as it indicates the refined clusters within each group. In particular, for two observations  $i$ and $i'$, if $t_{ji} \neq t_{ji'}$ then the local variable(s) in group $j$ refines the corresponding global clusters $k_{jt_{ji}}$ and $k_{jt_{ji'}}$ into two distinct sub-clusters. It is the dissimilarity in the local variable(s) for observations $i$ and $i'$ that leads to this refinement, which can aid in the understanding of the effect of the local variable(s) on the global variable clustering. With these two sets of latent indicators, we obtain an equivalent representation of the GLocal DP mixture via the following conditional distributions:
 \begin{equation}\label{eq:glocalDPmixturemodel}
     \begin{array}{ll}
         \bm{\beta}\mid \gamma \sim \text{GEM}(\gamma), & k_{jt}\mid \bm{\beta} \sim \bm{\beta}, \\
          {\bm{\pi}}_j \mid \alpha \sim \text{GEM}(\alpha), & t_{ji} \mid  {\bm{\pi}}_j \sim  {\bm{\pi}}_j,\\
        \phi_k \sim H, & \psi_{jt}^L \sim U_j,  \\
         \bm{x}_{ji}\mid (\phi_k)_{k=1}^{\infty}, (\psi_{jt}^L)_{t=1}^{\infty}, t_{ji}, &\hspace{-.1in}(k_{jt})_{t=1}^{\infty}  \sim F_1(\bm{x}_{ji}^L \mid \psi_{jt_{ji}}^L)F_2(\bm{x}_{ji}^G \mid \phi_{k_{jt_{ji}}}).  
     \end{array}
 \end{equation}
We remark that our clusters have hierarchical structure where the local-level clusters (given by $t_{ji}$) are nested within the global-level clusters (corresponding to $k_{jt_{ji}}$). This hierarchical nature of our clusters indicates that the local variables help refine the global clusters. In our motivating pan-cancer application, this plays an important role in the finer understanding of molecular subpopulations modified by cancer-specific clinical variables.  The Figure \ref{fig:GLocalDP_Graphical_Plot_subfig1} 
 in the Supplementary Material shows the graphical model representation of the GLocal DP mixture model. Marginalizing the local-level indicators, $t_{ji}$ yields the model, shown in the Figure \ref{fig:GLocalDP_Graphical_Plot_subfig2} in the Supplementary Material. Clearly, conditional on the data $\{\bm{x}_{ji}^L, \bm{x}_{ji}^G\}$, the marginalized global-level assignment of the observation $i$ in group $j$, $k_{ji}$, depends on the corresponding local variables $\bm{x}_{ji}^L$. Thus, the local variables can affect the global-level clustering.
 \subsection{The infinite limit of finite mixture models} \label{subsec:fmm_repr}
Alternatively to the stick-breaking representation, the GLocal DP mixture model in \eqref{eq:glocalDPmixturemodel} can be derived as the infinite limit of a finite mixture model. 
Specifically, consider the following finite mixture model, 
\begin{equation}
\label{eq:fmm_repr_1}
    \begin{array}{ll}
        \bm{\beta} \mid \gamma  \sim \mbox{Dir}(\gamma/L, \dots, \gamma/L), &  k_{jt}\mid \bm{\beta} \sim \bm{\beta}, \\
         {\bm{\pi}}_j \mid \alpha  \sim \mbox{Dir}(\alpha/T, \dots, \alpha/T), & t_{ji} \mid  {\bm{\pi}}_j \sim  {\bm{\pi}}_j,\\
        \phi_k  \sim H, & \psi_{jt}^L  \sim U_j,\\        \bm{x}_{ji}\mid (\phi_k)_{k=1}^{L}, (\psi_{jt}^L)_{t=1}^{T}, t_{ji}, (k_{jt})_{t=1}^{T} \sim &F_1(\bm{x}_{ji}^L \mid \psi_{jt_{ji}}^L)F_2(\bm{x}_{ji}^G \mid \phi_{k_{jt_{ji}}}), 
    \end{array}
\end{equation}
with $L\leq T$, where $\bm{\beta}$ is the global vector of mixing proportions, $\bm{ {\pi}}_j$ is the group-specific vector of mixing proportions, $L$ is the number of global mixture components, and $T$ is the number of local mixture components. Note that the truncation level notation $L$ is overloaded and does not relate to the superscript denoting local variables (or factors).

As $L\rightarrow \infty$, the infinite limit of this model is precisely the proposed GLocal DP mixture model.
The proof is provided in the Section~\ref{appendix_proof} of the Supplementary Material.
 Based on this finite mixture model approximation with large enough truncation levels $L$ and $T$, we develop an efficient posterior inference procedure of our model using a Metropolis-within-blocked-Gibbs sampler in Section \ref{sec:posterior inference}. 
 
\section{Posterior Inference}\label{sec:posterior inference}
 Consider the hierarchical representation in \eqref{eq:glocalDPmixturemodel} and the corresponding finite mixture model in \eqref{eq:fmm_repr_1}. Let $\bm{x} = (\bm{x}_j)_{j=1}^{J}$ denote the observations from all $J$ groups. Similarly, $\bm{t} = (\bm{t}_j)_{j=1}^{J}$ and $\bm{k} = (\bm{k}_j)_{j=1}^{J}$ denote the collection of all latent indicators. The collection of atoms are denoted by $\bm{\psi} = (\bm{\psi}_j)_{j=1}^{J}$  and $\bm{\phi} = (\bm{\phi}_k)_{k=1}^{L}$, with $\bm{\psi}_j = \left(\psi_{jt}^L\right)_{t=1}^{T}$. Let $f_1(.\mid \psi_{jt}^L)$ and $f_2(.\mid \phi_{k})$ be the density functions (with respect to some dominating measure) corresponding to the distributions $F_1(.  \mid \psi_{jt}^L)$ and $F_2( . \mid \phi_{k})$,  respectively. The augmented likelihood is then given by,
\begin{align*}
    \nonumber p(\bm{x} , \bm{t}, \bm{k}\mid  \bm{\psi}, \bm{\phi},  ( {\bm{\pi}}_j)_{j=1}^{J}, \bm{\beta}) = \left\{\prod_{j=1}^{J}\prod_{i=1}^{n_j} f_1(\bm{x}_{ji}^L\mid \psi_{jt_{ji}}^L) f_2(\bm{x}_{ji}^G\mid \phi_{k_{jt_{ji}}}) \right\} \times \\
    \phantom{ p(\bm{x} , \bm{t}, \bm{k}\mid  \bm{\psi}, \bm{\phi},  ( {\bm{\pi}}_j)_{j=1}^{J}, \bm{\beta}) = \prod_{j=1}^{J}\prod_{i=1}^{n_j}} \prod_{j=1}^{J}\prod_{i=1}^{n_j} \prod_{t=1}^{T} {\pi}_{jt}^{\mathds{1}(t_{ji}=t)}\prod_{j=1}^{J}\prod_{t=1}^{T}\prod_{k=1}^{L}\beta_k^{\mathds{1}(k_{jt}=k)}.
\end{align*}
The model parameters are $\{\bm{\psi} , \bm{\phi}, ( {\bm{\pi}}_j)_{j=1}^{J}, \bm{\beta}, \alpha, \gamma \}$, with the joint prior distribution given by,
\small{
\begin{equation*}
    p(\bm{\psi} , \bm{\phi}, ( {\bm{\pi}}_j)_{j=1}^{J}, \bm{\beta}, \alpha, \gamma) = \left\{\prod_{j=1}^{J}\prod_{t=1}^{T} p(\psi_{jt}^L)\right\}  \left\{\prod_{k=1}^{L} p(\phi_k) \right\} \left\{\prod_{j=1}^{J} p( {\bm{\pi}}_j|\alpha)\right\} p(\bm{\beta}| \gamma) p(\alpha) p(\gamma).
\end{equation*}}
We remark that $L$ and $T$ are the \emph{maximal} numbers of global and local clusters specified by the users. They should be large enough so that the numbers of sampled clusters are always strictly smaller than them over the course of the MCMC. Picking the maximal number of clusters in our algorithm is much more straightforward than picking the \emph{exact} number of clusters in many existing clustering algorithms. 
The detailed MCMC algorithm to sample the model parameters from the joint posterior distribution is provided in the Section~\ref{supp-posterior_inference} of the Supplementary Material. After MCMC, we use the least-squares method \citep{LeastSquares} to obtain a point estimate of the clustering using the posterior samples. More precisely, let $\bm{z}^{(b)} = (z_{1}^{(b)}, \dots, z_{n}^{(b)})$ be the clustering of $n$ observations obtained from the posterior sample $b=1,\dots,M$. For each clustering $\bm{z}$ in $\bm{z}^{(1)}, \dots, \bm{z}^{(M)}$, let $\delta(\bm{z})$ be an $n \times n$ co-clustering matrix with the $(i,j)$th element $\delta_{i, j}(\bm{z}) = \mathds{1}(z_i = z_j)$, where $\mathds{1}$ is the indicator function. The element-wise averaging of these co-clustering matrices yields the pairwise probability matrix of co-clustering, denoted by $\widehat{\bm{\Pi}}$. Then, the least squares point estimate of clustering is given by, 
\begin{equation}
\label{eq:Least Square Estimate}
    \bm{\widehat{z}}_{LS} = \arg \min_{\bm{z} \in \{\bm{z}^{(1)}, \dots, \bm{z}^{(M)}\}} \sum_{i=1}^{n} \sum_{j=1}^{n} \left( \delta_{i,j}(\bm{z}) - \widehat{\Pi}_{i,j} \right)^2.
\end{equation}
The proposed GLocal DP consists of the global-level and local-level clusters. The estimated global-level clusters are obtained using the least squares method by concatenating the global-level cluster labels across all groups $j = 1, \dots, J$ for each posterior sample. More precisely, the global-level clustering is obtained by considering $\bm{z}^{(b)} = (k_{1t_{11}}^{(b)}, \dots, k_{1t_{1n_1}}^{(b)}, k_{2t_{21}}^{(b)}, \dots, k_{Jt_{Jn_J}}^{(b)})$ in \eqref{eq:Least Square Estimate}. This ensures that global-level clusters are shared across groups. The point estimate of local-level clusters for group $j$ is obtained by considering $\bm{z}^{(b)} = (t_{j1}^{(b)}, \dots, t_{jn_j}^{(b)})$ in \eqref{eq:Least Square Estimate}, which provides the fine sub-clustering of the global-level clusters for group $j$. Furthermore, since GLocal DP reduces to HDP in the absence of local variables, our MCMC algorithm can be used for HDP sampling by simply setting $f_1(\ .\mid \psi_{jt}^L) = 1$ and skipping the sampling of $\psi_{jt}^L$ 
in Algorithm~\ref{alg:the_algorithm} in the Section~\ref{supp-posterior_inference} of the Supplementary Material. The blocked Gibbs sampler obtained as a by-product from Algorithm~\ref{alg:the_algorithm}, relies on the finite truncation and utilizes two latent indicators $t_{ji}$ and $k_{jt}$ to specify the underlying mixture component associated with the observation $x_{ji}$ by $z_{ji} = k_{jt_{ji}}$, which is a novel contribution to the HDP sampling algorithms. Contrarily, \citealp{das_etal} proposes a blocked Gibbs sampler relying on the finite truncation of HDP and uses one latent indicator, $z_{ji}$ in specifying the underlying mixture component associated with the observation $x_{ji}$. However, the blocked Gibbs sampler for HDP by \citealp{das_etal} cannot be extended to the GLocal DP, the reason for which is given in the Section~\ref{subsec:GLocal_HDP_algorithm_diff} of the Supplementary Material.

\section{Pan-Gastrointestinal Cancer Data Analysis}\label{sec:Real_data_analysis}
We recall that the motivation of the proposed GLocal DP stems from pan-cancer genomics. Integrated clustering analyses across cancers can objectively identify cancer subpopulations potentially beyond the tumor site of origin, which would improve our understanding of both within-tumor and between-tumor heterogeneity and potentially repurpose existing cancer treatments from one tumor site to another \citep{Schein2021, Rodrigues2022}. In databases like TCGA, genomic data are often accompanied by clinical data, providing largely orthogonal information regarding tumor heterogeneity, and some clinical data may be cancer-specific. Although there exist methods for clustering grouped data, they can only utilize a common set of variables and hence would have to discard important cancer-specific clinical variables. In this application, we aim to identify pan-cancer subpopulations using both shared and cancer-specific data in a coherent manner through GLocal DP. 
\par As mentioned in Section \ref{sec:mapgc}, we considered four cancers of the GI tract, i.e., esophageal, stomach, colon, and rectal cancer. We obtained the gene expression data for the four cancers from the publicly available TCGA database \citep{UCSC_Xena} along with their clinical data. The datasets consist of the log-transformed gene expression measurements for a common set of 60,483 genes in patients with the corresponding tumors. The gene expression data are available from 173, 407, 512, and 177 patients for esophageal, stomach, colon, and rectal cancer, respectively. The selection of clinical variables to include in our analysis is explained in the following. First, smoking has been identified as a major risk factor for esophageal cancer \citep{fan2008alcohol}. Second, CEA is an important prognostic marker for monitoring tumor progression in colorectal cancer. However, CEA is not collected for esophageal and stomach cancers. Third, recent studies have shown that several common cancers including colon cancer have been linked to obesity \citep{Cancer_obesity}. 
According to \citealp{BMI_colon}, measuring BMI is crucial for assessing the obesity-related risk of developing colon cancer. In conclusion, such scientifically relevant aspects led us to consider the number of cigarettes smoked per day as a local variable for esophageal cancer, pre-operative and pre-treatment CEA as the local variable for both colon and rectal cancers, and BMI as an additional variable specific to colon cancer. Note that no local variable was used for stomach cancer. Finally, we only considered patients having a non-missing clinical data for downstream analysis. This led to sample sizes of 92, 407, 173, and 120 for esophageal, stomach, colon, and rectal cancer respectively.\par 
Following the common practice, we performed UMAP on the combined gene expression data from the four cancers to reduce the data to two dimensions on a common manifold. The uniform manifold approximation and projection (UMAP, \citealp{McInnes2018}) has been a common practice for dimension reduction in many downstream analyses for genomic data \citep{luecken2019current, tonkin2019fast,  leelatian2020unsupervised, diaz2021review, zhang2021critical, app12094247}. Furthermore, a recent comparative study showed that UMAP considerably improved the performances of clustering algorithms \citep{10.1007/978-3-030-51935-3_34}. Aligning with the recommendations by \citealp{McInnes2018}, we tuned the hyperparameters of the UMAP algorithm such that the lower dimensional embeddings capture the global structure in the high-dimensional genomic data without losing the finer local features.

Our goal was to cluster the global variables (UMAP embeddings of gene expression data), while allowing the cancer-specific  clinical variables to aid in clustering. We considered truncation level $L = T = 20$ and the following sampling distributions, 
\begin{align*}
&F_1(\bm{x}_{ji}^L\mid \bm{\theta}_{ji}^L):=\mathcal{N}_{p_j}(\bm{x}_{ji}^L\mid \bm{\mu}_{jt_{ji}}, \sigma_{jt_{ji}}^2 \mathds{I}_{p_j}),\\
&F_2(\bm{x}_{ji}^G\mid \bm{\theta}_{ji}^G):=\mathcal{N}_{2}(\bm{x}_{ji}^G\mid \bm{\mu}_{k_{jt_{ji}}}, \sigma_{k_{jt_{ji}}}^2 \mathds{I}_{2}),
\end{align*}
where $\mathds{I}_{2}$ is a $2 \times 2$ identity matrix and $p_j$ is the dimension of the local variables in the population $j$ (i.e., $p_1 = 0, p_2 = 1$, $p_3 = 2$, and $p_4 = 1$ for stomach, esophageal, colon, and rectal cancers, respectively).
For hyperpriors, we assume
$\bm{\mu}_{jt_{ji}}\mid \sigma_{jt_{ji}}^2 \sim \mathcal{N}_{p_j}(0,\sigma_{jt_{ji}}^2 \mathds{I}_{p_j})$, $\bm{\mu}_{k_{jt_{ji}}} \mid \sigma_{k_{jt_{ji}}}^2\sim \mathcal{N}_2(0,\sigma_{k_{jt_{ji}}}^2 \mathds{I}_{2})$, and $\sigma_{jt_{ji}}^2,\sigma_{k_{jt_{ji}}}^2\, \alpha^{-1},\gamma^{-1}\sim \mathcal{IG}(0.1, 0.1)$. 

We have also considered a more general case (results not shown) where the covariance matrices have unequal variances (whenever applicable), which, however, did not yield a higher marginal likelihood than the simpler model. Furthermore, we have considered several choices of the dimension of UMAP embeddings and the truncation level of the GLocal DP, which shows our method is relatively robust; see Section~\ref{supp-real_data_analysis} of the Supplementary Material for details.

We ran $100,000$ iterations of our sampler, which took $<10$ minutes on a MacBook Pro with M1 chip and 16GB RAM. We discarded the first $25,000$ iterations as burn-in and retained every $75$th iteration of posterior samples. The traceplot of the log-posterior and autocorrelation function (ACF) plot are shown in the Figure~\ref{fig:LL_ReadData} and Figure~\ref{fig:ACF_ReadData}, respectively of the Supplementary Material. These plots do not show lack of convergence or poor mixing. Additionally, the traceplots of the concentration parameters, $\alpha$ and $\gamma$ are provided in the Figure~\ref{fig:Traceplots_concentration} of the Supplementary Material which also show good mixing.

\begin{figure}[!htp]
\centering
\begin{subfigure}{0.75\textwidth}
  \centering
    \includegraphics[width= 1\linewidth]{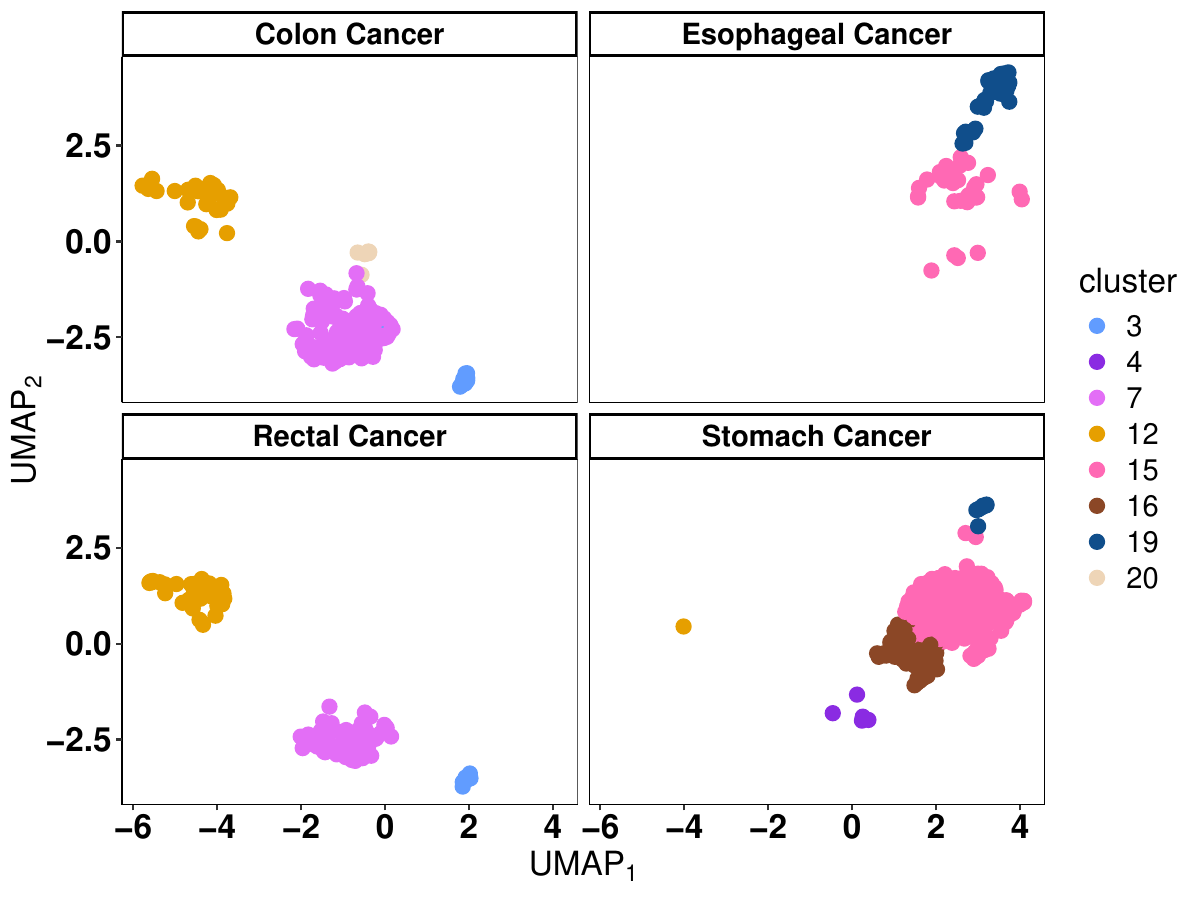}
    \caption{Global-level clusters.}
    \label{fig:global_level_clustering_RealData}
\end{subfigure}\par\bigskip
\begin{subfigure}{0.75\textwidth}
    \centering
    \includegraphics[width= 1\linewidth]{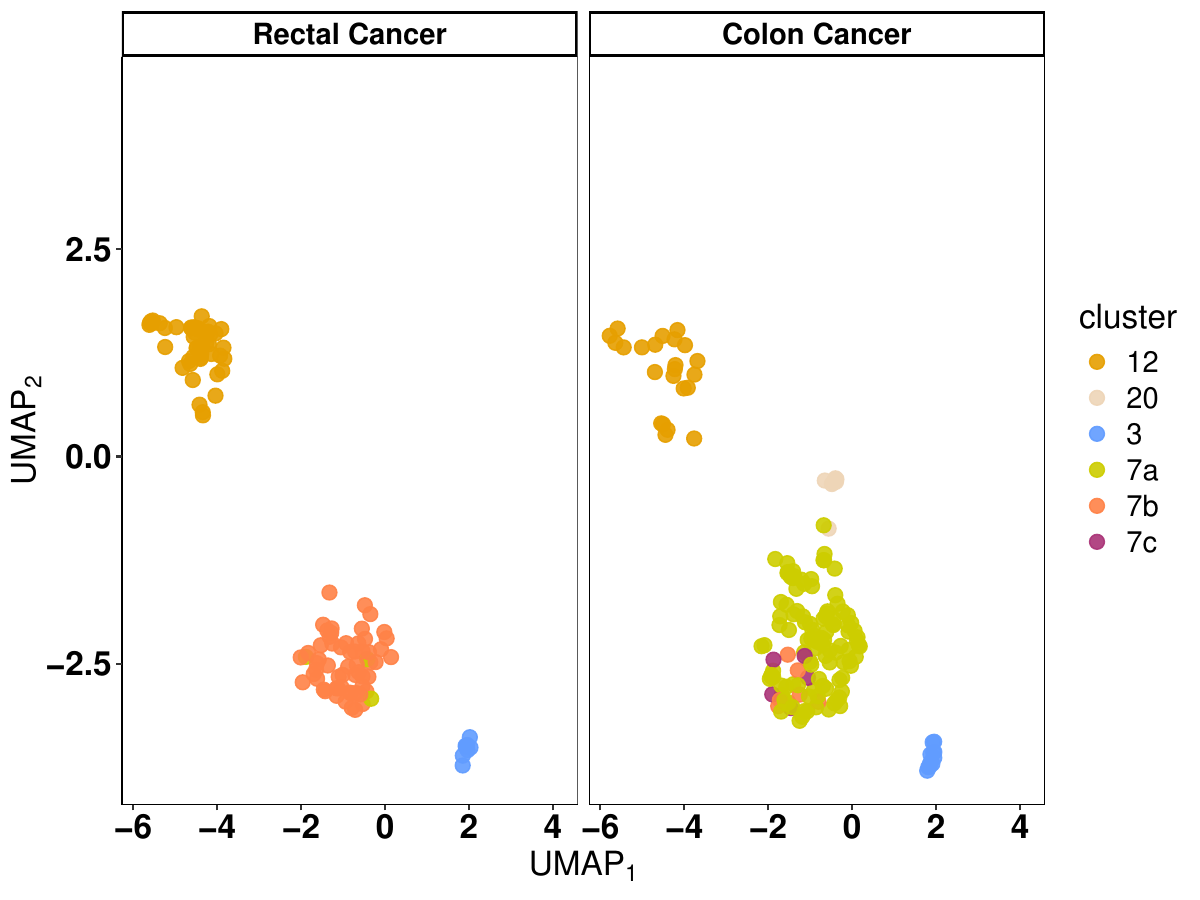}
    \caption{Local-level clusters.}
    \label{fig:local_level_clustering_RealData}
\end{subfigure}
\caption{Global variables. (a) The colors indicate global-level clusters estimated from GLocal DP. (b) The colors indicate the estimated local-level clusters.}
\label{fig:GLOCAL_clustering_RealData}
\end{figure}

\begin{figure}[!htp]
\centering
    \includegraphics[width= 0.7\linewidth]{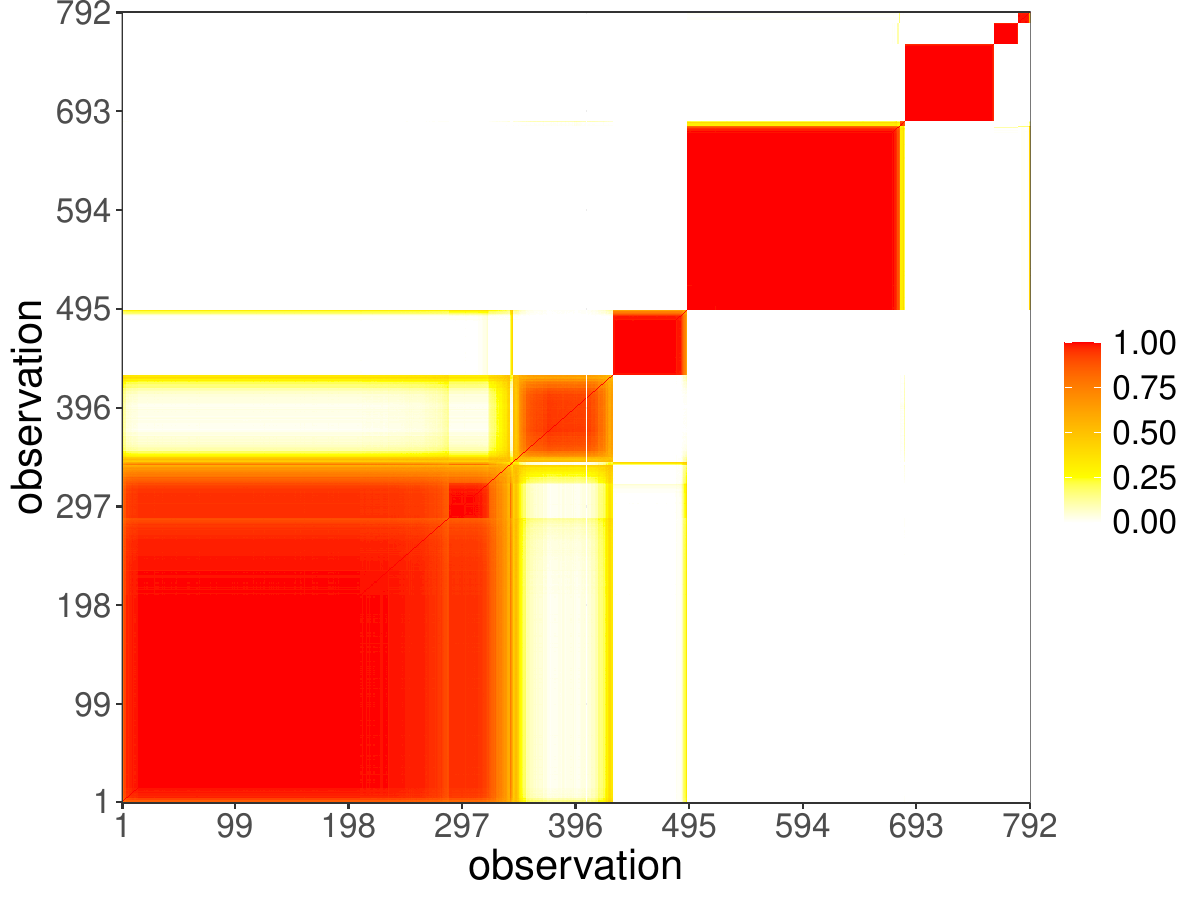}
\caption{Posterior co-clustering probabilities of observations assigned to global-level clusters.}
\label{fig:HeatmapGlobal}
\end{figure}

Figure \ref{fig:global_level_clustering_RealData} shows the global-level clusters obtained by the least-squares criterion applied to the posterior samples for all cancers. Figure \ref{fig:local_level_clustering_RealData} shows the local-level clusters for rectal and colon cancers. The heatmaps of the posterior co-clustering probabilities for both the global-level and local-level clustering are shown in Figure~\ref{fig:HeatmapGlobal} and Figure~\ref{fig:HeatmapLocal} respectively, in the Section~\ref{supp-real_data_analysis} of the Supplementary Material. Rectal and colon cancers were found to share three major global clusters, which is consistent with the known fact that these two cancers are similar to each other \citep{cancergenomeatlas2012comprehensive, CRC_similar}. 
 However, colon cancer has a unique subpopulation (cluster 20) that is not found in rectal cancer, for which the patients have moderate to high BMI. 
The two upper-GI cancers, stomach cancer and esophageal cancer, share some similarities through two shared clusters but are quite distinct from the lower-GI cancers. As stomach cancer has no local variable, it does not have local-level clusters.  Furthermore, the local variable of esophageal cancer did not generate sub-clusters. In summary, from the molecular point of view, there is little similarity between the upper-GI cancers and the lower-GI cancers whereas within the upper-GI cancers or within the lower-GI cancers, subpopulations may be defined beyond the tumor site of origin, more prominently within the lower-GI cancers.

\begin{figure}[!htp]
\centering
\includegraphics[width = 0.8\linewidth]{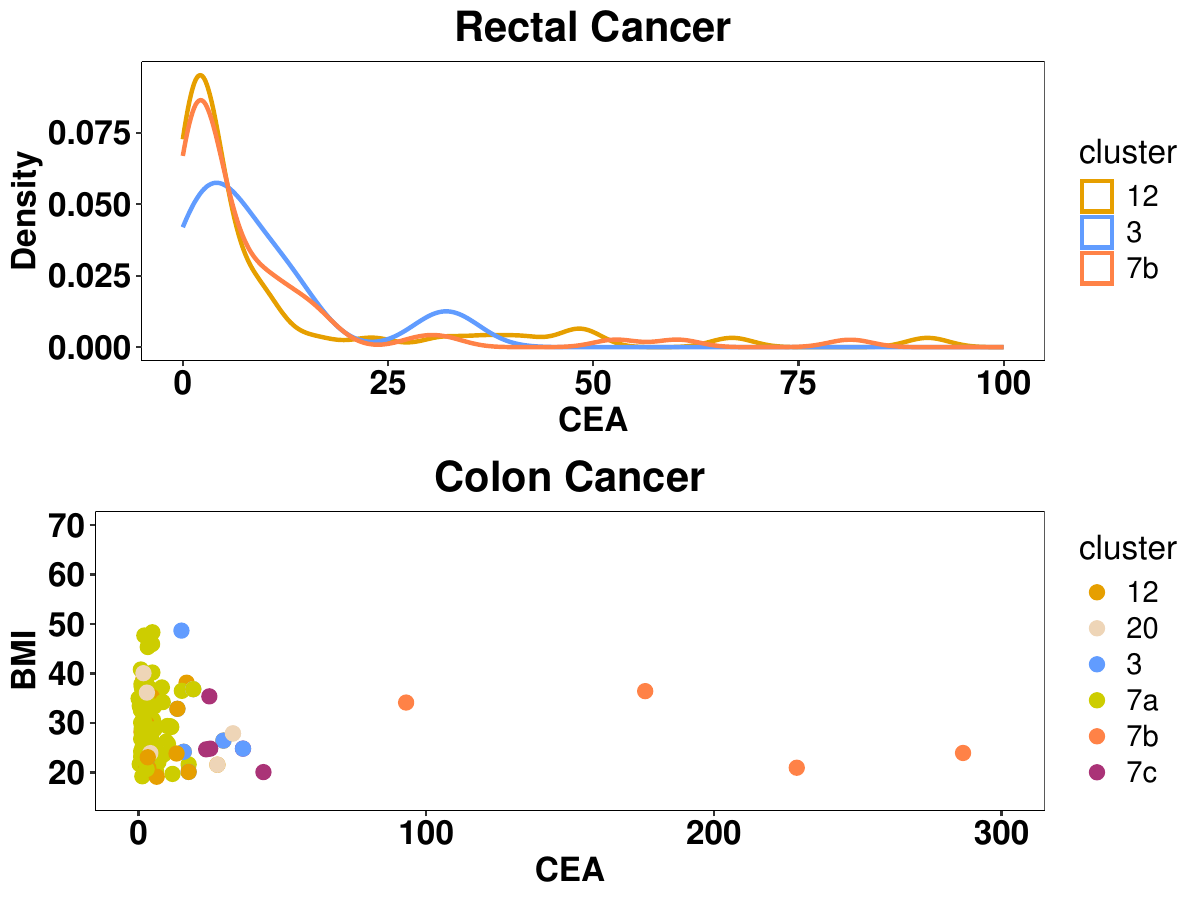}
\caption{Kernel density/scatter plot of local variables for rectal and colon cancers, colored by the estimated local-level clusters. 
}
\label{fig:Local_Variable_clustering_RealData}
\end{figure}

\begin{figure}[!htp]
\centering
\begin{subfigure}{0.95\textwidth}
  \centering
    \includegraphics[width= 1\linewidth]{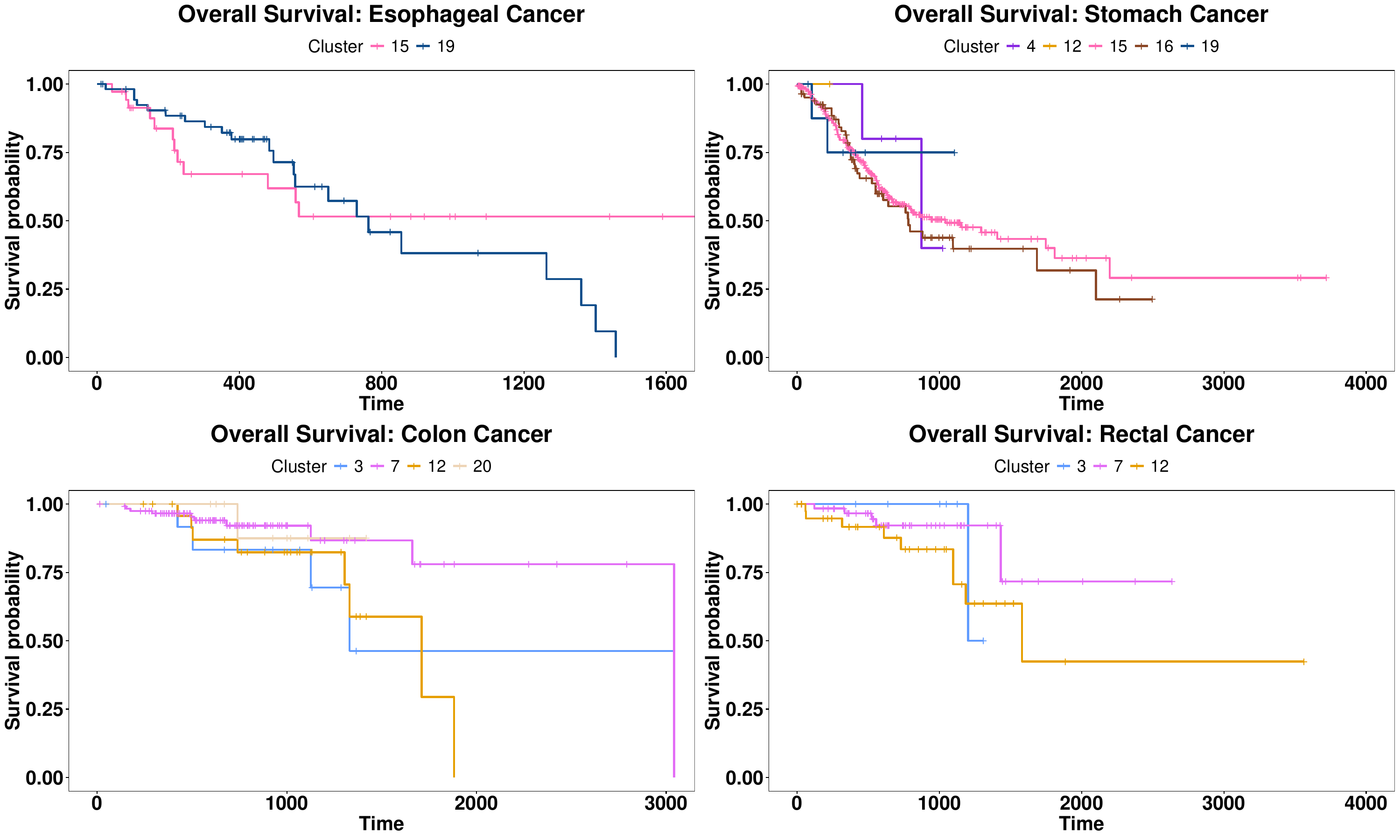}
    \caption{Survival curves according to global-level clusters.}
    \label{fig:global_KM_RealData}
\end{subfigure}\par\bigskip
\begin{subfigure}{0.95\textwidth}
    \centering
    \includegraphics[width= 1\linewidth]{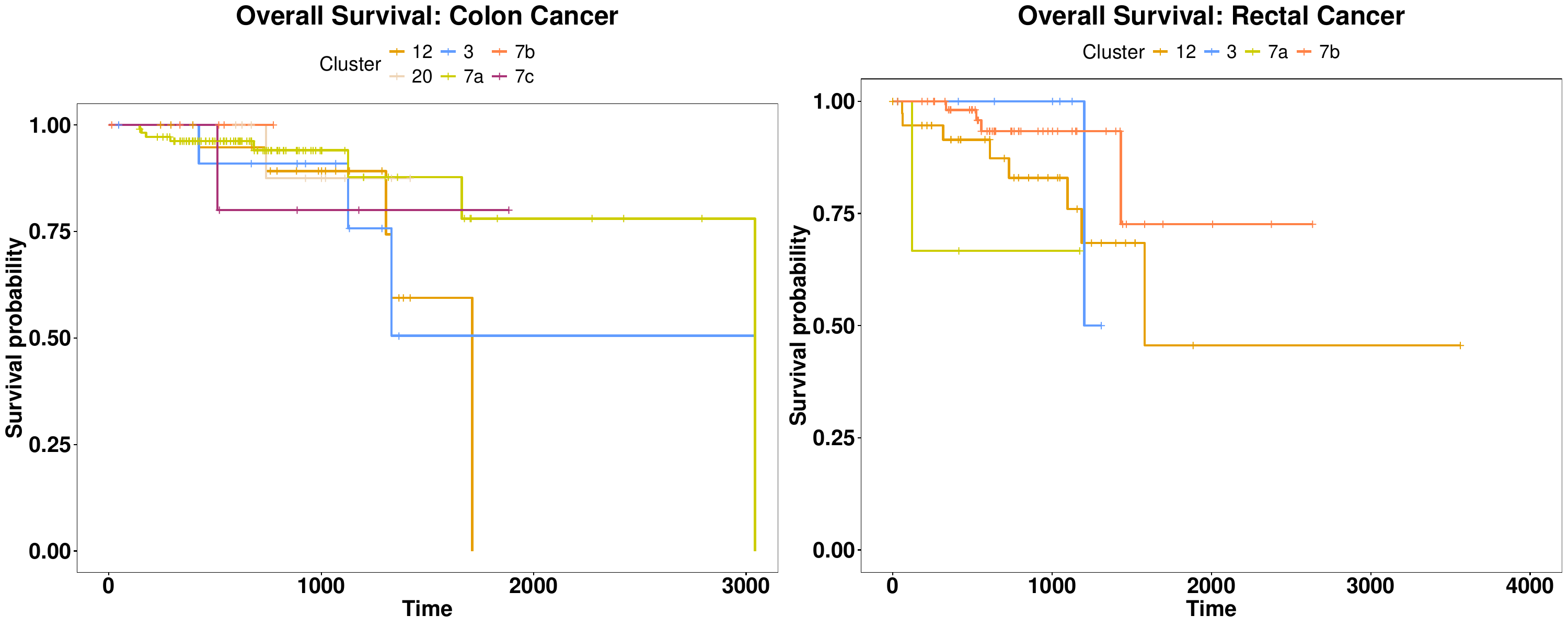}
    \caption{Survival curves according to local-level clusters.}
    \label{fig:local_KM_RealData}
\end{subfigure}
\caption{Kaplan-Meier survival curves by clusters estimated from GLocal DP.}
\label{fig:GLOCAL_KM_RealData}
\end{figure}

The local-level clusters (Figure \ref{fig:local_level_clustering_RealData})
refine the global-level clusters by utilizing the clinical information.  
Figure \ref{fig:Local_Variable_clustering_RealData} shows how the local-level clusters are influenced by the local variables. For example, colon cancer patients having high BMI or high CEA are labeled as clusters 7a and 7b, respectively, which cannot be identified with gene expression data alone. Moreover, the shared sub-clusters 7a and 7b between colon and rectal cancers correspond to patients with low and high preoperative CEA, respectively. To understand if the identified cancer subpopulations possibly inform cancer prognosis, we plotted the Kaplan-Meier survival curves for each of the identified cancer subpopulations in Figure \ref{fig:GLOCAL_KM_RealData}. The global-level cluster-specific survival curves (Figure \ref{fig:global_KM_RealData}) corresponding to the two major clusters for stomach cancer, i.e., clusters 15 and 16 exhibit some difference. In particular, the median survival time corresponding to cluster 15 is higher (1043) than that for cluster 16 (782). This highlights the possibility of some scientific connection between gene expression and the prognosis of cancer. By itself, it may be of scientific interest to understand the prognosis of cancer for patients having a particular gene expression and their response to cancer therapy. For esophageal cancer, the survival curves show even more significant differences. Particularly, patients in cluster 19, show a rapid decline in overall survival following a higher initial survival probability in comparison to the patients in cluster 15. 
Several studies have used preoperative CEA as a prognostic marker of colorectal cancer, with high preoperative CEA levels predicting poor overall survival and increased risk of recurrence \citep{dekker2019colorectal, sung2021global}. Accordingly, the local-level cluster-specific survival curves in Figure \ref{fig:local_KM_RealData} provide valuable insights into the prognosis of colorectal cancer with respect to preoperative CEA levels. For example, colon cancer patients belonging to cluster 7b in Figure \ref{fig:local_level_clustering_RealData} have extremely high CEA levels (median = 138 ng/mL) and high BMI (35.3). Their overall survival is shorter than that of patients in cluster 7a who have lower CEA (median = 2.75 ng/mL) and comparatively lower BMI (28.7). These findings are concurrent with existing scientific knowledge of high CEA and high BMI values being significant markers indicating poor cancer prognosis \citep{highCEA_colon, CEA_Rectal, BMI_colon}. Clustering based on gene expression data alone cannot discern the tumor heterogeneity from the prognostic perspective.

\begin{figure}[!htp]
\centering
\begin{subfigure}{0.49\textwidth}
  \centering
    \includegraphics[width= 1\linewidth]{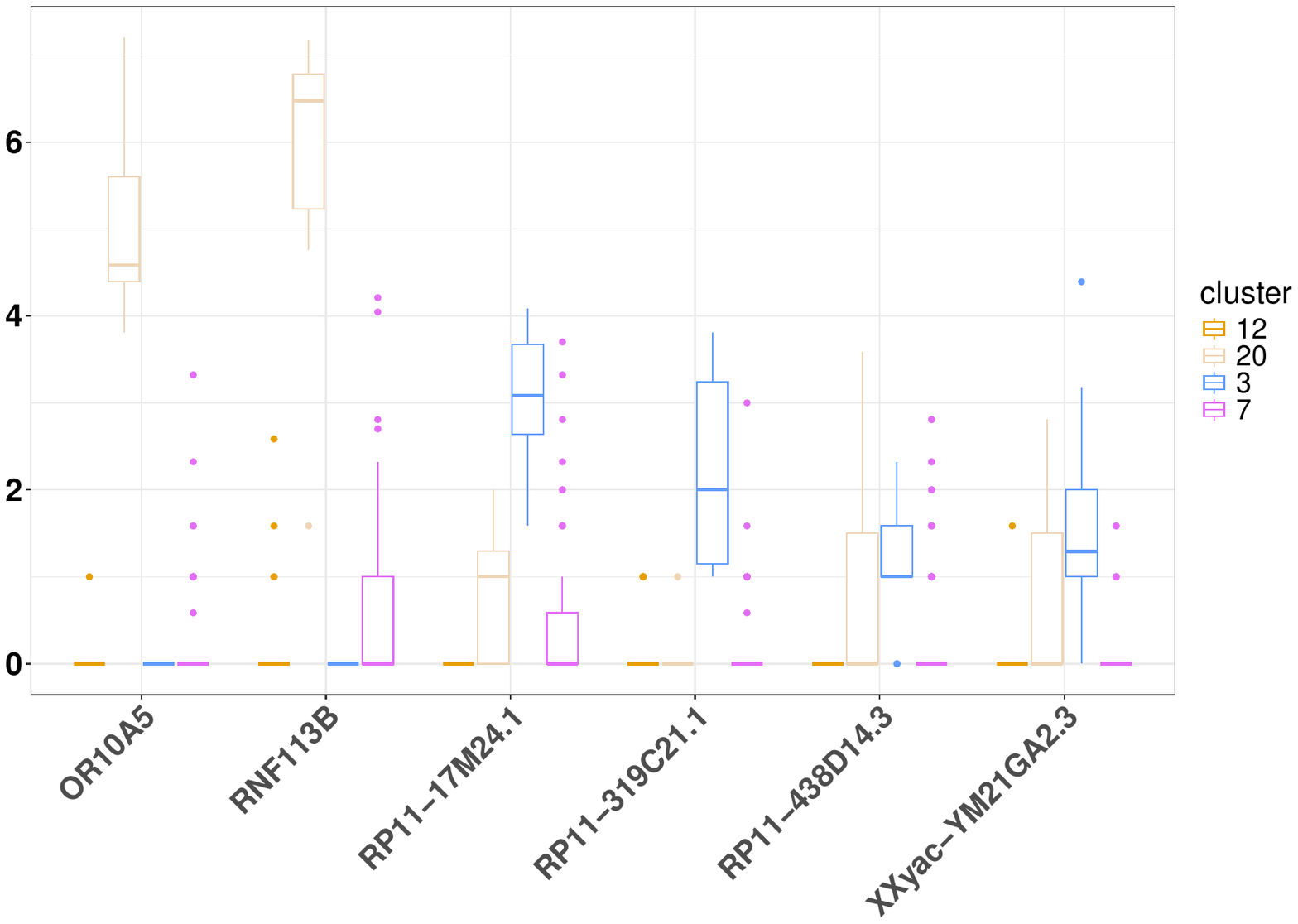}
    \caption{Colon}
    \label{fig:BoxplotDE_colon}
\end{subfigure}
\begin{subfigure}{0.49\textwidth}
    \centering
    \includegraphics[width= 1\linewidth]{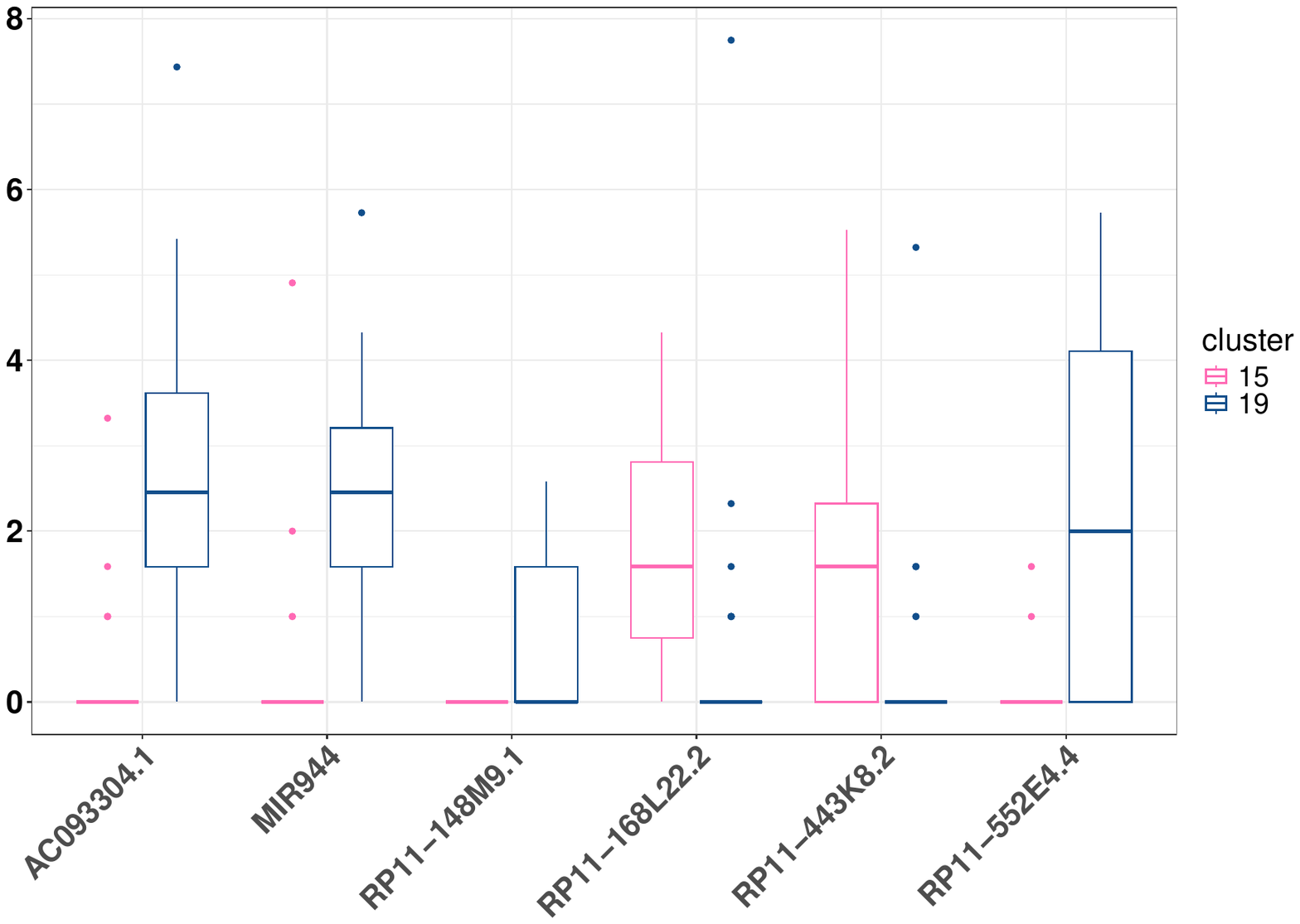}
    \caption{Esophageal}
    \label{fig:BoxplotDE_eso}
\end{subfigure}
\par
\begin{subfigure}{0.49\textwidth}
    \centering
    \includegraphics[width= 1\linewidth]{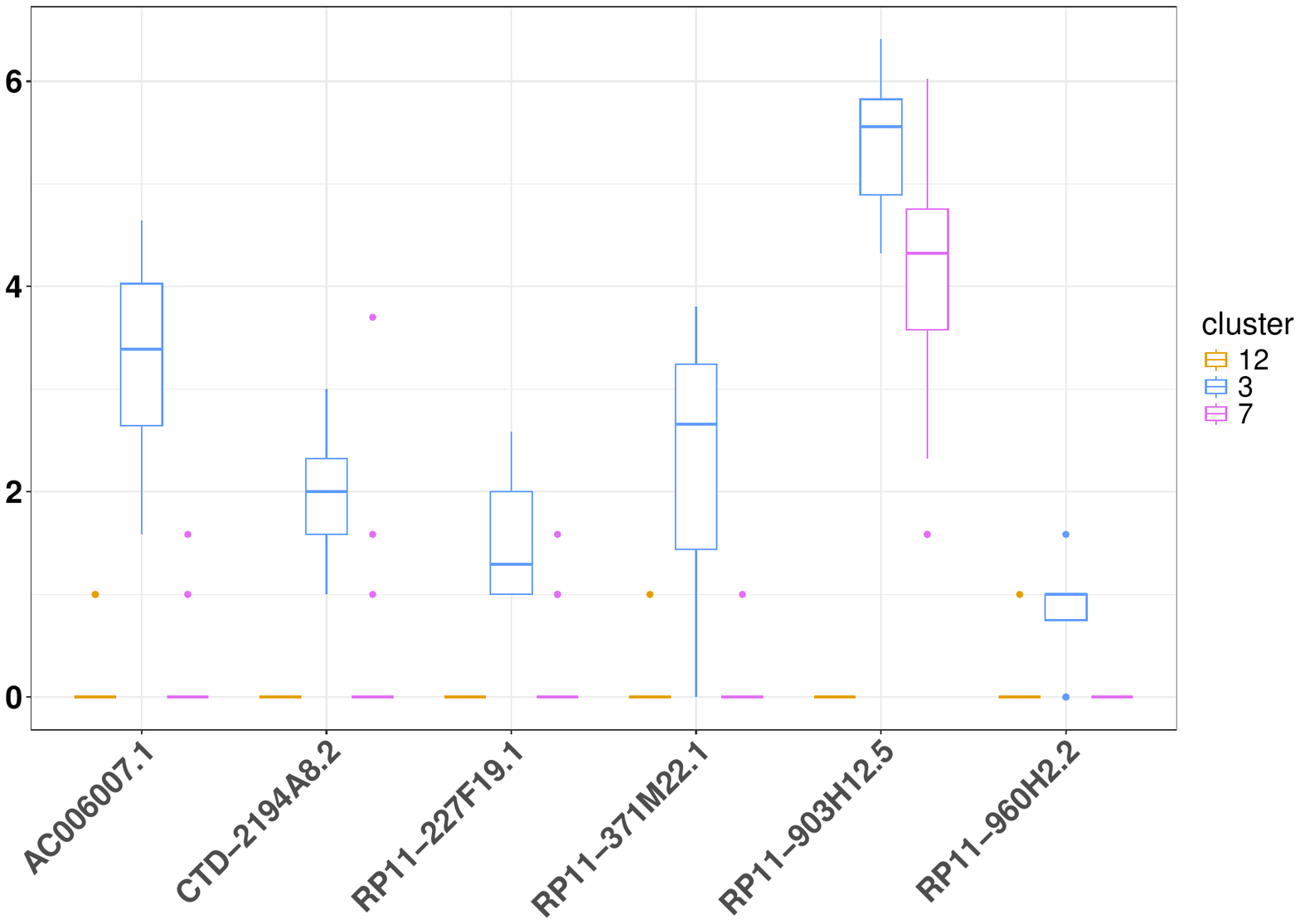}
    \caption{Rectal}
    \label{fig:BoxplotDE_rectal}
\end{subfigure}
\begin{subfigure}{0.49\textwidth}
    \centering
    \includegraphics[width= 1\linewidth]{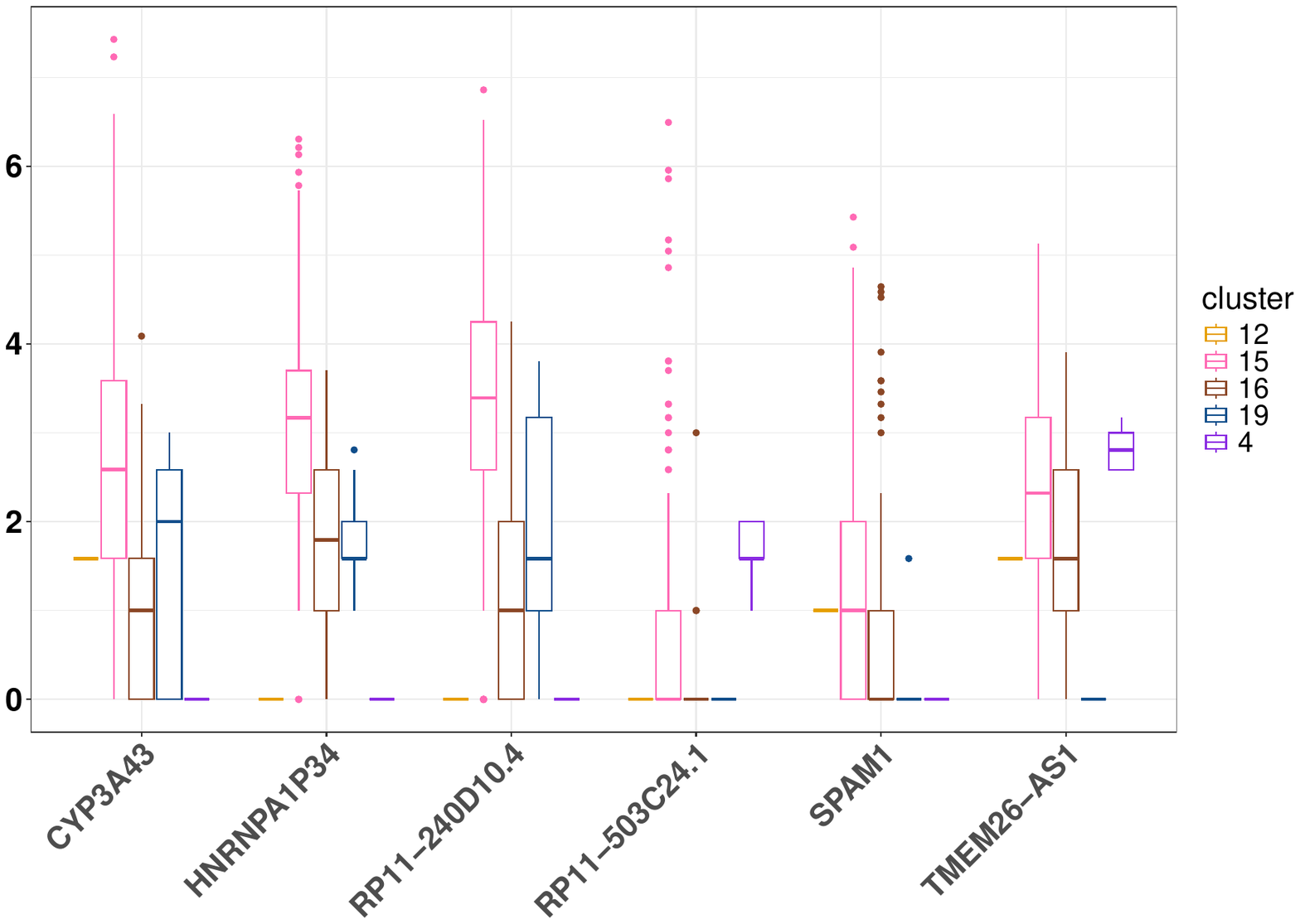}
    \caption{Stomach}
    \label{fig:BoxplotDE_stomach}
\end{subfigure}
\caption{Boxplot of gene expressions in the top 6 DE genes for each cancer in the different global-level clusters.}
\label{fig:BoxplotDE_Global}
\end{figure}

\begin{figure}[!htp]
\centering
\begin{subfigure}{0.49\textwidth}
  \centering
    \includegraphics[width= 1\linewidth]{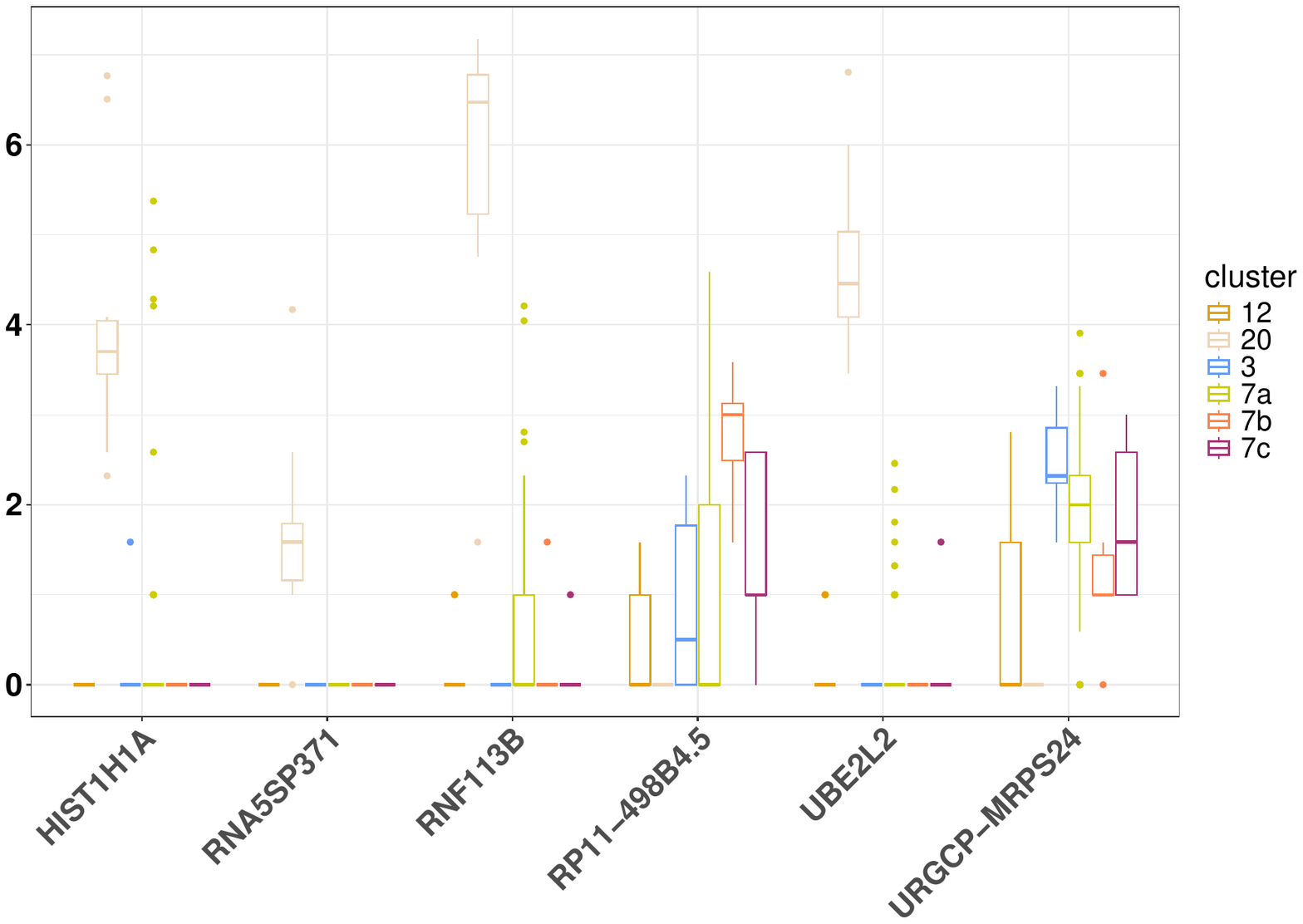}
    \caption{Colon}
    \label{fig:BoxplotDE_colon_local}
\end{subfigure}
\begin{subfigure}{0.49\textwidth}
    \centering
    \includegraphics[width= 1\linewidth]{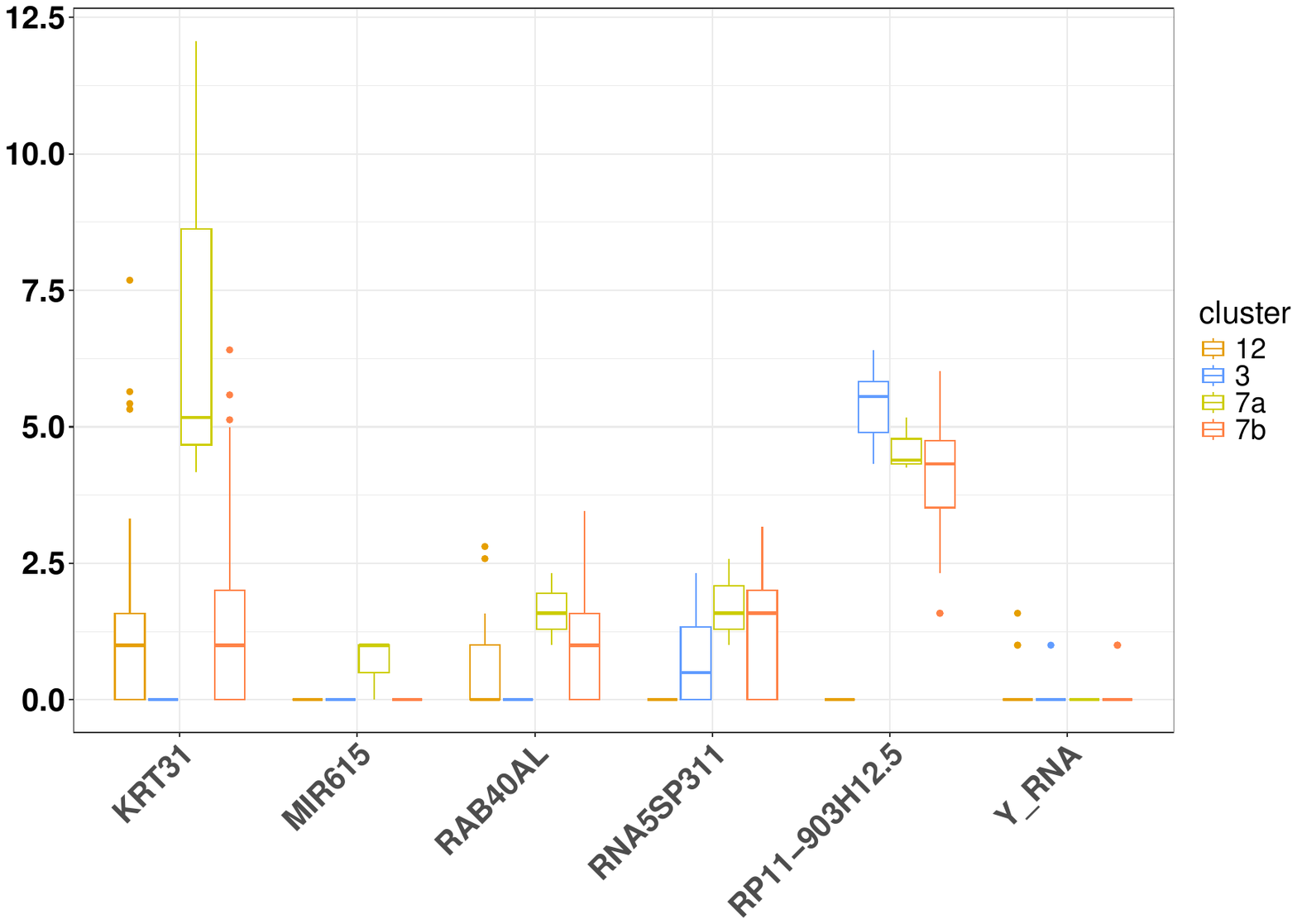}
    \caption{Rectal}
    \label{fig:BoxplotDE_rectal_local}
\end{subfigure}
\caption{Boxplot of gene expressions in the top 6 DE genes for (a) colon and (b) rectal cancers in the different local-level clusters.}
\label{fig:BoxplotDE_Local}
\end{figure}

After estimating the global- and local-level clusters, we identified the genes that best characterize the clusters. In particular, we identified the 6 most differently expressed (DE) genes for each cancer using the function \texttt{findMarkers} of the Bioconductor package \texttt{scran} \citep{scran} characterizing the global- and local-level clusters separately. Figure \ref{fig:BoxplotDE_Global} shows the distribution of the DE genes characterizing the global-level clusters for the different cancers. Figure \ref{fig:BoxplotDE_Local} shows the distribution of the DE genes characterizing the local-level clusters for colon and rectal cancers, which is different from the DE genes characterizing the global-level clusters. For example, Table \ref{tab:DE_gene_colon} shows the average within-cluster gene expressions corresponding to the global- and local-level clusters in the selected top 6 biomarkers for colon cancer. The gene RP11-498B4.5 shows higher mean expression, i.e., upregulation in cluster 7b (corresponding to patients with extremely high CEA values and BMI) in comparison to cluster 7a (corresponding to patients with low CEA values and lower BMI). This gene is a member of the \emph{Heat shock 70kDa protein 12A (HSPA12A)} class of genes. Recently, HSPA12A has been identified as a key driver in colorectal cancer \citep{Lu2023}. The upregulation of HSPA12A could significantly increase endothelial cell proliferation rates in colorectal cancer, which in turn is reflected by the patients' high CEA values. In summary, our analysis can be used to identify possible marker genes for the explanation of clinical characteristics of patients. 

\begin{table}[!htp]
\centering
\scalebox{0.85}{
\begin{tabular}[t]{c|c|cccccc}
\hline
Cluster-level & DE Genes & Cluster 3 & & Cluster 7 & & Cluster 12 & Cluster 20\\
\hline
 \multirow{6}*{Global} &RNF113B & 0.000 & &0.581 && 0.229 & 5.733\\
& XXyac-YM21GA2.3 & 1.624 && 0.059 && 0.059 & 0.739\\

& RP11-17M24.1 & 3.114 && 0.361 && 0.000 & 0.834\\

& OR10A5 & 0.000 && 0.144 && 0.037 & 4.978\\

& RP11-319C21.1 & 2.181 && 0.109 && 0.185 & 0.182\\

& RP11-438D14.3 & 1.148 && 0.274 && 0.000 & 0.901\\
\cline{2-8}
 \multirow{6}*{Local}  & DE Genes & Cluster 3 & Cluster 7a & Cluster 7b & Cluster 7c & Cluster 12 & Cluster 20 \\
\cline{2-8}
& HIST1H1A & 0.132 & 0.243 & 0.000 & 0.000 & 0.000 & 4.034\\
& UBE2L2 & 0.000 & 0.210 & 0.000 & 0.317 & 0.130 & 4.714 \\
& RNA5SP371 & 0.000 & 0.000 & 0.000 & 0.000 & 0.000 & 1.674 \\
& RP11-498B4.5 & 0.928 & 1.089 & 2.777 & 1.434 & 0.373 & 0.000 \\
& URGCP-MRPS24 & 2.466 & 1.792 & 1.341 & 1.834 & 0.782 & 0.000 \\
& RNF113B & 0.000 & 0.618 & 0.264 & 0.200 & 0.087 & 5.733 \\
\hline
\end{tabular}}
\caption{Average within-cluster gene expressions corresponding to global- and local-level clusters in the selected top 6 biomarkers for colon cancer.}
\label{tab:DE_gene_colon}
\end{table}

\section{Simulations}\label{sec:simulations}
Throughout the simulations, we assumed that there were three groups or populations.  
\subsection{Local variables available for all populations}\label{subsec:all_Local}
First, we considered a simulation setting in which all three populations had local variables. Specifically, there were one, two, and three local variables for populations 1, 2, and 3, respectively. We generated the data from, 
\begin{equation*}
      \bm{x}_{ji} \sim \left\{f_1(\bm{x}_{ji}^L\mid \psi_{jt}^L) f_2(\bm{x}_{ji}^G\mid \phi_{k}) \right\},
 \end{equation*}
where, 
 \begin{align*}
     f_1(\bm{x}_{ji}^L\mid \psi_{jt}^L) & = \sum_{t=1}^{L_{\ell_j}}  {\pi}_{jt}\mathcal{N}_{p_j}(\bm{x}_{ji}^L\mid \bm{\mu}_{jt}, \Sigma_{jt}),\\
     f_2(\bm{x}_{ji}^G\mid \phi_{k}) & = \sum_{k=1}^{L_g} \beta_k\mathcal{N}_{2}(\bm{x}_{ji}^G\mid \bm{\mu}_{k}, \Sigma_{k}),
 \end{align*}
\noindent with the diagonal matrices $\Sigma_{jt} =\text{Diag}(\sigma_{j1t}^2, \dots, \sigma_{jp_jt}^2)$ and $\Sigma_{k} =\text{Diag}(\sigma_{k1}^2, \sigma_{k2}^2)$. Here $p_1=1, p_2 = 2, p_3 = 3, L_{\ell_1} = 6, L_{\ell_2} = 7, L_{\ell_3} = 5$, and $L_g = 8$. The true parameters and the true mixture weights corresponding to the local variables are drawn from,\begin{align}\label{eq:simulation_local_parameter_generation}
     \sigma_{jlt}^2 & \sim \mathcal{IG}(2, 1), &\mu_{jlt} &\sim \mathcal{N}(0, \lambda_L^{-1}\sigma_{jlt}^2)\\
    \label{eq:simulation_local_mixture_weight_generation}
     \alpha & \sim Gamma(25, 1),  &\boldsymbol{ {\pi}}_j &\sim \mbox{Dir}(\alpha/L_{\ell_j}, \dots, \alpha/L_{\ell_j}),
 \end{align}
  for $j = 1, 2, 3$, $l = 1, \dots , p_j$, and  $t = 1, \dots, L_{\ell_j}$. 
The true local indicator $t_{ji}$ is drawn from a multinomial distribution with class probabilities $\boldsymbol{ {\pi}}_j$, for $j = 1, 2, 3$. Similarly, the true parameters and mixture weights corresponding to the global variables are drawn from,
 \begin{align}
 \label{eq:simulation_global_parameter_generation}
     \sigma_{kl}^2 & \sim \mathcal{IG}(2, 1),  &\mu_{kl} & \sim \mathcal{N}(0, \lambda_G^{-1}\sigma_{kl}^2),\\ 
     \label{eq:simulation_global_mixture_weight_generation}
     \gamma & \sim Gamma(25, 1),  &\boldsymbol{\beta}  & \sim \mbox{Dir}(\gamma/L_g, \dots, \gamma/L_g),
 \end{align}
 for $l = 1, 2$ and $k = 1, \dots , L_g$.
 The true latent indicator $k_{jt}$ is drawn from a multinomial distribution with the class probabilities $\boldsymbol{\beta}$, for $t = 1, \dots, L_g$. We considered the following sample sizes for the three populations, $n_1 = 100, n_2 = 110,$ and $n_3 = 115$.

\begin{figure}[!htp]
\centering
\begin{subfigure}{0.8\textwidth}
  \centering
    \includegraphics[width= 1\linewidth]{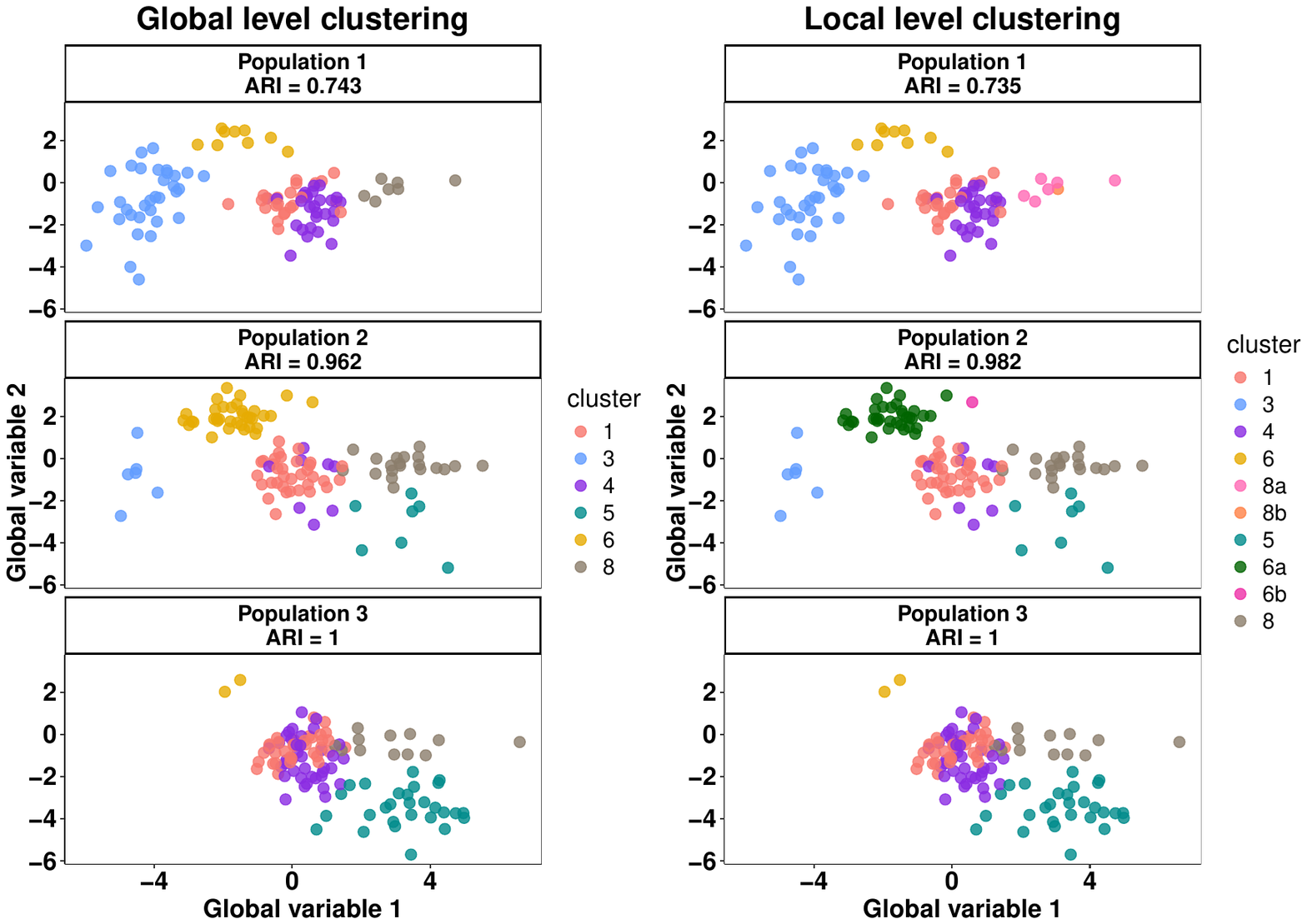}
    \caption{Global variables with global and local-level cluster labels.}
    \label{fig:global_clustering_overlapped}
\end{subfigure}
\begin{subfigure}{0.8\textwidth}
    \centering
    \includegraphics[width= 1\linewidth]{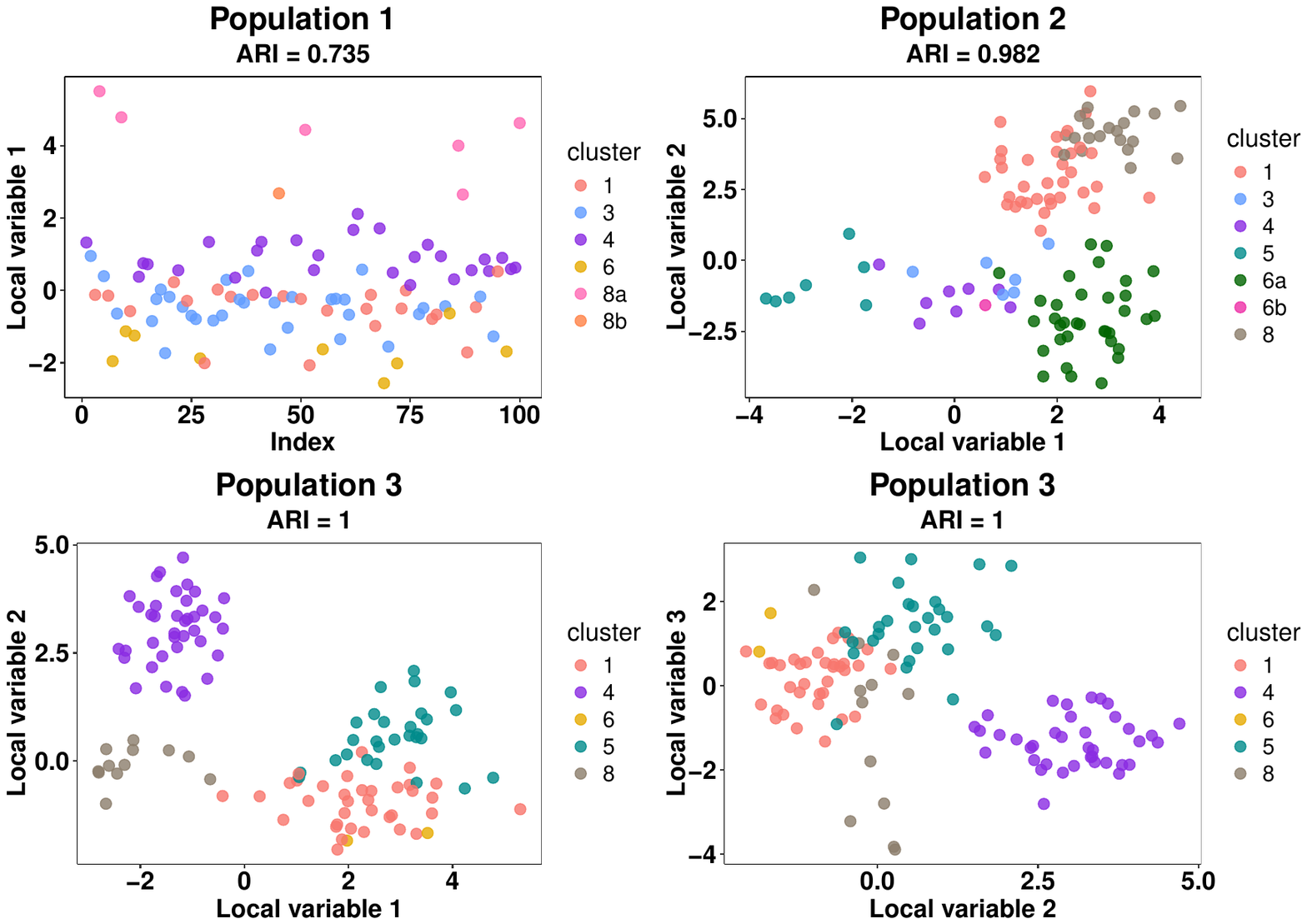}
    \caption{Local variables with local-level cluster labels.}
    \label{fig:local_clustering_overlapped}
\end{subfigure}
\caption{Clustering performance of GLocal DP when both the global and local variables are moderately separated. The colors indicate the estimated clusters.  Adjusted Rand index is reported at the top of each panel.}
\label{fig:GLOCAL_clustering_overlapped}
\end{figure}

We considered scenarios where the global and local variables were well separated and where those were moderately separated. We set $\lambda_L^{-1} = \lambda_G^{-1}= 0.1$ in \eqref{eq:simulation_local_parameter_generation} and \eqref{eq:simulation_global_parameter_generation} for the well-separated case and $\lambda_L^{-1} = \lambda_G^{-1} = 0.5$ for the moderately-separated case. 
To fit our model, we used a truncation level of $L = T = 10$. The priors are the same as in Section \ref{sec:Real_data_analysis}. We ran 50,000 iterations of our sampler, which took $<2$ minutes on a MacBook Pro with M1 chip and 16GB RAM. The first half of the iterations were discarded as burn-in, and posterior samples were retained at every 25th iteration after burn-in. We estimated the cluster labels by the least squares criterion. Figure \ref{fig:GLOCAL_clustering_overlapped} shows the clustering plot of both the global and local variables in the moderately-separated case. The adjusted Rand index \citep[ARI]{ARI} shows that our model was able to identify clusters with good accuracy. The accuracy was better as expected for the well-separated case (Figure~\ref{fig:GLOCAL_clustering_separated} in the Supplementary Material).

We also considered a case where the global variables are not separated but the local variables are well separated by setting $\lambda_G^{-1} = 1$ in \eqref{eq:simulation_global_parameter_generation} and $\lambda_L^{-1} = 0.01$ in \eqref{eq:simulation_local_parameter_generation}. All other data generating strategies were the same as before. 
We ran our sampler for 80,000 iterations with a burn-in of 50,000 samples and a thinning factor of 30. Figure~\ref{fig:GLOCAL_clustering_highoverlapped} in the Supplementary Material shows that even in this difficult scenario, when the global variables show no apparent clusters, the local variables help identify global clusters with very good accuracy.
\subsection{No local variable for one population}\label{subsec:main_no_local}
Next, we considered additional simulations in which a population has no local variable. The GLocal DP, by its very construction, can be used for clustering problems for which some or all populations have no local variables. In particular, we considered a simulation setting in which the population 1 has no local variables, and the populations 2 and 3 have two- and three-dimensional local variables as in Section \ref{subsec:all_Local}. The detailed simulation setting and results are shown in the Section~\ref{subsec:no_local} of the Supplementary Material. The clustering results show that GLocal DP can identify clusters with good accuracy in this scenario as well. Furthermore, the local variables in the populations 2 and 3 can identify finer sub-clusters. 

\subsection{Comparison with HDP}\label{subsec:HDP_comparison_two_dim}
Lastly, we compared the proposed GLocal DP with HDP, which only accounts for global variables. We considered two-dimensional global variables while fixing the dimension of the local variables to be one, two, and three for the three populations. We varied the degree of separation in the local variables for the three populations by varying the local-level precision parameter $\lambda_L = 0.5, 0.1, 0.01$ in \eqref{eq:simulation_local_parameter_generation}. All the other simulation details are the same as in Section \ref{subsec:all_Local}. HDP was applied to the global variables only whereas GLocal DP was applied to both global and local variables. 
 All simulations were replicated 50 times. 
    
 \begin{figure}[!ht]
     \centering
     \includegraphics[width= 0.9\linewidth]{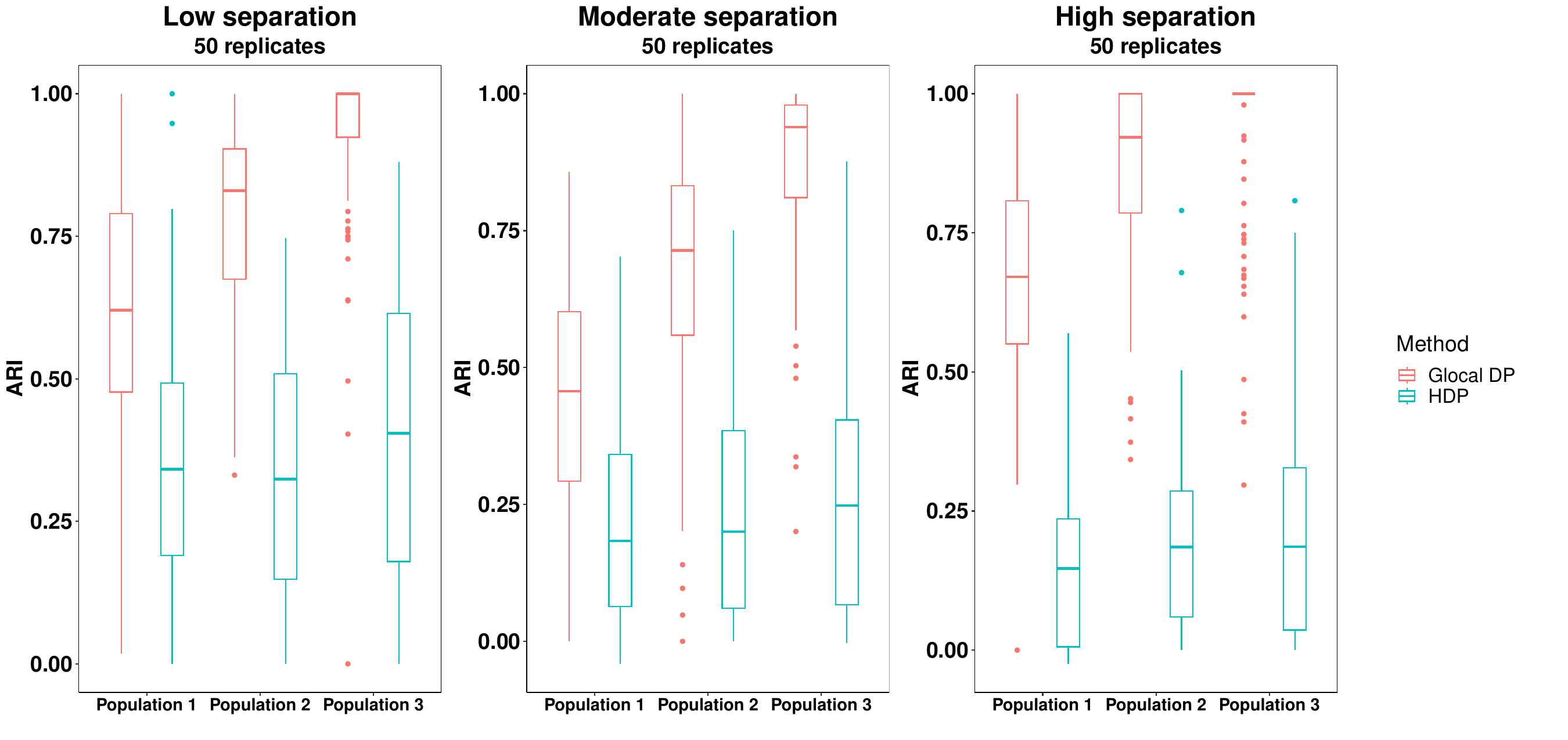}
     \caption{Comparison of clustering performance of GLocal DP with HDP for varying separation of local variables.}
     \label{fig:HDP_Comp_2_dim}
 \end{figure}
    \par Figure \ref{fig:HDP_Comp_2_dim} clearly shows that the clustering performance of the proposed GLocal DP was better than HDP. Furthermore, the clustering performance of our method clearly improves with the increasing separation in the local variables. 
    
    In the Section~\ref{subsec:HDP_comparison} of the Supplementary Material, we performed additional simulations with varying dimensions of the global variables and compared the GLocal DP with HDP in these scenarios. In summary, the clustering performance of GLocal DP shows significant improvement over HDP as the local variables become more separated regardless of the dimension of the global variables.

\section{Conclusion}\label{sec:discussion}
We have introduced the GLocal DP as a stochastic process for modeling a group of random measures that account for varying variable sets of the underlying grouped data. We have also introduced the corresponding infinite mixture model and presented how the GLocal DP mixture model can be used for clustering grouped data incorporating group-specific local variables. We have characterized the GLocal
DP using the stick-breaking representation and the representation
as a limit of a finite mixture model, which led to an efficient posterior sampling algorithm. We illustrated our method using both simulations and an application to a pan-cancer dataset, including shared gene expression data and cancer-specific clinical data. We identified global clusters shared across cancers as well as finer cancer-specific sub-clusters using local variables, which would not have been possible using existing methods. Our simulations highlight the importance of incorporating local variables, when available, in achieving superior clustering performance. The real data analysis underscores the importance of local variables (cancer prognostic markers) in the understanding of cancer prognosis. Particularly, the local variables help in the classification of survival patterns of cancer patients with the levels of associated risk factors, concurrent with existing scientific knowledge. Moreover, our analysis exclusively shows a disparate differentially expressed gene set characterizing the sub-clusters, which cannot be found by existing grouped-clustering methods that only identify the shared clusters. The upregulation of marker genes in cancer subpopulations and its corresponding effect on the prognostic clinical biomarkers were identified, which is further corroborated by existing literature. Furthermore, the application of the proposed model is not only limited to the field of cancer genomics. The proposed method can be used for a general grouped clustering framework, wherein the available data consists of important group-specific variables apart from shared variables. \par
There are a few possible future directions for this work. First, it may be possible to design a more efficient collapsed sampler that avoids the sampling of global and local atoms. This might improve the mixing properties of the sampler in high dimensions. This can possibly be applied to the high dimensional genomic data without resorting to dimension reduction. Alternatively, it might be interesting to consider a variational Bayes algorithm for scalable inference. Second, it will be interesting to consider the theoretical properties of the proposed GLocal DP. It might be possible to look at the posterior convergence rates of the GLocal DP mixing measure under various conditions on the geometry of the support of the underlying true base measure. Third, it maybe possible to extend our model to incorporate the group-clustering feature of the nested DP along with the cluster-sharing feature of the HDP \citep{beraha2021semi, balocchi2022clusteringarealunitsmultiple, hidden_HDP} or take the advantage of common atoms or shared atoms nested models \citep{denti2023common, BNPnestedDP, dangelodenti2024}. This can possibly provide insights on similar cancer subtypes apart from clustering shared observations across the tumor subtypes, while the cancer-specific clinical variables can help refine the clusters shared across cancers.
\bibliographystyle{apalike}
\bibliography{references} 
\newpage 
\bigskip
\begin{center}
{\Large \bf Supplementary Materials for ``Global-Local Dirichlet Processes for Identifying Pan-Cancer Subpopulations Using Both Shared and Cancer-Specific Data''}
\end{center}
\beginsupplement

\section{Proof of the Infinite Limit of Finite Mixture Model}
\label{appendix_proof}
\numberwithin{equation}{section}
The finite mixture model representation of the GLocal DP is given by, 
\begin{equation}
\label{app-eq:fmm_repr_1}
    \begin{aligned}
        \bm{\beta} & \sim \mbox{Dir}(\gamma/L, \dots, \gamma/L), &  k_{jt} &\sim \bm{\beta}, \\
         {\bm{\pi}}_j& \sim \mbox{Dir}(\alpha/T, \dots, \alpha/T), & t_{ji} &\sim  {\bm{\pi}}_j,\\
        \phi_k & \sim H, & \psi_{jt}^L & \sim U_j,\\
        \bm{x}_{ji} &\sim F_1(\bm{x}_{ji}^L \mid \psi_{jt_{ji}}^L)F_2(\bm{x}_{ji}^G \mid \phi_{k_{jt_{ji}}}), & & 
    \end{aligned}
\end{equation}
where $\bm{\beta}$ is the global vector of mixing proportions, $\bm{ {\pi}}_j$ is the group-specific vector of mixing proportions, $L$ is the number of global mixture components, and $T \geq L$ is the number of local mixture components. 
Further, as $L\rightarrow \infty$, the infinite limit of this model is the proposed GLocal DP mixture model. 
\begin{proof}
 Consider the random probability measure 
\begin{equation*}
    V^{L} = \sum_{k=1}^{L}\beta_{k}\delta_{\phi_k},
\end{equation*}
where $\bm{\beta} = (\beta_k)_{k=1}^{L} \sim \mbox{Dir}(\gamma/L, \dots, \gamma/L)$ and $\phi_k \overset{iid}{\sim} H, \ \ k = 1,\dots, L$  independent of $\bm{\beta}$. \citealp{ishwaran_zarepour} shows that for every measurable function $g$, integrable with respect to $H$, we have, as $L\rightarrow \infty$
\begin{equation}\label{eq:convergence_of_V}
    \int g(\theta)dV^L(\theta) \overset{\mathcal{D}}{\rightarrow}
    \int g(\theta)dV(\theta). 
\end{equation}
Further, for $T\geq L$, define
\begin{equation*}
    G_j^{T, L} = \sum_{t=1}^{T} {\pi}_{jt}\delta_{\psi_{jt}},
\end{equation*}
  where $ {\bm{\pi}}_j=( {\pi}_{jt})_{t=1}^{T}  \sim \mbox{Dir}(\alpha/T, \dots, \alpha/T)$ and $\psi_{jt} = (\psi_{jt}^L, \psi_{jt}^G) \overset{iid}{\sim} U_j\otimes V^L$ independent of $ {\bm{\pi}}_j$. Let $B_j \times C$ be an arbitrary measurable subset of $\Theta_j \times \Omega$. Then,
    \begin{align}
    \nonumber G_j^{T, L}(B_j\times C) & = \sum_{t=1}^{T} {\pi}_{jt}\mathds{1}_{B_j} (\psi_{jt}^L) \mathds{1}_C (\psi_{jt}^G)\\
    \label{eq:finite_mixtue_proof_DP_eqn1} & = \sum_{t=1}^{T} \sum_{k=1}^{L} {\pi}_{jt}\mathds{1}_{B_j} (\psi_{jt}^L) \mathds{1}_C (\phi_k)   
\end{align}
Here the indicator function $\mathds{1}_A(x) = 1$ if $x \in A$ and is $0$ otherwise. The second equality follows since for $T<\infty$ and any fixed $t$, $\psi_{jt}^G = \phi_k$, for some $k = 1, \dots, L$. Since \eqref{eq:finite_mixtue_proof_DP_eqn1} holds for any arbitrary measurable $B_j \times C$, we have
\begin{equation}\label{eq:convergence_of_Gj}
    G_j^{T, L} \sim \mbox{DP}(\alpha, U_j\otimes V^L).
\end{equation} 
It is clear from \eqref{eq:convergence_of_V} and \eqref{eq:convergence_of_Gj}, that as $L\rightarrow \infty$, $T\rightarrow \infty$, and the marginal distribution that the finite mixture model induces on the observations approaches the proposed GLocal DP mixture model.
\end{proof}

\section{Posterior Inference for the GLocal DP}
\label{supp-posterior_inference}
\begin{figure}[!htp]
\centering
\begin{adjustbox}{minipage=\linewidth,scale=0.85}
\begin{subfigure}[t]{1\linewidth}
    \centering
\begin{tikzpicture}[node distance={15mm}, thick, main/.style = {draw, circle, minimum size=1cm, scale=0.5, transform shape}] 
\node[obs, rectangle] (xL) {$\bm{x}_{ji}^L$};%
 \node[latent, above = of xL] (t) {$t_{ji}$}; %
 \node[latent, above =of t] (pi) {$\bm{\pi}_j$};
  \node[latent, above = of pi, yshift = 1.25cm] (alpha) {$\alpha$}; %
 \node[latent, left = of t, xshift = -0.5cm, yshift = 0.5cm] (psi) {$\bm{\psi}_{j}^{L}$}; %

 \node[latent, rectangle, left = of alpha, xshift = -0.55cm, rounded corners] (Uj) {$U_j$}; %

  \node[obs, rectangle, right = of xL, xshift = 0.5cm] (xG) {$\bm{x}_{ji}^G$};%
  \node[latent, above = of xG, draw = none] (h1) {};%
 \node[latent, above = of h1] (k) {$\bm{k}_{j}$}; %
 \node[latent, above right = of k, xshift = 1cm] (b1) {$\bm{\beta}$}; %
 \node[latent, above = of b1] (gamma) {$\gamma$}; %
 \node[latent, below = of b1, yshift = -1.5cm] (phi) {$\bm{\phi}$}; %
 \node[latent, rectangle, above = of phi, rounded corners] (H) {$H$}; %
% plate
\node[plate=$J$, inner sep=35pt, fit=(psi)(xL)(t)(pi)(xG)(k)] (plate1) {};
\node[plate=$n_j$, inner sep=18pt, outer sep=-2pt, fit=(t)(xL)(xG)] (plate2) {};
% edges
 \edge {pi}  {t} 
 \edge {t}  {xL} 
 \edge {psi}  {xL} 
 \edge {b1}  {k} 
 \edge {k}  {xG} 
 \edge {t}  {xG} 
  \edge {phi}  {xG}
  \edge {alpha}  {pi}
  \edge {gamma}  {b1}
  \edge {Uj}  {psi}
  \edge {H}  {phi}
\end{tikzpicture} 
\caption{GLocal DP mixture model.}
\label{fig:GLocalDP_Graphical_Plot_subfig1}
    \end{subfigure}
    \par\bigskip
    \begin{subfigure}[t]{1\linewidth}
    % \ContinuedFloat
    \centering
    \begin{tikzpicture}[node distance={15mm}, thick, main/.style = {draw, circle, minimum size=1cm, scale=0.5, transform shape}] 
\node[obs, rectangle] (xL) {$\bm{x}_{ji}^L$};%
 \node[latent, above = of xL, draw = none] (t) {}; %
  \node[latent, right = of t, xshift = -0.6cm] (kji) {$k_{ji}$}; %
 \node[latent, above =of t] (pi) {$\bm{\pi}_j$};
  \node[latent, above = of pi, yshift = 1.25cm] (alpha) {$\alpha$}; %
 \node[latent, left = of t, xshift = -0.5cm, yshift = 0.5cm] (psi) {$\bm{\psi}_{j}^{L}$}; %

 \node[latent, rectangle, left = of alpha, xshift = -0.55cm, rounded corners] (Uj) {$U_j$}; %

  \node[obs, rectangle, right = of xL, xshift = 0.5cm] (xG) {$\bm{x}_{ji}^G$};%
  \node[latent, above = of xG, draw = none] (h1) {};%
 \node[latent, above = of h1, draw = none] (k) {}; %
 \node[latent, above right = of k, xshift = 1cm] (b1) {$\bm{\beta}$}; %
 \node[latent, above = of b1] (gamma) {$\gamma$}; %
 \node[latent, below = of b1, yshift = -1.5cm] (phi) {$\bm{\phi}$}; %
 \node[latent, rectangle, above = of phi, rounded corners] (H) {$H$}; %
% plate
\node[plate=$J$, inner sep=35pt, fit=(psi)(xL)(t)(pi)(xG)(k)] (plate1) {};
\node[plate=$n_j$, inner sep=18pt, outer sep=-2pt, fit=(t)(xL)(xG)] (plate2) {};
% edges
 \edge {pi}  {kji} 
 \edge {kji}  {xL} 
 \edge {psi}  {xL} 
 \edge {b1}  {kji} 
 \edge {kji}  {xG} 
  \edge {phi}  {xG}
  \edge {alpha}  {pi}
  \edge {gamma}  {b1}
  \edge {Uj}  {psi}
  \edge {H}  {phi}
\end{tikzpicture}
\caption{GLocal DP mixture model after marginalization.}
\label{fig:GLocalDP_Graphical_Plot_subfig2}
    \end{subfigure}
    \caption{Graphical representation of GLocal Dirichlet process mixture model. Each node in the graph is associated with a random variable, where shaded rectangle denotes an observed variable. Rectangular plates denote replication of the model within the rectangle.}
    \label{fig:GLocalDP_Graphical_Plots}   
    \end{adjustbox}
\end{figure}
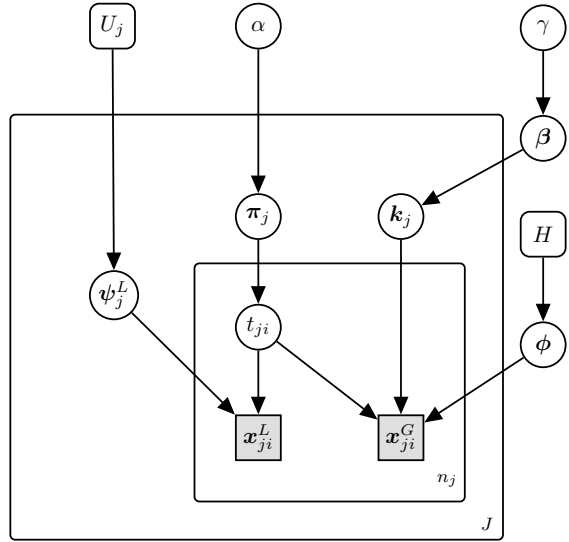
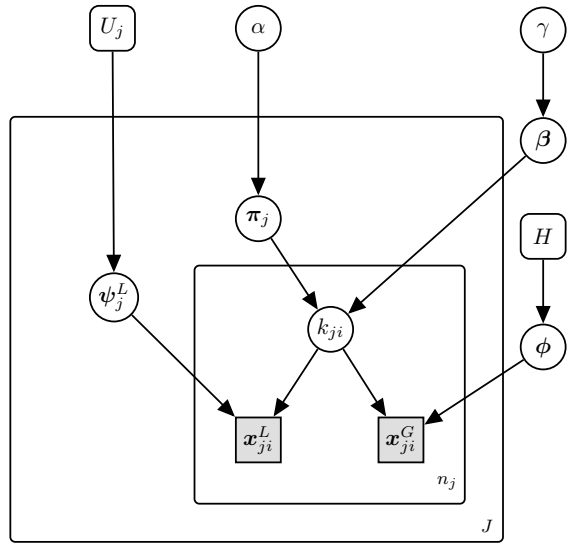

In this section, we present the detailed posterior inference algorithm for the GLocal DP. Consider the finite mixture model representation of the GLocal DP,
\begin{equation}
\label{supp-eq:fmm_repr_1}
    \begin{aligned}
        \bm{\beta} & \sim \mbox{Dir}(\gamma/L, \dots, \gamma/L), &  k_{jt} &\sim \bm{\beta}, \\
         {\bm{\pi}}_j& \sim \mbox{Dir}(\alpha/T, \dots, \alpha/T), & t_{ji} &\sim  {\bm{\pi}}_j,\\
        \phi_k & \sim H, & \psi_{jt}^L & \sim U_j,\\
        \bm{x}_{ji} &\sim F_1(\bm{x}_{ji}^L \mid \psi_{jt_{ji}}^L)F_2(\bm{x}_{ji}^G \mid \phi_{k_{jt_{ji}}}), & & 
    \end{aligned}
\end{equation}
Recall that we let $\bm{x} = (\bm{x}_j)_{j=1}^{J}$ to denote the observations from all $J$ groups. Similarly, $\bm{t} = (\bm{t}_j)_{j=1}^{J}$ and $\bm{k} = (\bm{k}_j)_{j=1}^{J}$ denotes the collection of all local-level and global-level latent indicators respectively. The collection of all local  atoms are denoted by $\bm{\psi} = (\bm{\psi}_j)_{j=1}^{J}$, with $\bm{\psi}_j = \left(\psi_{jt}^L\right)_{t=1}^{T}$ denoting the local atoms of group $j$. Similarly, the collection of global atoms are given by $\bm{\phi} = (\bm{\phi}_k)_{k=1}^{L}$. The graphical model representation of the GLocal DP mixture model is presented in Figure \ref{fig:GLocalDP_Graphical_Plots}.

Furthermore, recall that $F_1(.  \mid \psi_{jt}^L)$ and $F_2( . \mid \phi_{k})$ denotes the conditional distribution of the local  and global variables respectively, conditional on the local and global parameters. Let $f_1(.\mid \psi_{jt}^L)$ and $f_2(.\mid \phi_{k})$ be the density functions (with respect to some dominating measure) corresponding to the distributions $F_1(.  \mid \psi_{jt}^L)$ and $F_2( . \mid \phi_{k})$,  respectively. The augmented likelihood is then given by,
\begin{align}
    \label{supp-eq:augmented_likelihood}
    \nonumber p(\bm{x} , \bm{t}, \bm{k}\mid  \bm{\psi}, \bm{\phi},  ( {\bm{\pi}}_j)_{j=1}^{J}, \bm{\beta}) = \left\{\prod_{j=1}^{J}\prod_{i=1}^{n_j} f_1(\bm{x}_{ji}^L\mid \psi_{jt_{ji}}^L) f_2(\bm{x}_{ji}^G\mid \phi_{k_{jt_{ji}}}) \right\} \times \\
    \phantom{ p(\bm{x} , \bm{t}, \bm{k}\mid  \bm{\psi}, \bm{\phi},  ( {\bm{\pi}}_j)_{j=1}^{J}, \bm{\beta}) = \prod_{j=1}^{J}\prod_{i=1}^{n_j}} \prod_{j=1}^{J}\prod_{i=1}^{n_j} \prod_{t=1}^{T} {\pi}_{jt}^{\mathds{1}(t_{ji}=t)}\prod_{j=1}^{J}\prod_{t=1}^{T}\prod_{k=1}^{L}\beta_k^{\mathds{1}(k_{jt}=k)}
\end{align}
We use the general notation of $p(\cdot)$ to denote the prior distribution of any parameter, explicitly conditioning on model parameters wherever applicable. 
The joint prior distribution is given by,
\small{
\begin{equation}
    \label{supp-eq:joint_prior}
    p(\bm{\psi} , \bm{\phi}, ( {\bm{\pi}}_j)_{j=1}^{J}, \bm{\beta}, \alpha, \gamma) = \left\{\prod_{j=1}^{J}\prod_{t=1}^{T} p(\psi_{jt}^L)\right\}  \left\{\prod_{k=1}^{L} p(\phi_k) \right\} \left\{\prod_{j=1}^{J} p( {\bm{\pi}}_j| \alpha)\right\} p(\bm{\beta}|\gamma) p(\alpha) p(\gamma)
\end{equation}
}
Let $p(\cdot| - )$  be the generic notation for full conditional distribution. The full conditional distributions are straightforward to derive from \eqref{supp-eq:augmented_likelihood} and \eqref{supp-eq:joint_prior}. The Blocked Gibbs sampler consists of sampling from the full conditional distributions. The steps  of the proposed MCMC algorithm for posterior inference is outlined in Algorithm \ref{alg:the_algorithm}.

\begin{algorithm}
\caption{Blocked Gibbs Sampler for the GLocal DP}
\label{alg:the_algorithm}
\begin{algorithmic}[1]
    \State Sample $ {\bm{\pi}}_j$ from the Supplementary \eqref{eq:MCMC_update_pi} 
    \State Sample $\bm{\beta}$ from the Supplementary \eqref{eq:MCMC_update_beta} 
    \State Sample $\phi_k$ from the Supplementary \eqref{eq:MCMC_update_global_atoms}
    \State Sample $\psi_{jt}^L$ from the Supplementary \eqref{eq:MCMC_update_local_atoms} 
    \State Sample $t_{ji}$ from the Supplementary \eqref{eq:MCMC_calulate_local_probabilities} and the Supplementary \eqref{eq:MCMC_calulate_local_sampling}
    \State Sample $k_{jt}$ from the Supplementary \eqref{eq:MCMC_calulate_global_probabilities} and the Supplementary \eqref{eq:MCMC_calulate_global_sampling}
    \State Sample $\alpha$ from the Supplementary \eqref{eq:MCMC_update_alpha} and from the Supplementary \eqref{eeq:MCMC_update_alpha_MH}
    \State Sample $\gamma$ from the Supplementary \eqref{eq:MCMC_update_gamma} and from the Supplementary \eqref{eeq:MCMC_update_gamma_MH}
\end{algorithmic} 
\end{algorithm} 
\subsection*{Posterior distribution of the group-specific weights}
The conditional posterior for the group-specific weights, $ {\bm{\pi}}_j$ are given by,
\begin{align}
\label{eq:MCMC_update_pi}
    p( {\bm{\pi}}_j\mid -) & \sim \mbox{Dir}( {m}_{j1} + \alpha/T, \dots,  {m}_{jT} + \alpha/T), & \text{where $ {m}_{jt} = \sum_{i=1}^{n_j} \mathds{1}(t_{ji} = t)$}.
\end{align}
\subsection*{Posterior distribution of the global weights}
The conditional posterior for the global weights, $\bm{\beta}$ are given by,
\begin{align}
\label{eq:MCMC_update_beta}
    p(\bm{\beta}\mid -) & \sim \mbox{Dir}(d_1 + \gamma/L, \dots, d_L + \gamma/L), &  \text{where $ {d}_k = \sum_{j=1}^{J}\sum_{t=1}^{T} \mathds{1}(k_{jt} = k)$}.
\end{align}
\subsection*{Posterior distribution of the global atoms}
The updates for the global atoms, $\phi_k$ are obtained from the full conditional distribution,
\begin{align}
\label{eq:MCMC_update_global_atoms}  
    p(\phi_k\mid -) &\propto \left\{ \prod_{j=1}^{J}\prod_{\substack{i=1 \ni \\ k_{jt_{ji} = k}}}^{n_j}f_2(\bm{x}_{ji}^G\mid \phi_k)\right\}p(\phi_k) & k = 1, \dots, L.
\end{align}
For any given likelihood, assuming conjugate priors for the global atoms, yield conjugate Gibbs updates for $\phi_k$.
\subsection*{Posterior distribution of the local atoms}
The conditional posterior distribution for the local atoms are given by,
\begin{align}
\label{eq:MCMC_update_local_atoms}     
    p(\psi_{jt}^L\mid -) &\propto \left\{ \prod_{\substack{i=1 \ni \\ t_{ji} = t}}^{n_j}f_1(\bm{x}_{ji}^L\mid \psi_{jt}^L)\right\}p(\psi_{jt}^L), & t=1, \dots, T;\  j = 1, \dots, J.
\end{align}
Similarly, we may assume conjugate priors for the local atoms, which yields conjugate Gibbs updates.
\subsection*{Posterior distribution of the local-level latent indicators}
To update the local-level latent indicator variables, we first update the corresponding multinomial class probabilities. The conditional posterior probabilities are given by,
\begin{align}
 \label{eq:MCMC_calulate_local_probabilities}    
    Pr(t_{ji} = t\mid -) &\propto  {\pi}_{jt}f_1(\bm{x}_{ji}^L\mid \psi_{jt}^L)f_2(\bm{x}_{ji}^G\mid \phi_{k_{jt}}), & t=1, \dots, T; \ i = 1, \dots, n_j; \ j = 1, \dots, J.
\end{align}
The local-level latent indicators are then sampled from a multinomial distribution with probabilities given by  \eqref{eq:MCMC_calulate_local_probabilities}. In particular, if $p_t^{ji} = Pr(t_{ji} = t\mid -)$, then the local-level latent variables are updated by sampling 
\begin{equation}
\label{eq:MCMC_calulate_local_sampling} 
    t_{ji} \sim \mbox{multinomial}(p_1^{ji}, \dots, p_T^{ji}), \ \ i = 1, \dots, n_j, \ j=1,\dots, J.
\end{equation}
\subsection*{Posterior distribution of the global-level latent indicators}
Similarly to the local-level latent variables, we first update the corresponding multinomial class probabilities. The conditional posterior probabilities to update the global-level latent variables are given by,
\begin{align}
\label{eq:MCMC_calulate_global_probabilities}
    Pr(k_{jt} = k\mid -) &\propto \beta_k \prod_{\substack{i = 1\\ \ni t_{ji} = t}}^{n_j} f_2(\bm{x}_{ji}^G\mid \phi_k), & k=1, \dots, L; \ t = 1, \dots, T; \ j = 1, \dots, J.
\end{align}
Furthermore, if $p_k^{jt} = Pr(k_{jt} = k\mid -)$, then the global-level latent variables are updated by sampling 
\begin{equation}
\label{eq:MCMC_calulate_global_sampling}
    k_{jt} \sim \mbox{multinomial}(p_1^{jt}, \dots, p_L^{jt}),\ \ t = 1, \dots, T, \ j=1,\dots, J.
\end{equation}
\subsection*{Posterior distribution of the concentration parameters}
The conditional posterior for $\alpha$ is given by,
\begin{align}
\label{eq:MCMC_update_alpha}
    p(\alpha\mid -) & \propto  \frac{\{\Gamma(\alpha)\}^J}{\{\Gamma(\alpha/T)\}^{JT}} \prod_{j=1}^{J}\prod_{t=1}^{T} {\pi}_{jt}^{\alpha/T - 1} \, p(\alpha).
\end{align}
We assume a non-informative gamma prior for $\alpha$, i.e., $p(\alpha)\equiv \mbox{gamma}(a_{\alpha}, b_{\alpha})$, where $a_{\alpha}$ and $b_{\alpha}$ are known hyperparameters (usually 0.1 or 0.01). We update $\alpha$ using a Metropolis-Hastings (MH) step with a gamma proposal distribution. In particular, we choose the proposal distribution $q(\alpha)$ to be the same as the prior distribution, which we found to work pretty well in all our simulations. Letting $g(\alpha)$ to denote the target distribution (same as \eqref{eq:MCMC_update_alpha}), the MH step accepts a new proposed value of $\alpha$ at iteration $t$, say $\alpha_t$ with probability 
\begin{equation}
\label{eeq:MCMC_update_alpha_MH}
    \min\left\{1, \frac{g(\alpha_t)q(\alpha_{t-1})}{g(\alpha_{t-1})q(\alpha_{t})}\right\},
\end{equation}
where $\alpha_{t-1}$ denotes the value of $\alpha$ at iteration $t-1$.\\
\noindent Similarly, the conditional posterior distribution of $\gamma$ is given by,
\begin{align}
\label{eq:MCMC_update_gamma}
    p(\gamma\mid -) & \propto  \frac{\Gamma(\gamma)}{\{\Gamma(\gamma/L)\}^L} \prod_{k=1}^{L}\beta_k^{\gamma/L - 1} \, p(\gamma).
\end{align}
As before, we assume a non-informative gamma prior for $\gamma$ (say, $Gamma(a_{\gamma}, b_{\gamma})$) and we adopted an MH step for its update. The proposal distribution $q(\gamma)$ was taken to be the same as the prior distribution as in the previous case. The MH step accepts a new proposed sample at iteration $t$, $\gamma_t$ with probability 
\begin{equation}
\label{eeq:MCMC_update_gamma_MH}
    \min\left\{1, \frac{g(\gamma_t)q(\gamma_{t-1})}{g(\gamma_{t-1})q(\gamma_{t})}\right\},
\end{equation}
where $\gamma_{t-1}$ denotes the value of $\gamma$ at iteration $t-1$ and $g(\gamma)$ denotes the target distribution in \eqref{eq:MCMC_update_gamma}.
\subsection{GLocal DP vs. HDP posterior inference algorithm}
\label{subsec:GLocal_HDP_algorithm_diff}
Recall that in the absence of local variables for all the groups GLocal DP reduces to HDP and our MCMC algorithm can be directly used for HDP sampling. In particular, letting $f_2(\ . \mid \phi_k )$ denote the density of the shared variables, setting  $f_1(\ .\mid \psi_{jt}^L) = 1$, and wiping out the sampling of $\psi_{jt}^L$, 
% \eqref{eq:GLocalDPSampling algorithm} 
Algorithm \ref{alg:the_algorithm} in the Supplementary Section \ref{supp-posterior_inference}
reduces to a blocked-Gibbs sampling algorithm for HDP. Furthermore, the blocked Gibbs algorithm arising as a special case of our proposed sampler is a novel contribution to the HDP sampling algorithms. Contrarily, the sampling algorithm for HDP is not applicable for the GLocal DP, despite HDP being a special case, the rationale for which is as follows.
Consider the HDP in \eqref{hdp eq1} of the main manuscript. The stick-breaking representation of the group-specific random measure $G_j$ is given by, $G_j = \sum_{t = 1}^{\infty}  {\pi}_{jt}\delta_{\psi_{jt}}$, where $ {\bm{\pi}}_j=( {\pi}_{jt})_{t=1}^{\infty} \sim \text{GEM}(\alpha_0)$ and $\psi_{jt} \overset{iid}{\sim} G_0$ independent of $ {\bm{\pi}}_j$. Similarly, the base measure $G_0$ is represented as, $ G_0 = \sum_{k = 1}^{\infty} \beta_k\delta_{\phi_k}$, where $\bm{\beta}=(\beta_k)_{k=1}^{\infty} \sim \text{GEM}(\gamma)$ and $\phi_k \overset{iid}{\sim} H$ independent of $\bm{\beta}$. Letting $\pi_{jk} = \sum_{t\in I_{jk}^*}  {\pi}_{jt}$, where $I_{jk}^* = \{t : \psi_{jt} = \phi_k\}$, we have the equivalent representation 
\begin{align*}
    G_j = \sum_{t = 1}^{\infty}  {\pi}_{jt} \delta_{\psi_{jt}} \equiv \sum_{k = 1}^{\infty} \sum_{t \in I_{jk}^*}  {\pi}_{jt}\delta_{\psi_{jt}} = \sum_{k = 1}^{\infty} \pi_{jk}\delta_{\phi_{k}}.
\end{align*}
This collapsed representation further relates the group-specific weights $\bm{\pi}_j$ with the global weights $\bm{\beta}$ as
\begin{equation*}
    \bm{\pi}_j \sim \mbox{DP}(\alpha_0, \bm{\beta}),
\end{equation*}
where $\bm{\pi}_j = (\pi_{jk})_{k=1}^{\infty}$ and $\bm{\beta}$ are probability measures on positive integers. Hence an equivalent representation of HDP mixture model is given by,
 \begin{equation}\label{eq:HDPmixturemodel}
     \begin{aligned}
         \bm{\beta}\mid \gamma &\sim \text{GEM}(\gamma) &  & \\
         \bm{\pi}_j\mid \alpha_0, \bm{\beta} &\sim \mbox{DP}(\alpha_0, \bm{\beta})  &z_{ji}\mid \bm{\pi}_j &\sim \bm{\pi}_j \\
         \phi_k\mid H & \sim H    &\bm{x}_{ji}\mid z_{ji}, (\phi_k)_{k=1}^{\infty} &\sim F(\bm{x}_{ji} \mid \phi_{z_{ji}}).
     \end{aligned}
 \end{equation}
The blocked Gibbs sampler for HDP \citep{das_etal} rely on this representation. Consider the stick-breaking representation of $G_j$ for our GLocal DP in \eqref{eq:stick-breaking_local} of the main manuscript. Similarly, we define $I_{jk} = \{t : \psi_{jt}^G = \phi_k\}$ and $\pi_{jk} = \sum_{t\in I_{jk}}  {\pi}_{jt}$. However, due to the presence of the local factors $\psi_{jt}^L$ in GLocal DP we have,
 \begin{equation}
        G_j = \sum_{k = 1}^{\infty} \sum_{t \in I_{jk}}  {\pi}_{jt}\delta_{(\psi_{jt}^L, \psi_{jt}^G)}.
 \end{equation}
Consequently, we do not get a collapsed representation for $G_j$ as in HDP. In particular, it is straightforward  to see that
\begin{equation}
    \bm{\pi}_j \sim \mbox{DP}(\alpha, U_j \otimes \bm{\beta}),
\end{equation}
where $\bm{\pi}_j$ is not a probability measure on positive integers unlike HDP. Accordingly, the blocked Gibbs sampler for HDP is not applicable for the GLocal DP.

\section{Real Data Analysis}
\label{supp-real_data_analysis}

\underline{Sensitivity.} In the main manuscript, we presented the analysis by performing UMAP on the combined gene expression data from the four cancers to reduce the data to two dimensions on a common manifold. Also, we considered the truncation levels $T=L = 20$ for the GLocal DP. To study the effect of the number of dimensions in downstream cluster analysis, we considered 2-, 3- and 5-dimensional UMAP embeddings as global variables, with the same local variables as before. Simultaneously, we varied truncation levels $L = T = 10, 20, 30, 40,$ and 
the following sampling distributions, 
\begin{align*}
&F_1(\bm{x}_{ji}^L\mid \bm{\theta}_{ji}^L):=\mathcal{N}_{p_j}(\bm{x}_{ji}^L\mid \bm{\mu}_{jt_{ji}}, \sigma_{jt_{ji}}^2 \mathds{I}_{p_j}),\\
&F_2(\bm{x}_{ji}^G\mid \bm{\theta}_{ji}^G):=\mathcal{N}_{2}(\bm{x}_{ji}^G\mid \bm{\mu}_{k_{jt_{ji}}}, \sigma_{k_{jt_{ji}}}^2 \mathds{I}_{p}),
\end{align*}
where $\mathds{I}_{p}$ is a $p \times p$ identity matrix with $p = 2, 3, 5$ corresponding to the dimension of UMAP embeddings, and $p_j$ is the dimension of the local variables in the population $j$ (i.e., $p_1 = 0, p_2 = 1$, $p_3 = 2$, and $p_4 = 1$ for stomach, esophageal, colon, and rectal cancers, respectively).
For hyperpriors, we assume
$\bm{\mu}_{jt_{ji}}\mid \sigma_{jt_{ji}}^2 \sim \mathcal{N}_{p_j}(0,\sigma_{jt_{ji}}^2 \mathds{I}_{p_j})$, $\bm{\mu}_{k_{jt_{ji}}} \mid \sigma_{k_{jt_{ji}}}^2\sim \mathcal{N}_2(0,\sigma_{k_{jt_{ji}}}^2 \mathds{I}_{2})$, and $\sigma_{jt_{ji}}^2,\sigma_{k_{jt_{ji}}}^2\, \alpha^{-1},\gamma^{-1}\sim \mathcal{IG}(0.1, 0.1)$. 
We considered 100 independent replications to study the sensitivity of the estimated number of global and local clusters with various truncation levels of GLocal DP and different dimensional UMAP embeddings. For each replication, we considered 50,000 iterations of our sampler and retained every 25th posterior sample post burn-in of 25,000. We estimated the global- and local-level clusters by the least-squares method.  Table \ref{tab:sensitivity_truncation} shows the mean number of global and local clusters along with the standard deviation (s.d). Our method is quite robust with respect to the truncation level, especially for $L=T=20,30,40$, and the dimension. 

\begin{table}
            \centering
            \resizebox{\columnwidth}{!}{
            \begin{tabular}[t]{c|c|c|c|c|c}
            \hline
            Dimension & Truncation level ($L=T$) & 10 & 20 & 30 & 40\\
            \hline
            \multirow{2}*{2} & Number of global clusters & 7.91 (0.351) & 8.02 (0.348) & 7.96 (0.374) & 8.13 (0.485)\\
            & Number of local clusters  & 9.66  (0.476) & 11.77 (0.737) & 12.20 (0.985) & 12.32 (1.162)\\
            \hline
            \multirow{2}*{3} & Number of global clusters & 8.00 (0.100) & 8.00 (0.000) & 8.03 (0.171) & 8.01 (0.100)\\
            & Number of local clusters  & 9.66 (0.476) & 11.10 (0.302) & 11.10 (0.302) & 11.34 (0.623)\\
            \hline
            \multirow{2}*{5} & Number of global clusters & 7.04 (0.197) & 7.17 (0.403) & 7.19 (0.394) & 7.25 (0.500)\\
            & Number of local clusters  & 9.73 (0.446) & 11.75 (1.533) & 12.07 (1.849) & 12.69 (2.356)\\
            \hline
            \end{tabular}}
            \caption{The estimated number of global and local clusters against the truncation levels of GLocal DP for 2-, 3-, and 5-dimensional UMAP embeddings as the global variables. We report the mean (s.d.) over 100 independent replications.}
            \label{tab:sensitivity_truncation}
            \end{table}
            
\begin{figure}[!htp]
\centering
    \includegraphics[width= 0.75\linewidth]{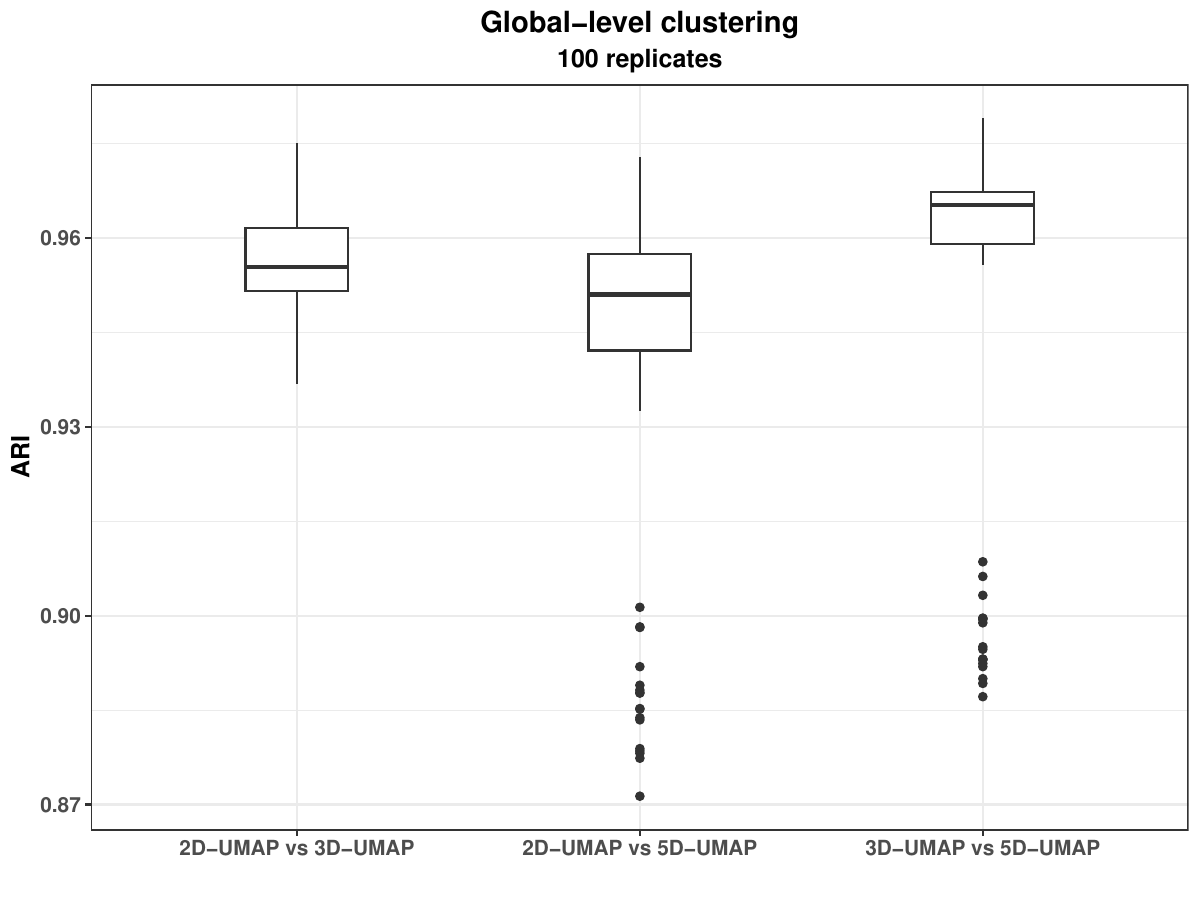}
\caption{The pairwise boxplot of adjusted Rand Index for the global-level clusters to assess the agreement of estimated clusters across different dimensional UMAP embeddings as global variables.}
\label{fig:ARI_2D_3D_5D}
\end{figure}

We further looked at the pairwise boxplots of adjusted Rand Index \citep{ARI} between the 2-, 3-, and 5-dimensional UMAP embeddings as global variables to assess the robustness of estimated clustering across different dimensional embeddings obtained from UMAP. Figure \ref{fig:ARI_2D_3D_5D} shows high agreement in the global-level clustering across the different dimensions of global variables. These analyses led us to choose the 2-dimensional UMAP embeddings and truncation levels $L = T = 20$ for downstream clustering using GLocal DP, as reported in the main manuscript.\par
\begin{figure}[!htp]
            \centering
                \includegraphics[width= 0.75\linewidth]{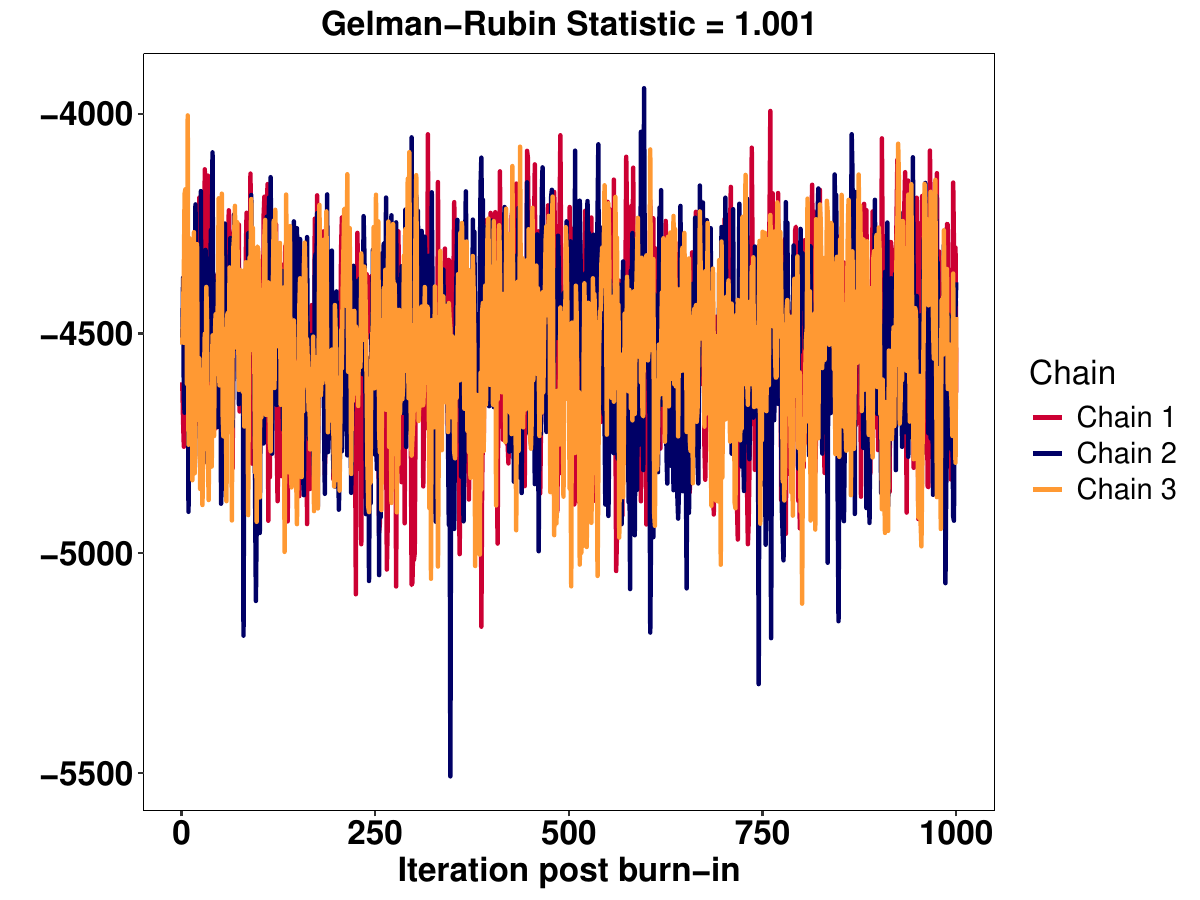}
            \caption{The traceplot of log-posterior post burn-in and thinning for the three independent chains of our sampler. The Gelman-Rubin statistic value is reported at the top of the figure. }
            \label{fig:LogPosteriorMultiChain}
            \end{figure}
\begin{figure}[!ht]
        \centering
        \begin{subfigure}{0.48\textwidth}
          \centering
            \includegraphics[width= 1\linewidth]{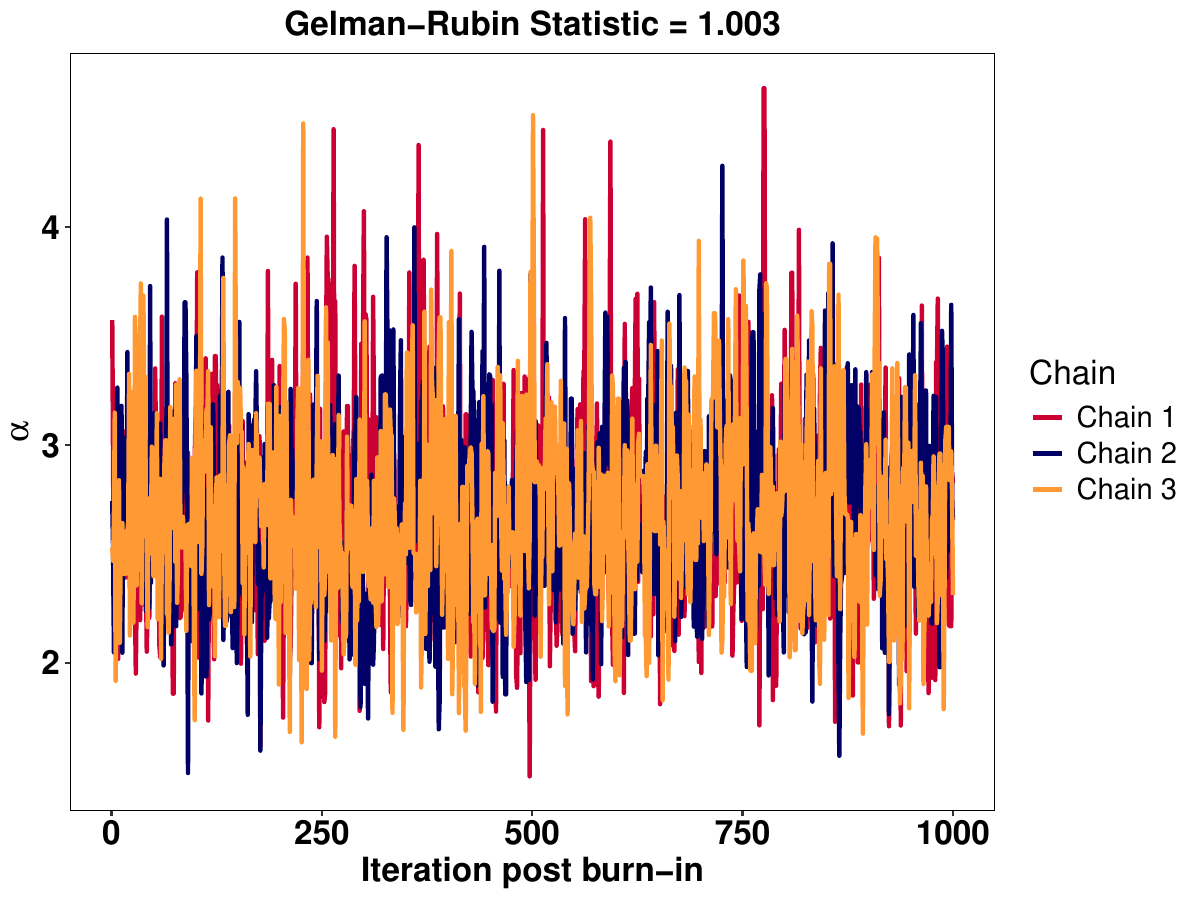}
            \caption{Traceplot of $\alpha$.}
            \label{fig:Traceplot_alphaMulti}
        \end{subfigure}
        \begin{subfigure}{0.48\textwidth}
            \centering
            \includegraphics[width= 1\linewidth]{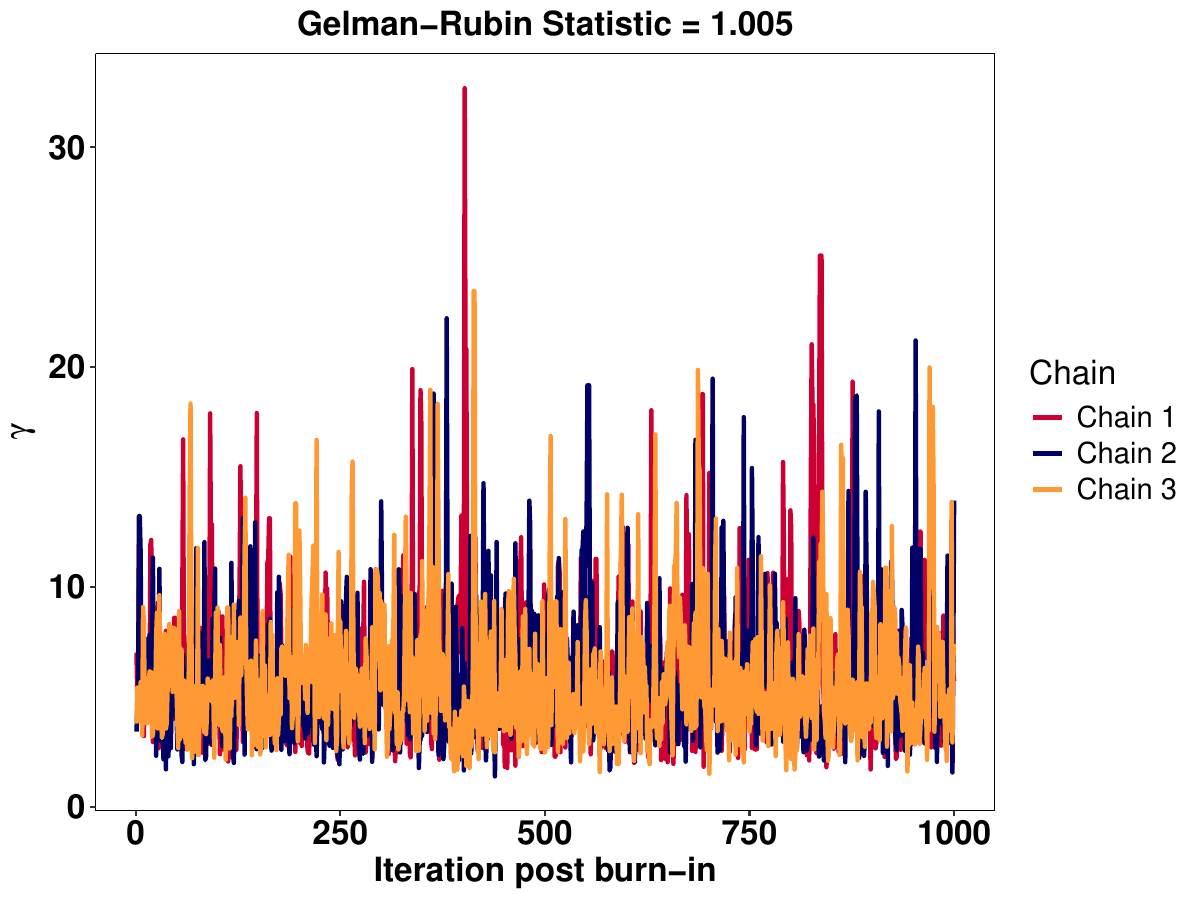}
            \caption{Traceplot of $\gamma$.}
            \label{fig:Traceplot_gammaMulti}
        \end{subfigure}
        \caption{The traceplots of the concentration parameters post burn-in and thinning for the three independent chains of our sampler. The Gelman-Rubin statistic value is reported at the top of the figure.}
        \label{fig:Traceplots_concentrationMulti}
        \end{figure}

\underline{MCMC convergence and mixing.} With the 2-dimensional UMAP embeddings as the global variables, the same local variables as in the main manuscript, and truncation levels $L = T = 20$, we  considered three independent MCMC chains for our sampler. For each independent chain, we ran our MCMC for $100,000$ iterations, discarded the first $25,000$ iterations as burn-in, and retained every $75$th posterior sample. We looked at the Gelman and Rubin's convergence diagnostic \citep{GelmanRubin} for the log-posterior from the three independent chains to quantitatively assess the convergence of our sampler. Figure \ref{fig:LogPosteriorMultiChain} shows the traceplots of the log-posterior for these chains along with the Gelman-Rubin statistic value (reported at the top of the figure). Clearly, the Gelman-Rubin statistic indicates no lack of convergence of our sampler. Figure \ref{fig:Traceplots_concentrationMulti} shows additional traceplots of the concentration parameters $\alpha$ and $\gamma$ from the three independent chains along with the corresponding Gelman-Rubin statistic values, which also demonstrates good mixing.

\begin{figure}[!ht]
\centering
\begin{subfigure}{0.48\textwidth}
  \centering
    \includegraphics[width= 1\linewidth]{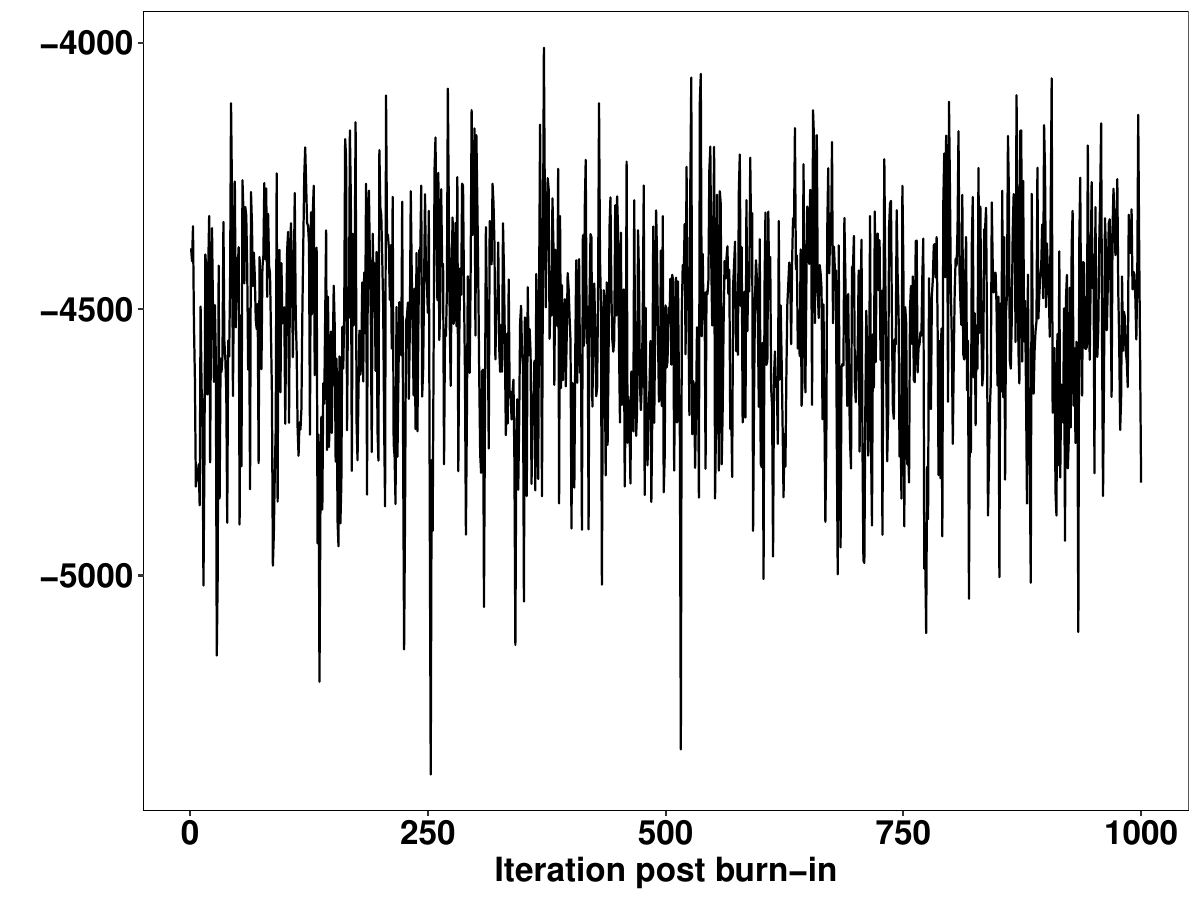}
    \caption{Traceplot of log-posterior.}
    \label{fig:LL_ReadData}
\end{subfigure}
\begin{subfigure}{0.48\textwidth}
    \centering
    \includegraphics[width= 1\linewidth]{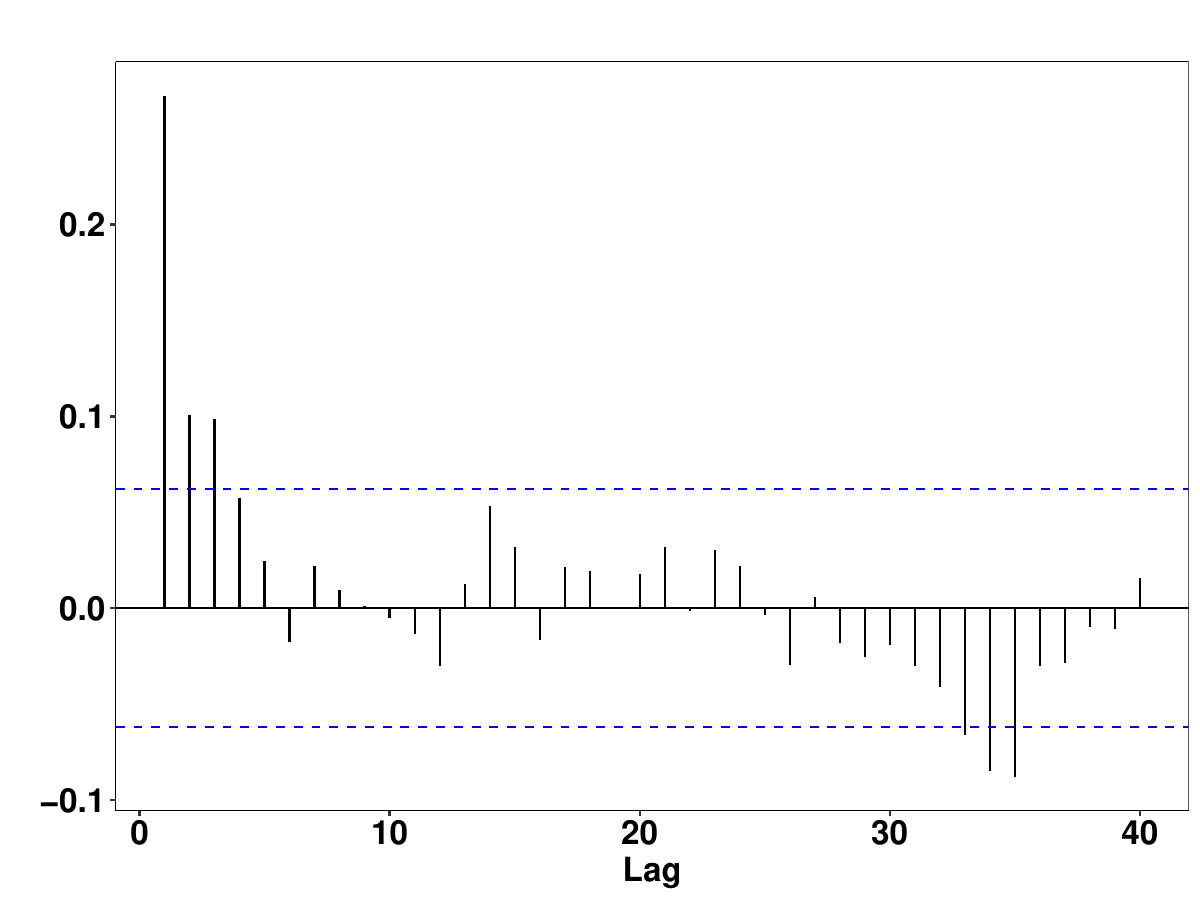}
    \caption{ACF plot of log-posterior.}
    \label{fig:ACF_ReadData}
\end{subfigure}
\caption{The traceplot and ACF of log-posterior post burn-in and thinning.}
\label{fig:Diagnostics_RealData}
\end{figure}

\begin{figure}[!ht]
\centering
\begin{subfigure}{0.48\textwidth}
  \centering
    \includegraphics[width= 1\linewidth]{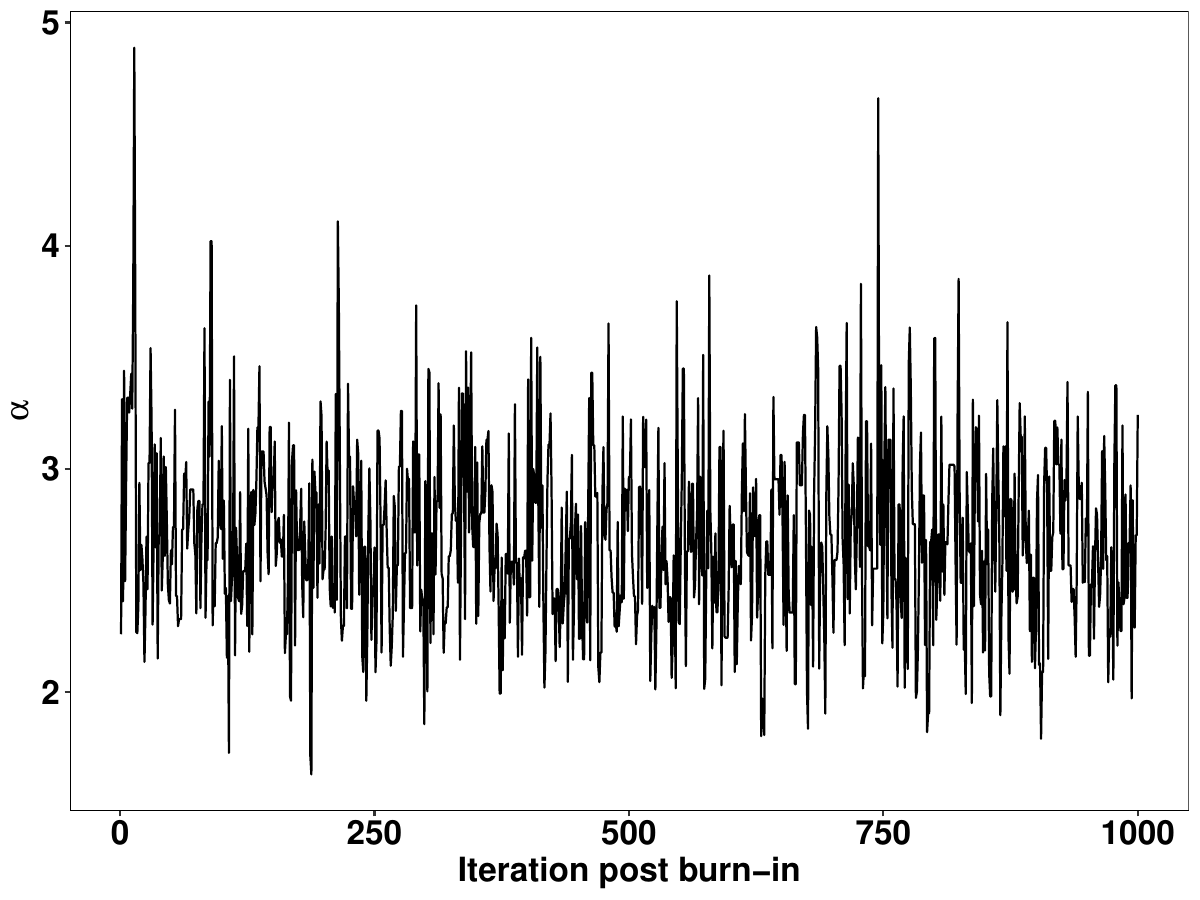}
    \caption{Traceplot of $\alpha$.}
    \label{fig:Traceplot_alpha}
\end{subfigure}
\begin{subfigure}{0.48\textwidth}
    \centering
    \includegraphics[width= 1\linewidth]{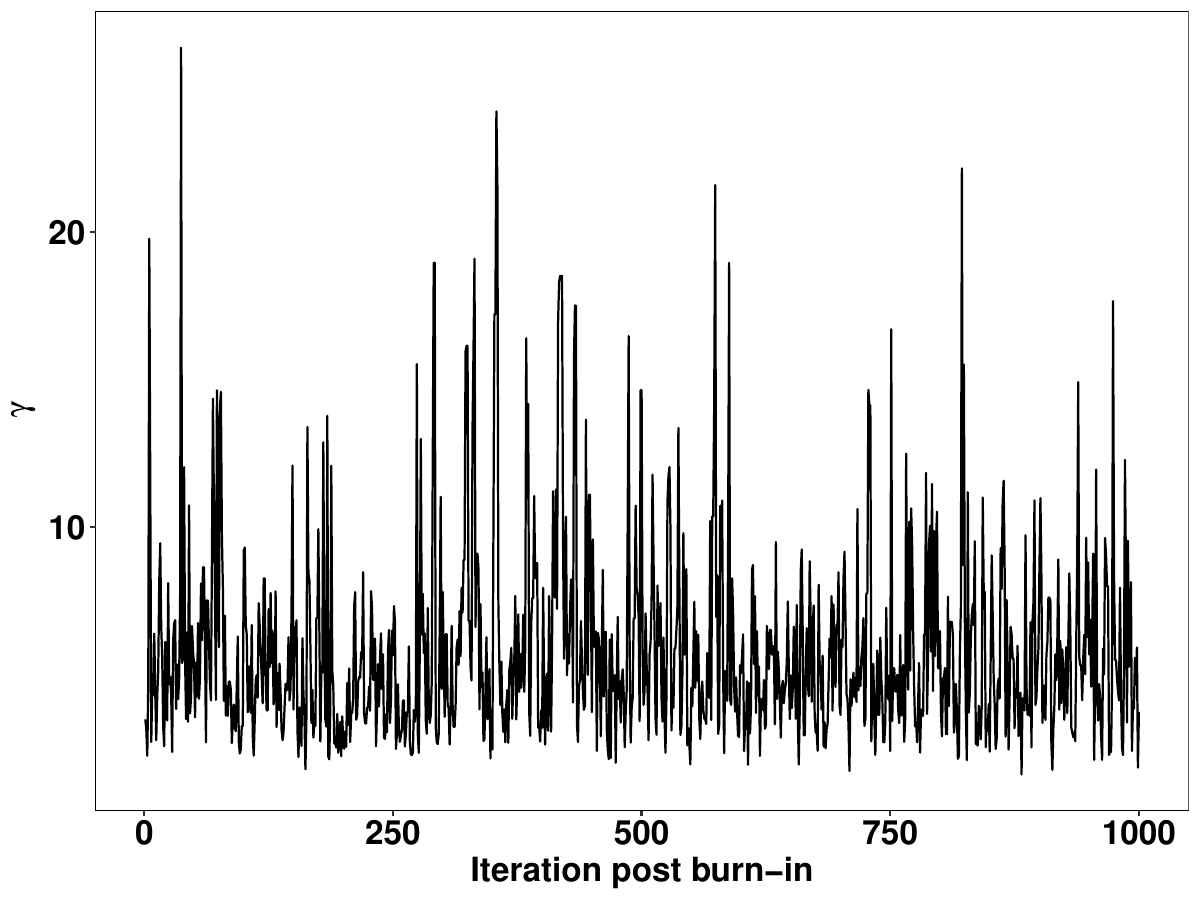}
    \caption{Traceplot of $\gamma$.}
    \label{fig:Traceplot_gamma}
\end{subfigure}
\caption{The traceplots of the concentration parameters post burn-in and thinning.}
\label{fig:Traceplots_concentration}
\end{figure}

 \begin{figure}[!ht]
\centering
\begin{subfigure}{0.48\textwidth}
  \centering
    \includegraphics[width= 1\linewidth]{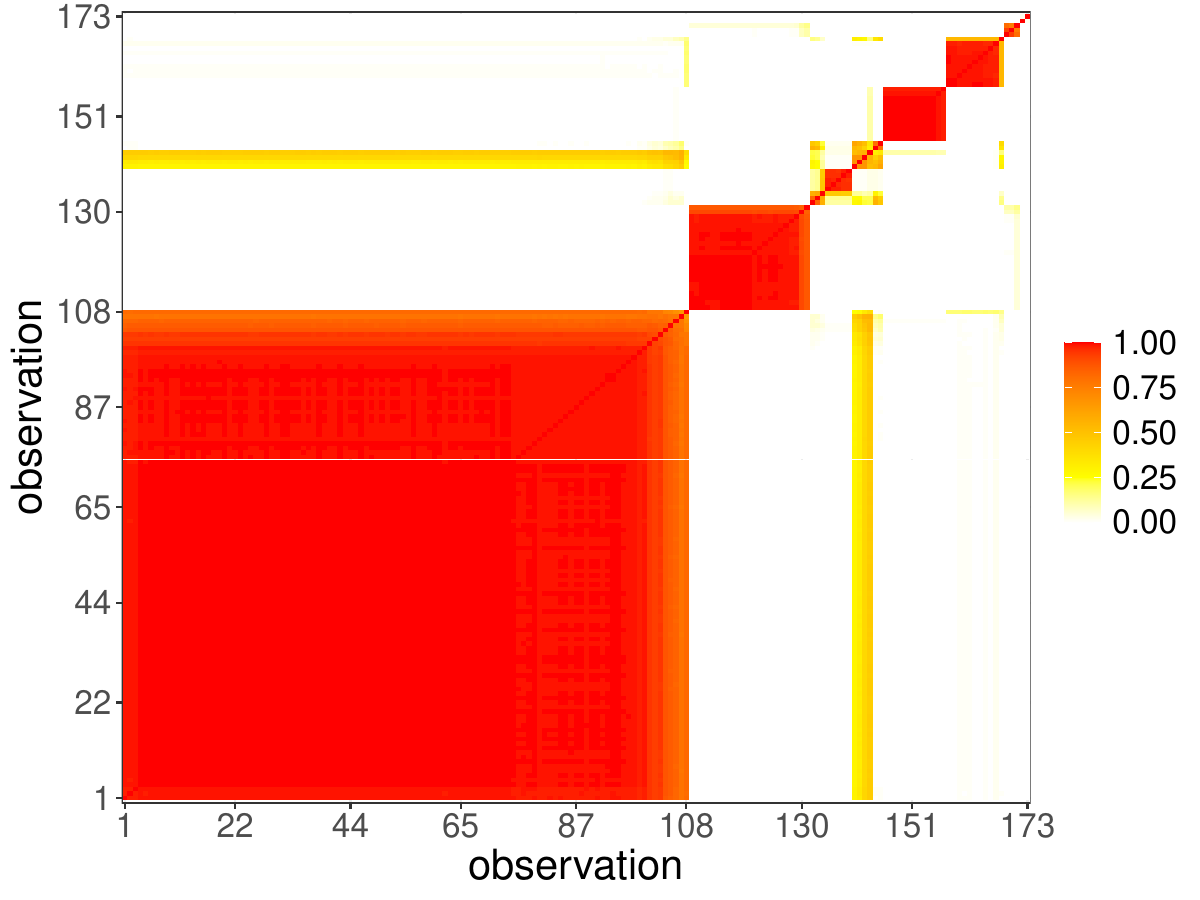}
    \caption{Colon cancer.}
    \label{fig:HeatmapLocalColon}
\end{subfigure}
\begin{subfigure}{0.48\textwidth}
    \centering
    \includegraphics[width= 1\linewidth]{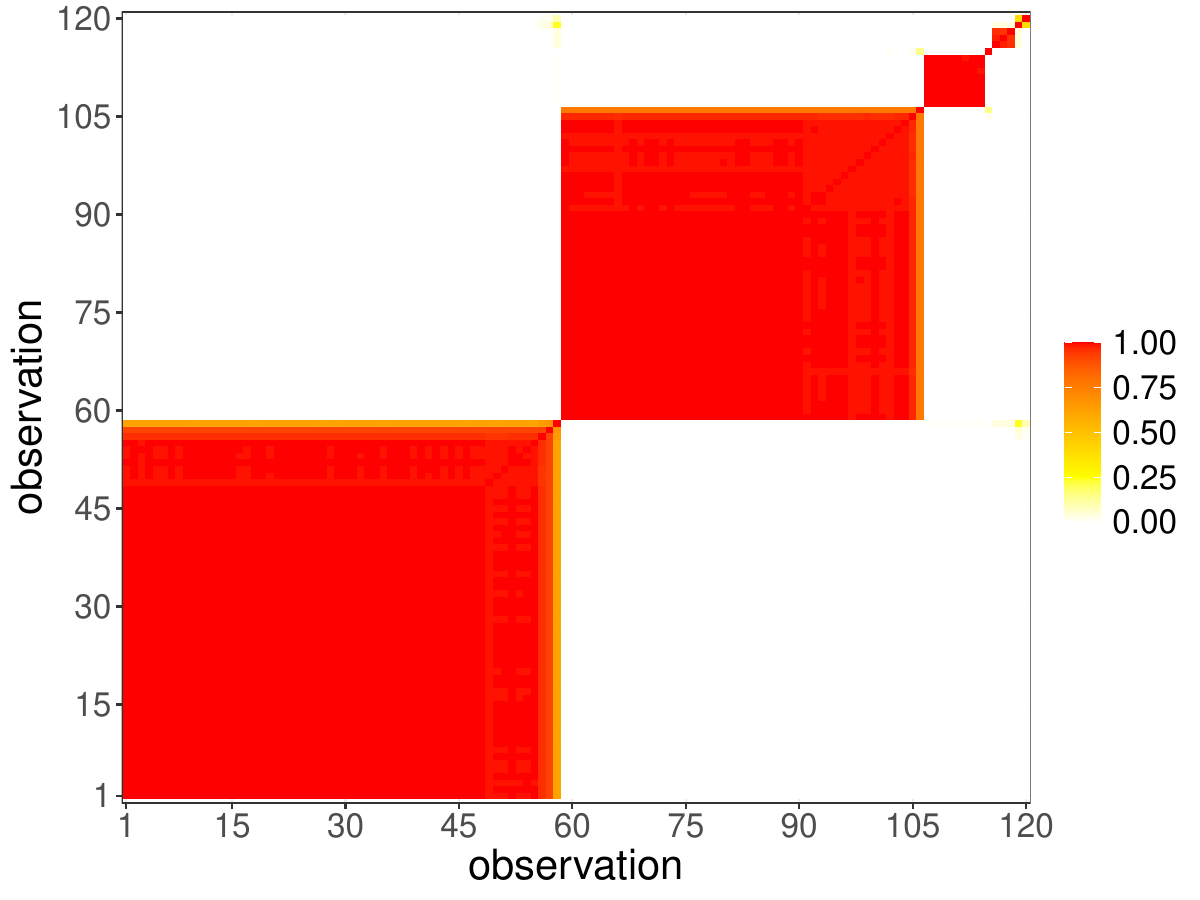}
    \caption{Rectal cancer.}
    \label{fig:HeatmapLocalRectal}
\end{subfigure}
\caption{Posterior co-clustering probabilities of observations assigned to local-level clusters for (a) colon cancer and (b) rectal cancer.}
\label{fig:HeatmapLocal}
\end{figure}
\section{Additional Simulations}
\subsection{Local variables for all populations}
\label{supp-subsec:all_Local}

\begin{figure}[!ht]
\centering
\begin{subfigure}{0.48\textwidth}
  \centering
    \includegraphics[width= 0.9\linewidth]{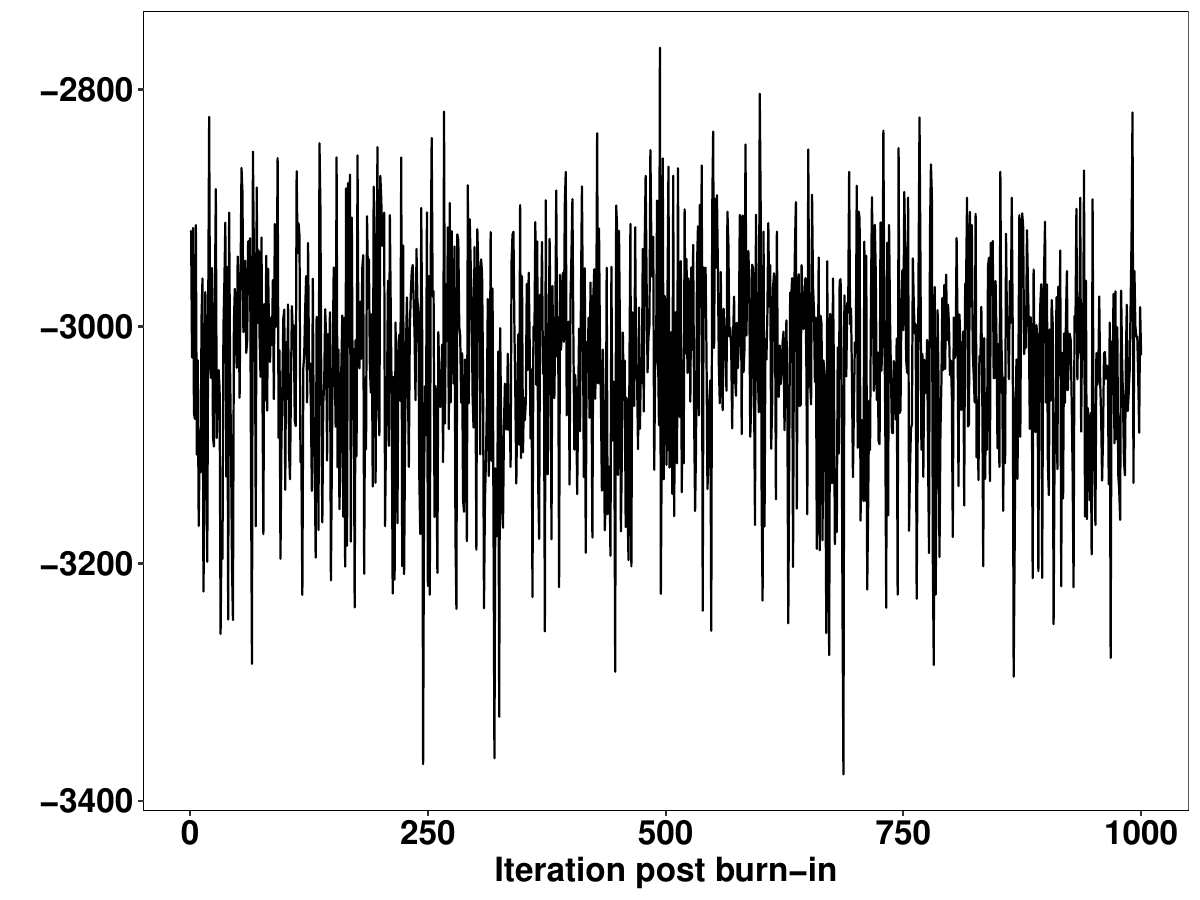}
    \caption{Traceplot of log-posterior.}
    \label{fig:LL_separated}
\end{subfigure}
\begin{subfigure}{0.48\textwidth}
    \centering
    \includegraphics[width= 0.9\linewidth]{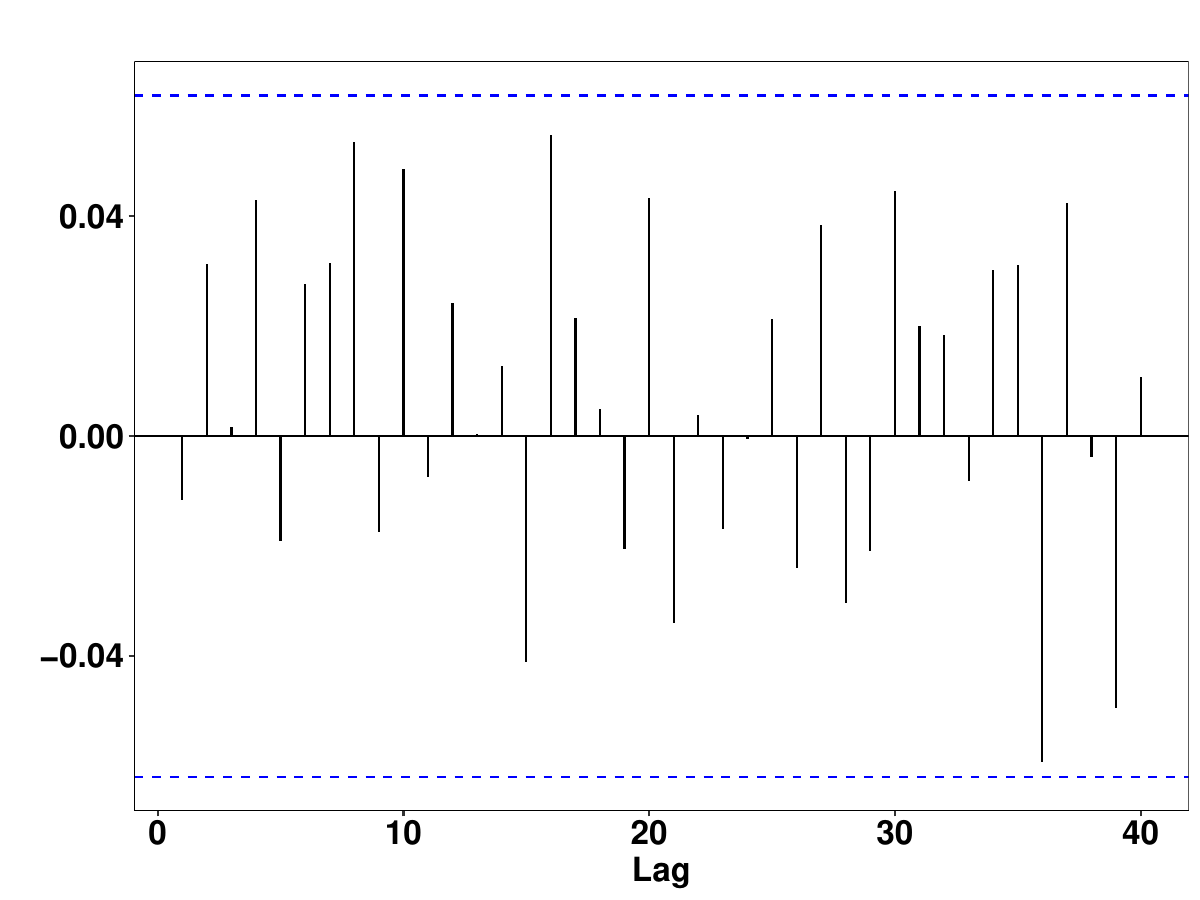}
    \caption{ACF plot of log-posterior.}
    \label{fig:ACF_separated}
\end{subfigure}
\caption{The traceplot and ACF of log-posterior post burn-in and thinning. The corresponding data has high separation in both global and local variables.}
\end{figure}
\begin{figure}[!htp]
\centering
\begin{subfigure}{0.75\textwidth}
  \centering
    \includegraphics[width= 1\linewidth]{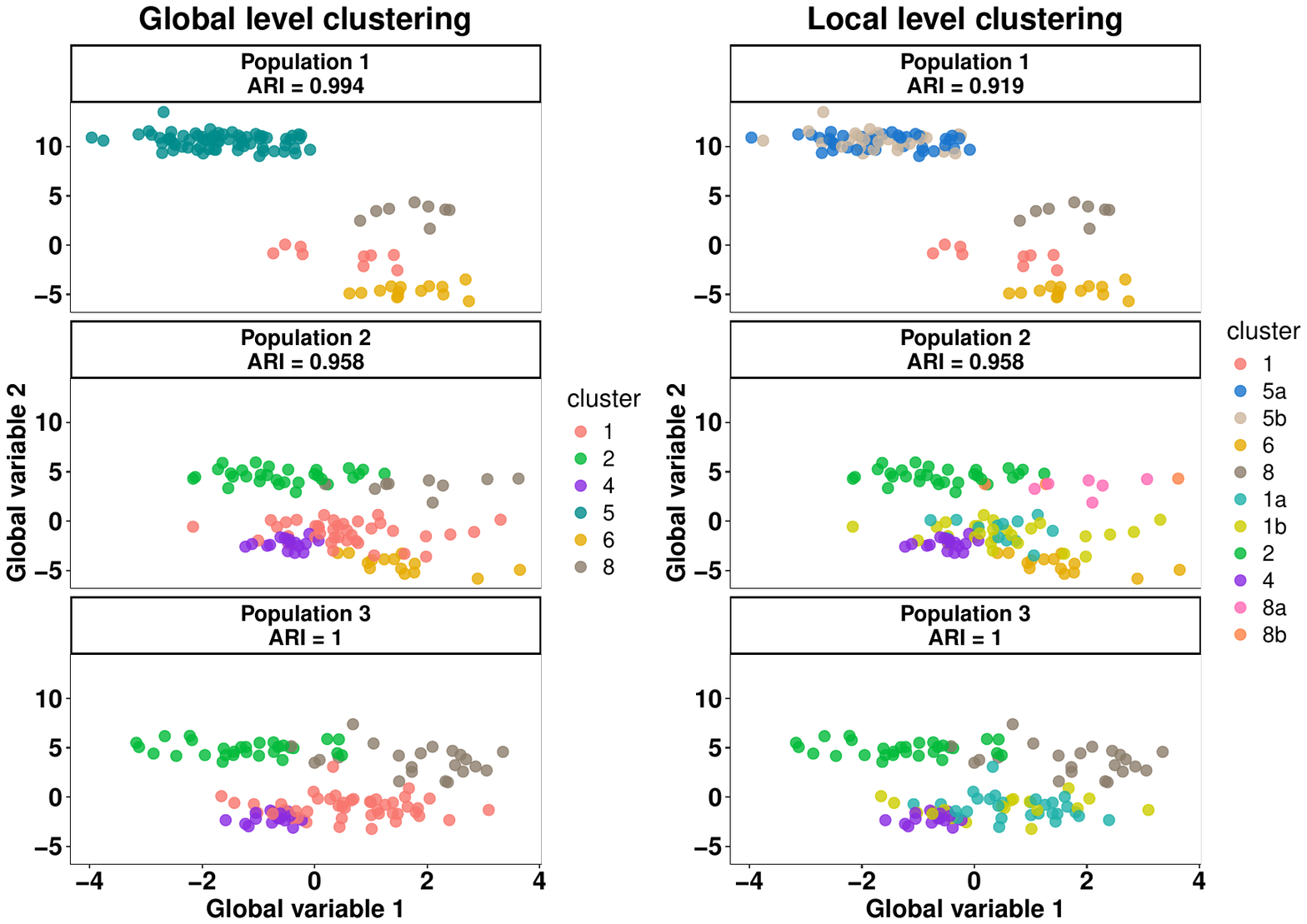}
    \caption{Global variables colored by global and local-level clusters.}
    \label{fig:global_clustering_separated}
\end{subfigure}
\begin{subfigure}{0.75\textwidth}
    \centering
    \includegraphics[width= 1\linewidth]{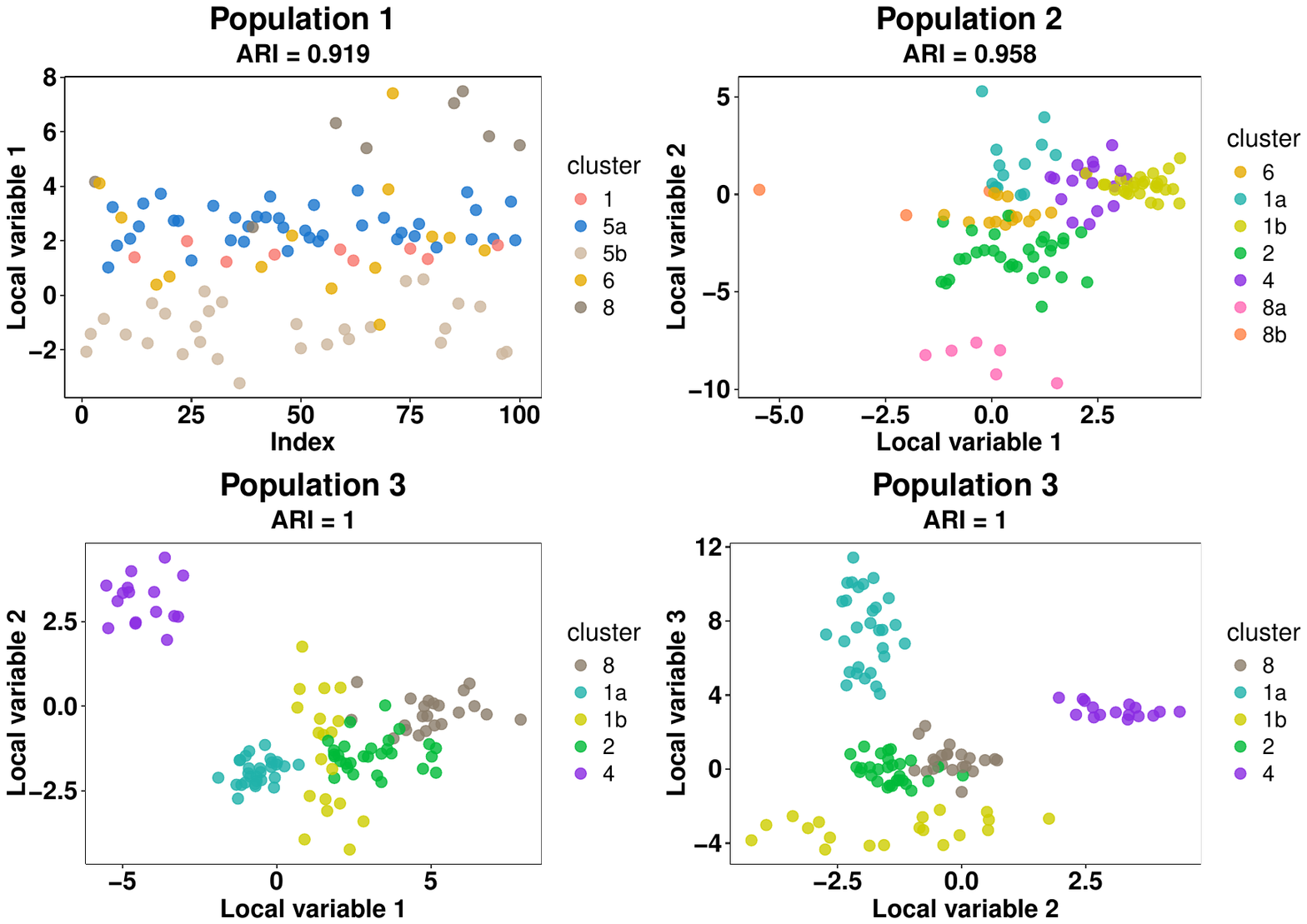}
    \caption{Local variables colored by local-level clusters.}
    \label{fig:local_clustering_separated}
\end{subfigure}
\caption{Clustering performance of GLocal DP when both the global and local variables are well separated. The colors indicate the estimated clusters.  Adjusted Rand index is reported at the top of each panel.}
\label{fig:GLOCAL_clustering_separated}
\end{figure}
\begin{figure}[!htp]
\centering
\begin{subfigure}{0.75\textwidth}
  \centering
    \includegraphics[width= 1\linewidth]{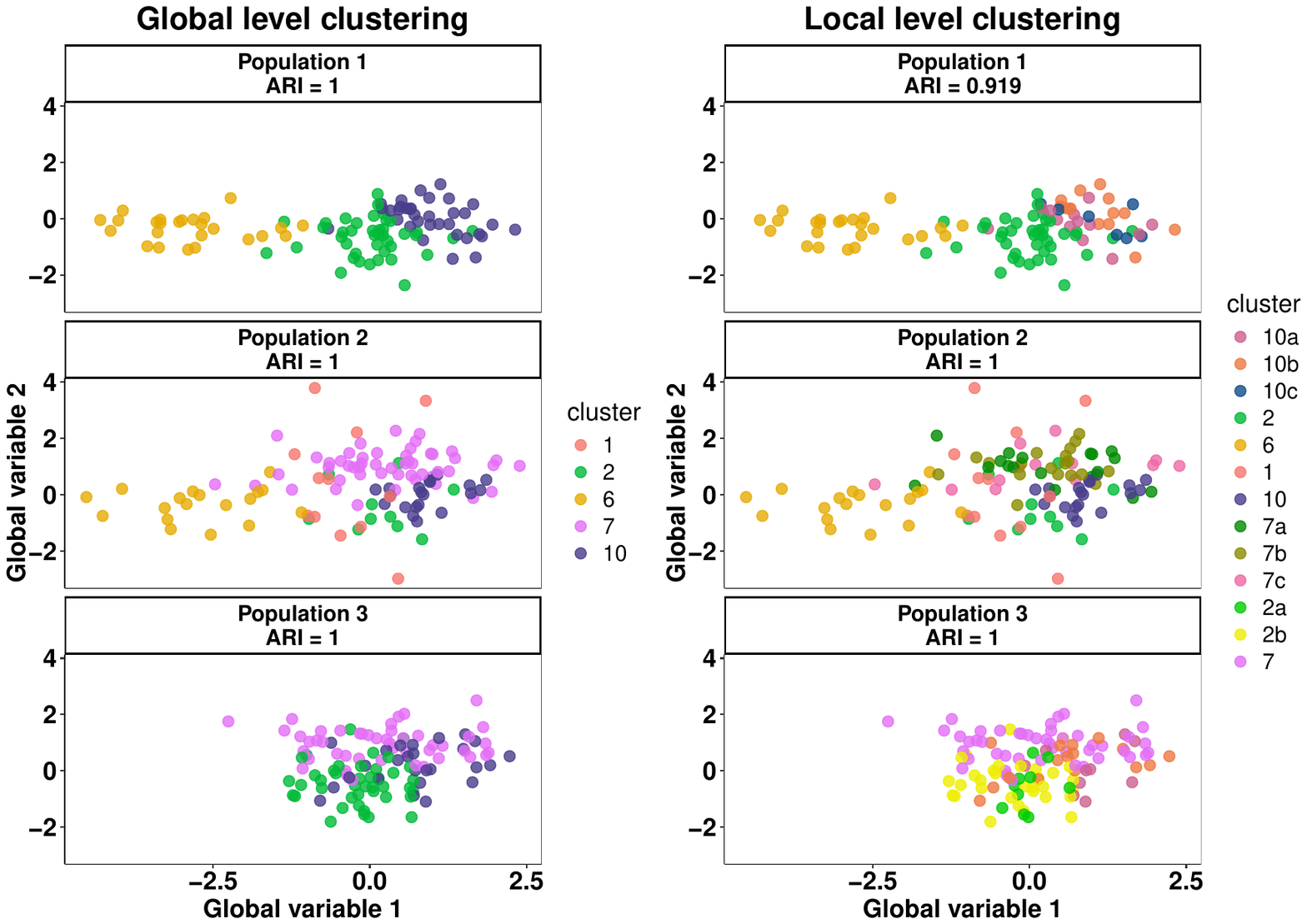}
    \caption{Global variables with global and local-level cluster labels.}
    \label{fig:global_clustering_highoverlapped}
\end{subfigure}
\begin{subfigure}{0.75\textwidth}
    \centering
    \includegraphics[width= 1\linewidth]{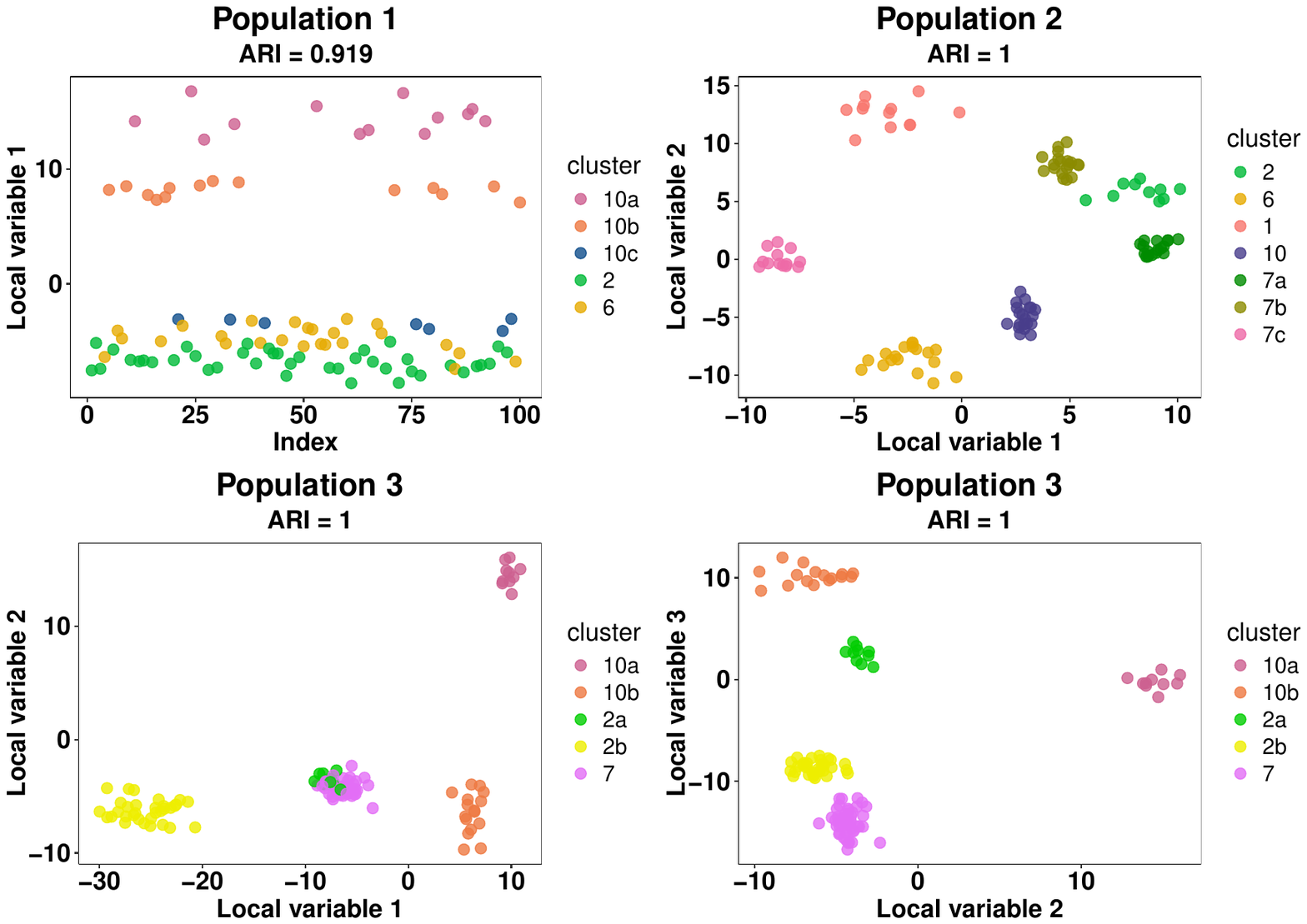}
    \caption{Local variables with local-level cluster labels.}
    \label{fig:local_clustering_highoverlapped}
\end{subfigure}
\caption{Clustering performance of GLocal DP when the global variables are highly overlapped, but the local variables are separated. The colors indicate the estimated clusters.  Adjusted rand index is reported at the top of each panel.}
\label{fig:GLOCAL_clustering_highoverlapped}
\end{figure}

In the main manuscript we presented the clustering performance of the GLocal DP when the global variables are moderately separated. Here, we present the clustering results when the global variables are well separated in Figure \ref{fig:GLOCAL_clustering_separated}. Furthermore, we also present the plot when the global variables are not separated but the local variables are separated in Figure \ref{fig:GLOCAL_clustering_highoverlapped}.

\subsection{No local variable for one population}
\label{subsec:no_local}
\begin{figure}[!ht]
\centering
\begin{subfigure}{0.48\textwidth}
  \centering
    \includegraphics[width= 0.9\linewidth]{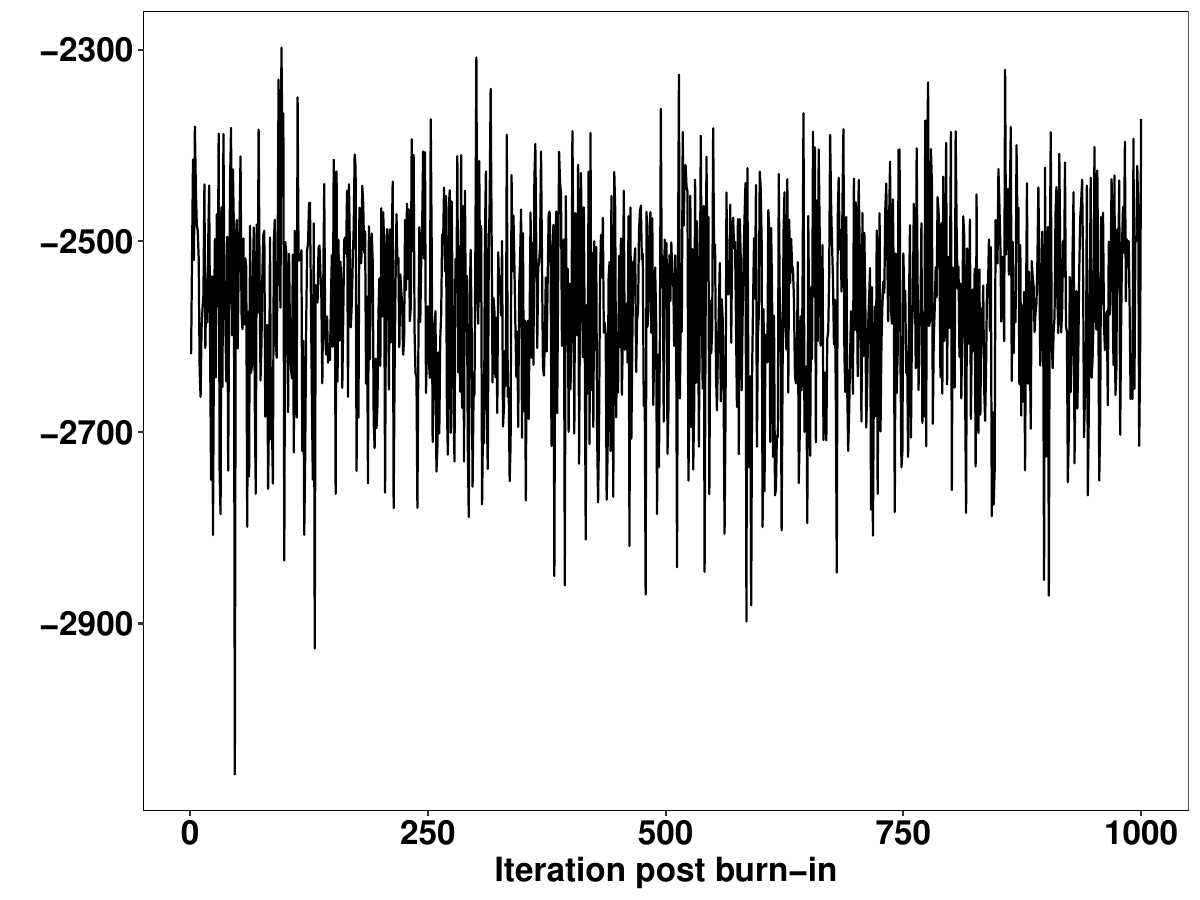}
    \caption{Traceplot of log-posterior.}
    \label{fig:LL_NoLocal}
\end{subfigure}
\begin{subfigure}{0.48\textwidth}
    \centering
    \includegraphics[width= 0.9\linewidth]{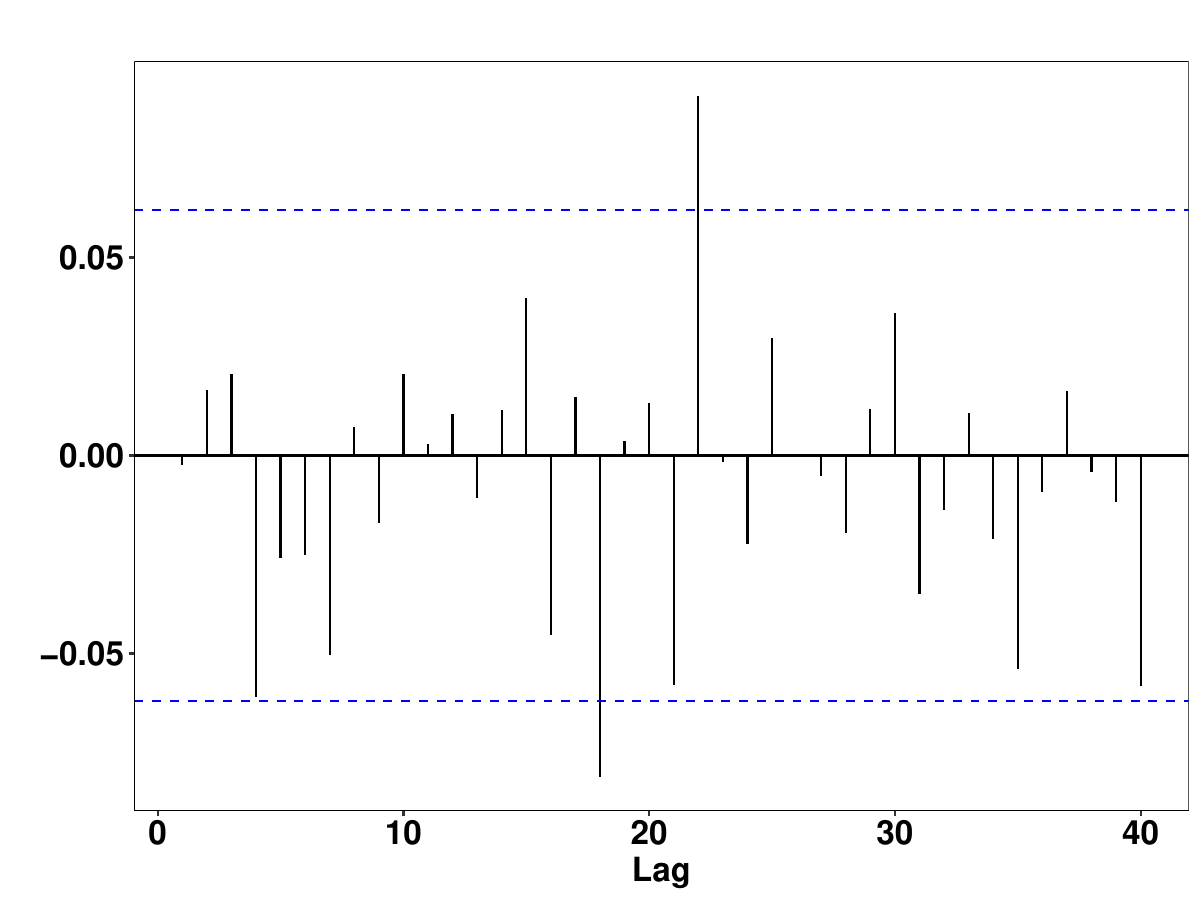}
    \caption{ACF plot of log-posterior.}
    \label{fig:ACF_NoLocal}
\end{subfigure}
\caption{The traceplot and ACF of log-posterior post burn-in and thinning. The population 1 lacks any local variable.}
\label{fig:Diagnostics_NoLocal}
\end{figure}

\begin{figure}[!htp]
\centering
\begin{subfigure}{0.75\textwidth}
  \centering
    \includegraphics[width= 1\linewidth]{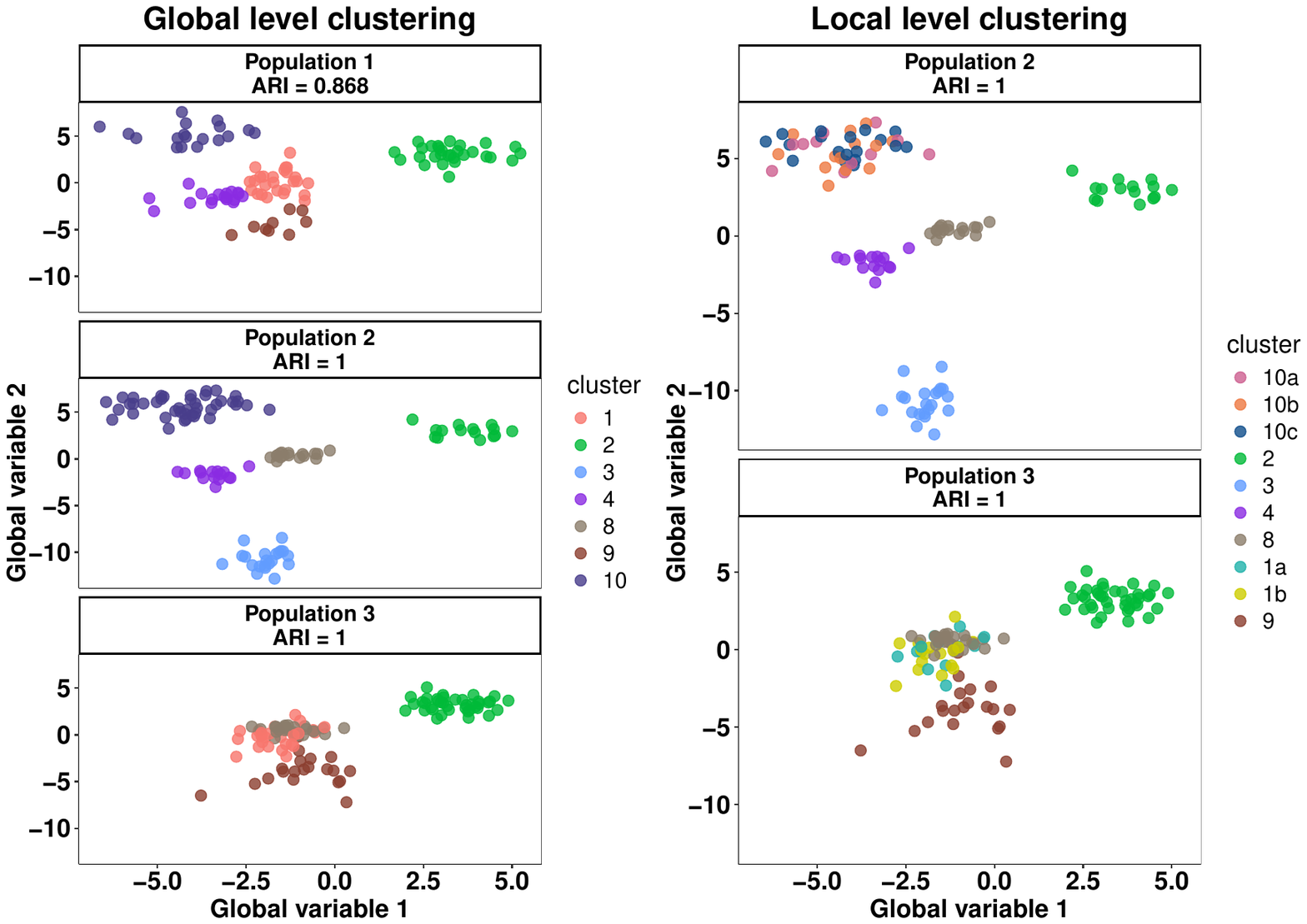}
    \caption{Global variables with global and local-level cluster labels.}
    \label{fig:global_clustering_NoLocal}
\end{subfigure}
\begin{subfigure}{0.95\textwidth}
    \centering
    \includegraphics[width= 1\linewidth]{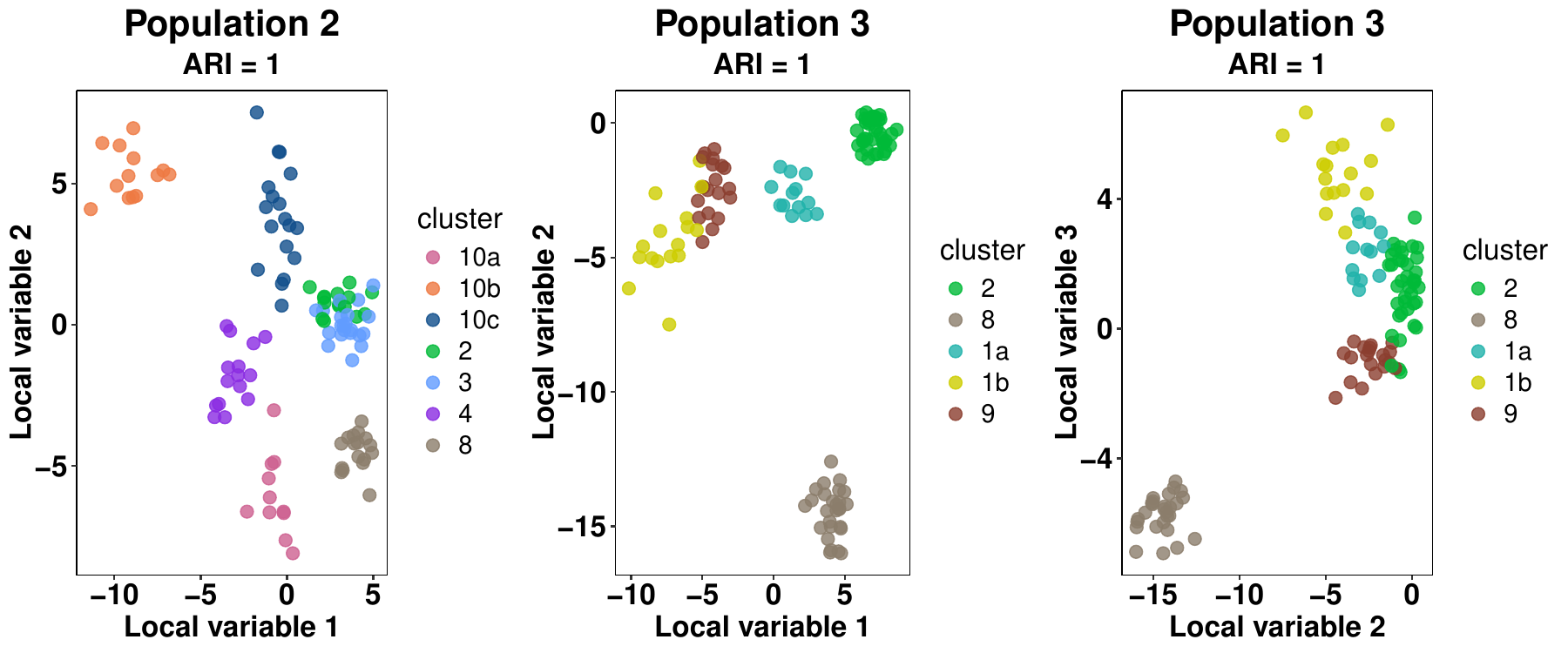}
    \caption{Local variables with local-level cluster labels.}
    \label{fig:local_clustering_NoLocal}
\end{subfigure}
\caption{Clustering performance of GLocal DP when the population 1 lacks local variable. The colors indicate the estimated clusters.  Adjusted Rand index is reported at the top of each panel.}
\label{fig:GLOCAL_clustering_NoLocal}
\end{figure}

 In this subsection, we considered the case where one of the three populations has no local variable. In particular, populations 1, 2, and 3 have 0, 2, and 3 local variables, respectively. All other simulation specifications are same as in Section~\ref{subsec:all_Local} of the main manuscript. 
As before, we considered 50,000 iterations of our sampler and considered a burn-in of 25,000 and thinning by a factor of 25. The traceplot and the ACF plot of the log-posterior are shown in Figure \ref{fig:Diagnostics_NoLocal}, which show no lack of convergence of our sampler and no significant auto-correlation. 
The clustering plots in Figure \ref{fig:GLOCAL_clustering_NoLocal} again show that our model can identify clusters with very good accuracy even when a population lacks local variables.

\subsection{Comparison with HDP}
\label{subsec:HDP_comparison}
  We performed additional simulations to compare the proposed GLocal DP with HDP. We considered varying dimensions of the global variables while fixing the dimension of the local variables to be one, two, and three for the three populations. As before, we varied the degree of separation in the local variables for the three populations by varying the local-level precision parameter $\lambda_L = 0.5, 0.1, 0.01$. All the other simulation details are the same as in the main manuscript. Furthermore, HDP was applied to the global variables only whereas GLocal DP was applied to both global and local variables. All simulations were replicated 50 times. 
  
 \begin{figure}[!ht]
    \centering
        \begin{subfigure}{0.85\textwidth}
        \centering
        \includegraphics[width = 1\linewidth]{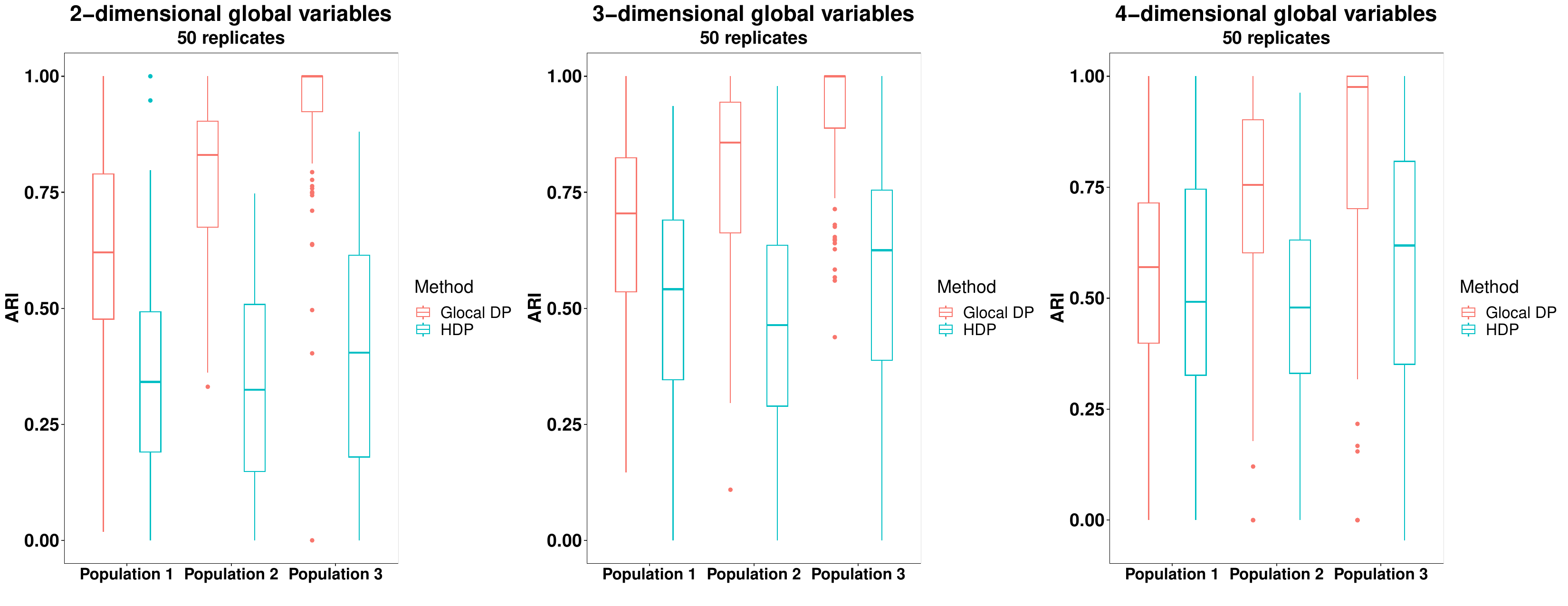}
        \caption{Low separation in the local variables.}
        \label{fig:HDP_comp_low}
        \end{subfigure}\par
        \centering
         \begin{subfigure}{0.85\textwidth}
         \centering
        \includegraphics[width = 1 \linewidth]{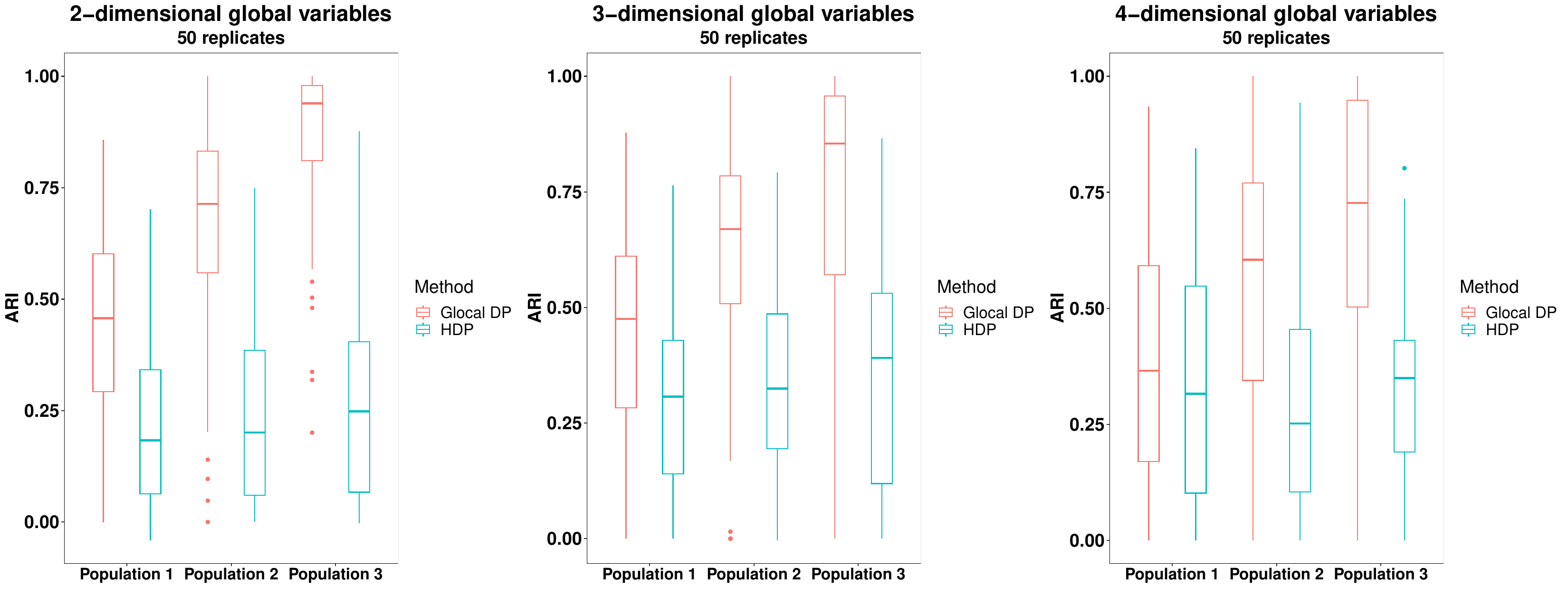}
        \caption{Moderate separation in the local variables.}
        \label{fig:HDP_comp_moderate}
        \end{subfigure}\par
            \centering
        \begin{subfigure}{0.85\textwidth}
        \centering
        \includegraphics[width = 1 \linewidth]{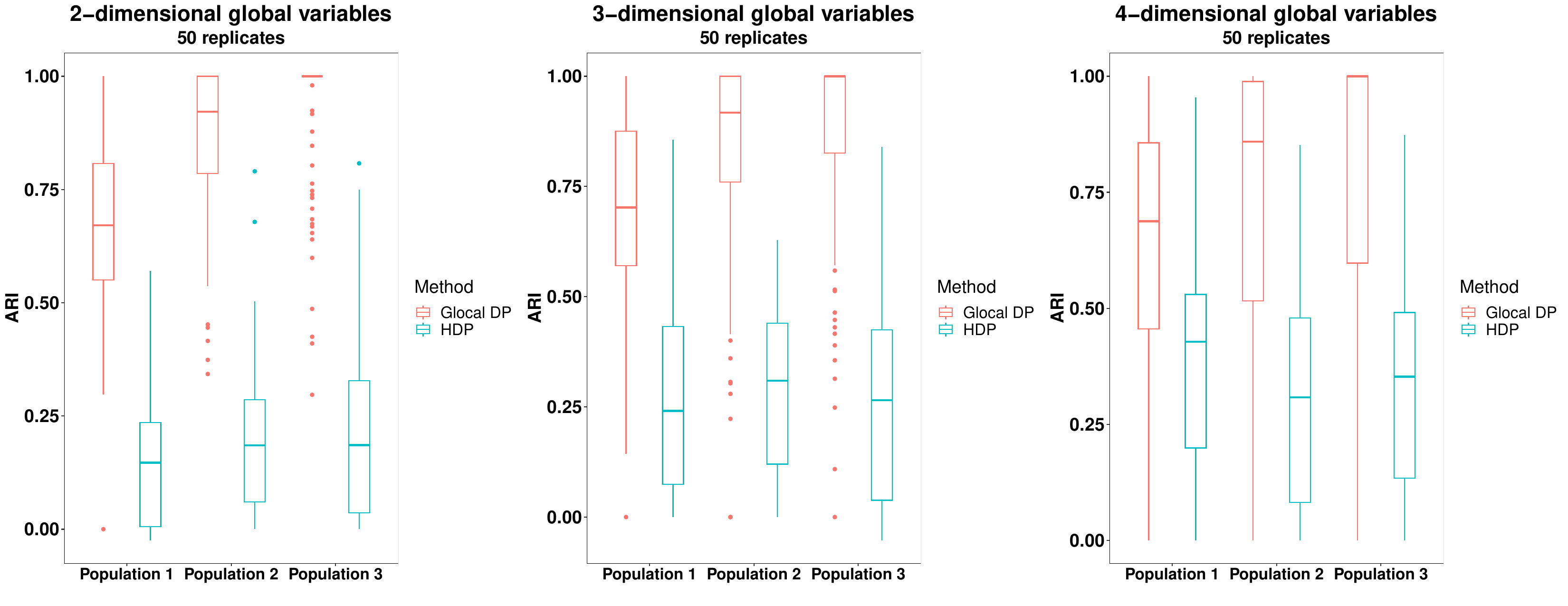}
        \caption{High separation in the local variables.}
        \label{fig:HDP_comp_high}
        \end{subfigure}
            \caption{Comparison of clustering performance of GLocal DP with HDP for varying separation of local variables and dimension of global variables.}
    \label{fig:HDP_comparisons}
    \end{figure}

    \par Figure \ref{fig:HDP_comparisons} shows that for 2-dimensional global variables, the clustering performance of our GLocal DP is uniformly better than HDP. For 3-dimensional global variables, the clustering performance of our method is still better than HDP and it improves with increasing separability in local variables. For 4-dimensional global variables, the clustering performance of our method clearly improves with increasing separation in the local variables. 
    In summary, the clustering performance of GLocal DP shows significant improvement over HDP as the local variables become more separated regardless of the dimension of the global variables.\par
\begin{figure}[!htp]
\centering
\begin{subfigure}{0.75\textwidth}
  \centering
    \includegraphics[width= 1\linewidth]{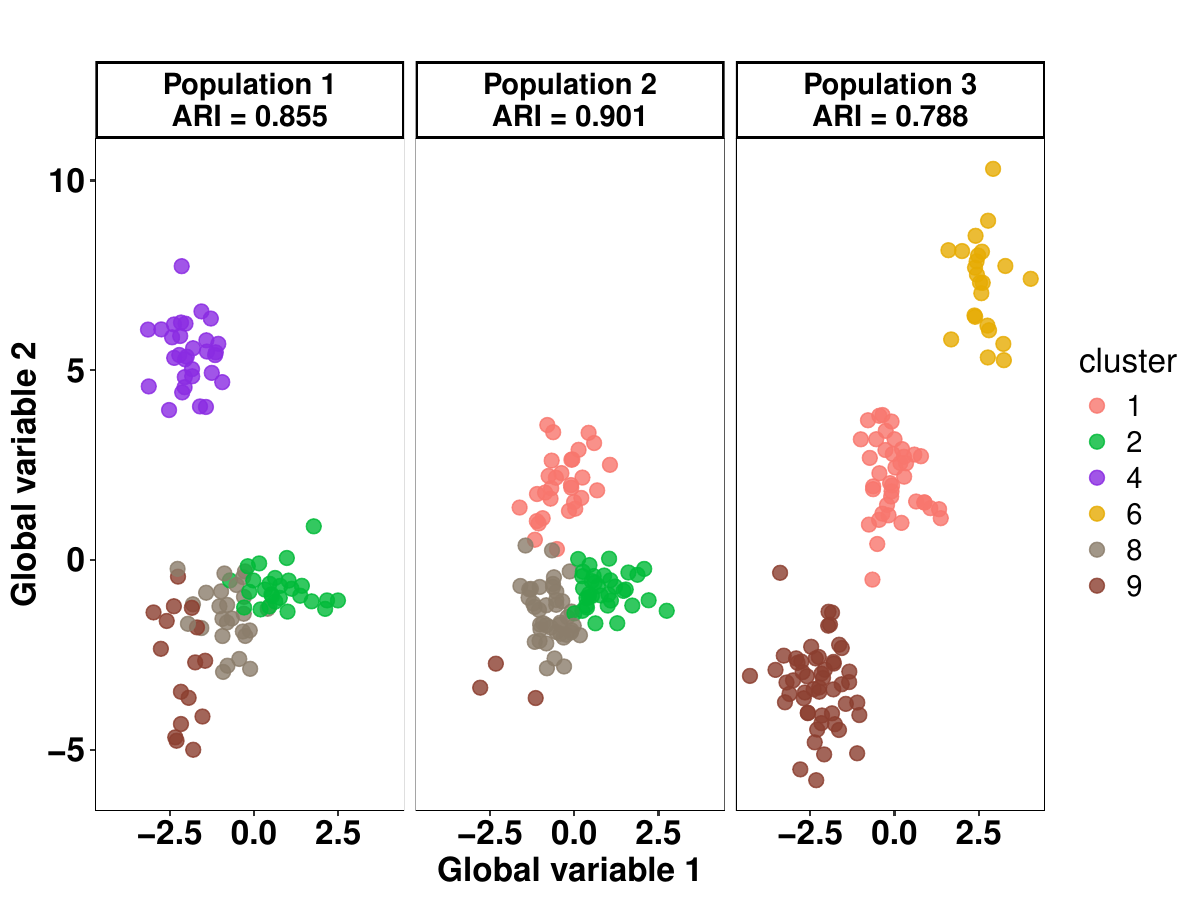}
    \caption{GLocal DP.}
    \label{fig:GLocalDP_Informative_One_Local}
\end{subfigure}
\begin{subfigure}{0.75\textwidth}
    \centering
    \includegraphics[width= 1\linewidth]{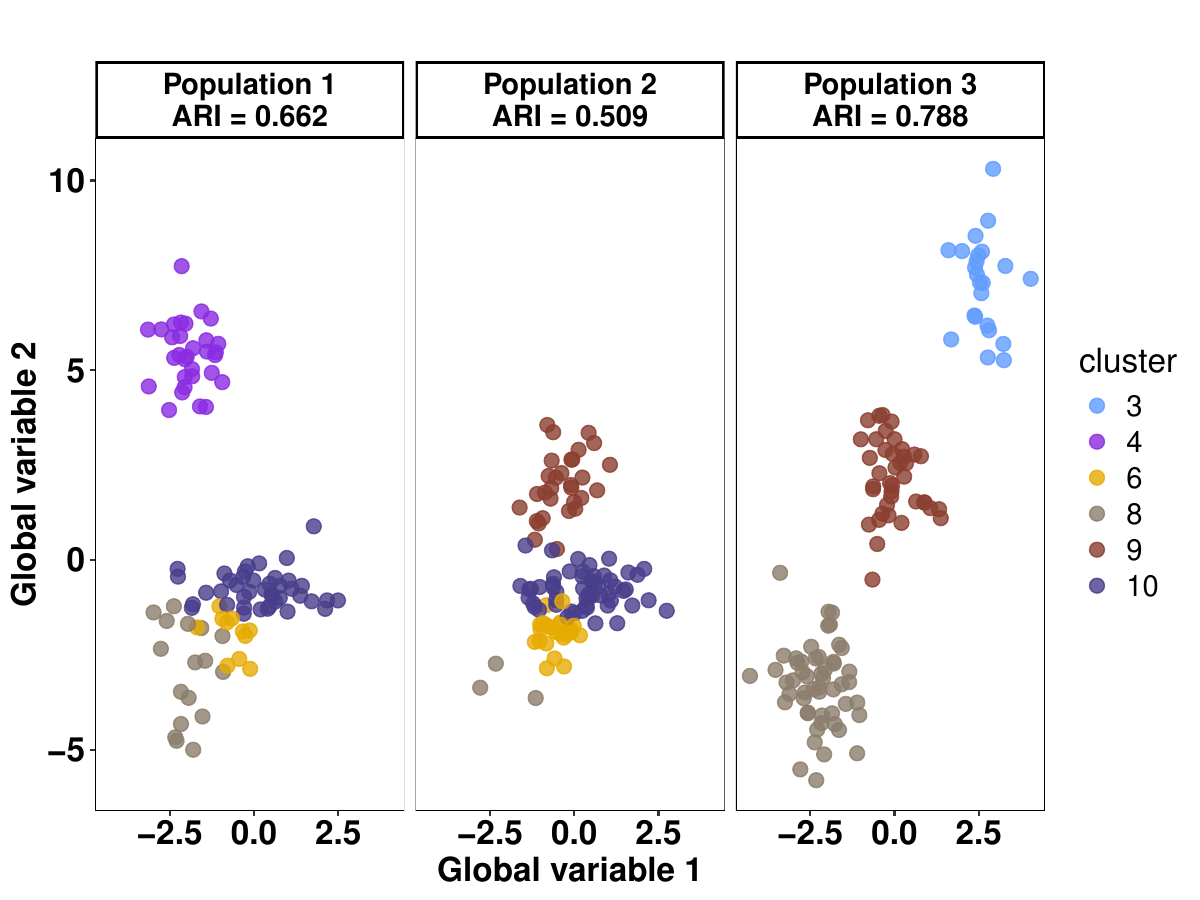}
    \caption{HDP.}
    \label{fig:HDP_Informative_One_Local}
\end{subfigure}
\caption{Clustering performance of GLocal DP and HDP when only the population 1 has informative local variable. The colors indicate the estimated clusters.  Adjusted Rand index is reported at the top of each panel.}
\label{fig:GLOCAL_HDP_Better}
\end{figure}
\begin{figure}[!htp]
\centering
\begin{subfigure}{0.75\textwidth}
  \centering
    \includegraphics[width= 1\linewidth]{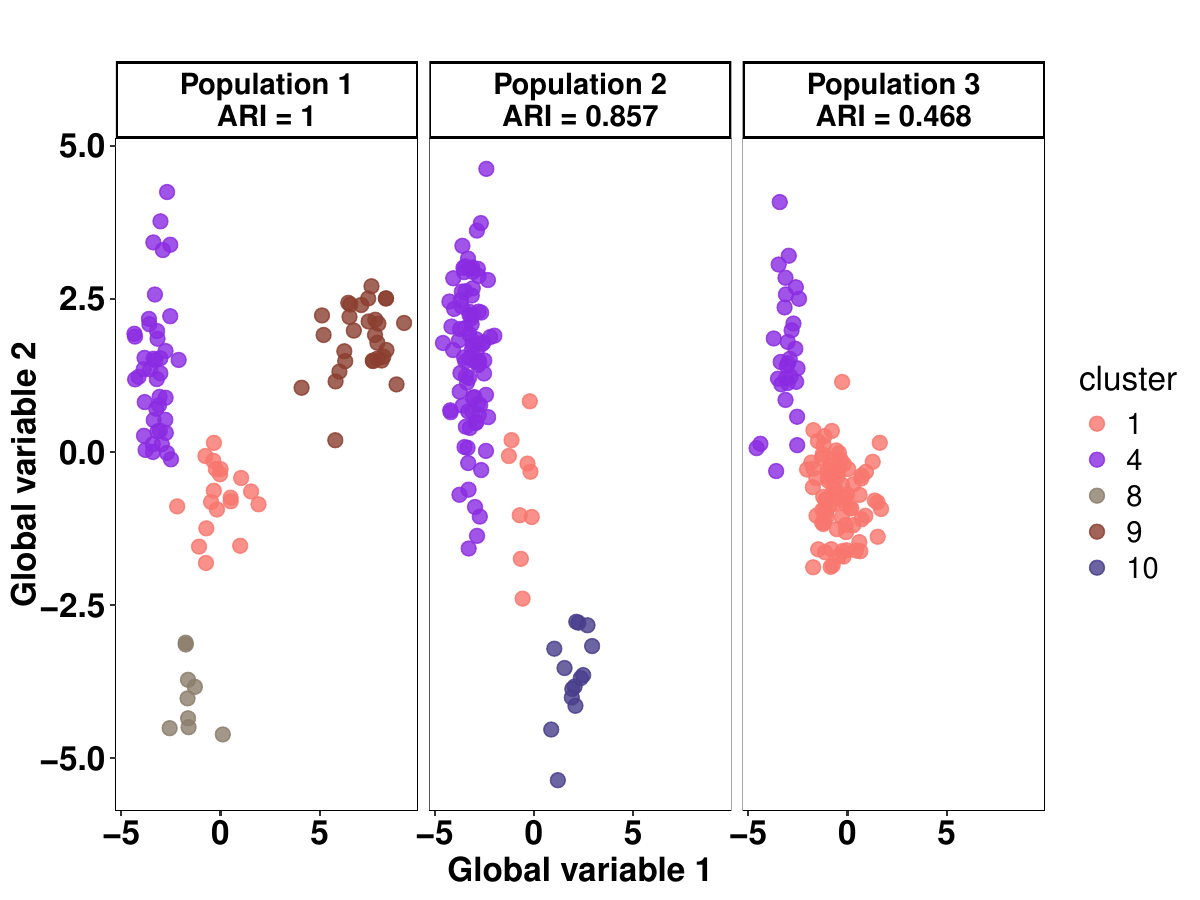}
    \caption{GLocal DP.}
    \label{fig:GLocalDP_NonInformative_One_Loca}
\end{subfigure}
\begin{subfigure}{0.75\textwidth}
    \centering
    \includegraphics[width= 1\linewidth]{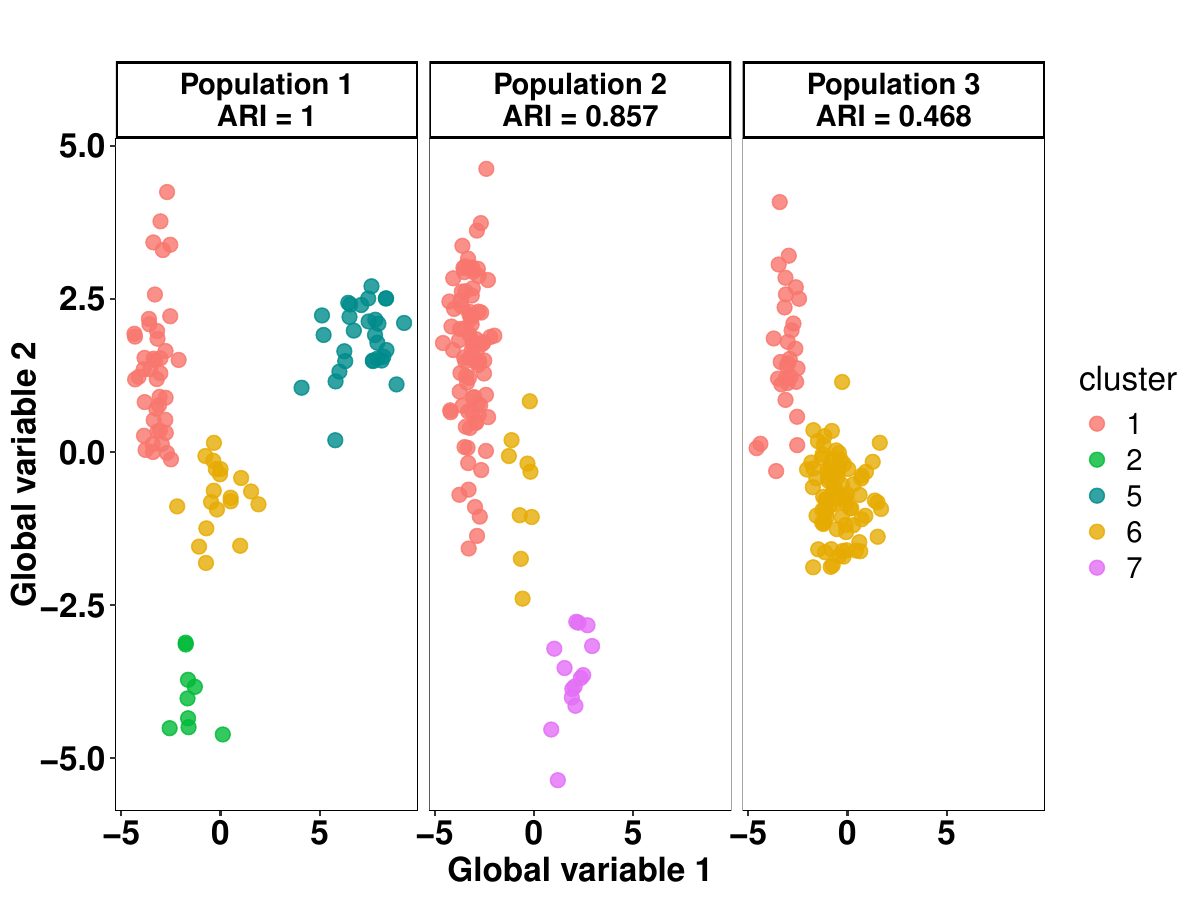}
    \caption{HDP.}
    \label{fig:HDP_NonInformative_One_Local}
\end{subfigure}
\caption{Clustering performance of GLocal DP and HDP when local variable provides no additional information. The colors indicate the estimated clusters.  Adjusted Rand index is reported at the top of each panel.}
\label{fig:GLOCAL_HDP_Same}
\end{figure}

    We also considered simulations to understand the impact of local variables on the clustering performance of GLocal DP and HDP. In particular, we consider the case where only population 1 has a local variable and populations 2 and 3 are devoid of local variables. First, we consider a scenario in which the local variable in population 1 was drawn from a 6 component univariate Gaussian mixture. The global variables for all other groups are drawn from an 8 component Gaussian mixture distribution. All other simulation strategies are same as in Section~\ref{subsec:all_Local} of the main manuscript. Note that populations 2 and 3 only consist of the two-dimensional global variables. This scenario corresponds to the case where only one population has an informative local variable. Second, we consider a scenario in which the local variable in population 1 was simply a Gaussian noise, i.e, this corresponds to the scenario where the local variable provides no information in the clustering. For both cases, we considered 50,000 iterations of the GLocal DP and HDP. As before, after discarding the first half of the iterations and retaining every 25th posterior sample therein, we looked at the clustering results. The clustering plots for the two scenarios are presented in Figures \ref{fig:GLOCAL_HDP_Better} and \ref{fig:GLOCAL_HDP_Same} respectively. Clearly, from Figure \ref{fig:GLocalDP_Informative_One_Local}, we see that if the local variable is informative, GLocal DP improves the clustering performance, not only in the group including the local variable, but across the populations in comparison to HDP (Figure \ref{fig:HDP_Informative_One_Local}). Furthermore, Figure \ref{fig:GLOCAL_HDP_Same} shows that in absence of additional information from the local variable, clustering performance of GLocal DP and HDP are equivalent.

\begin{figure}[!ht]
    \centering
    \includegraphics[width= 1\linewidth]{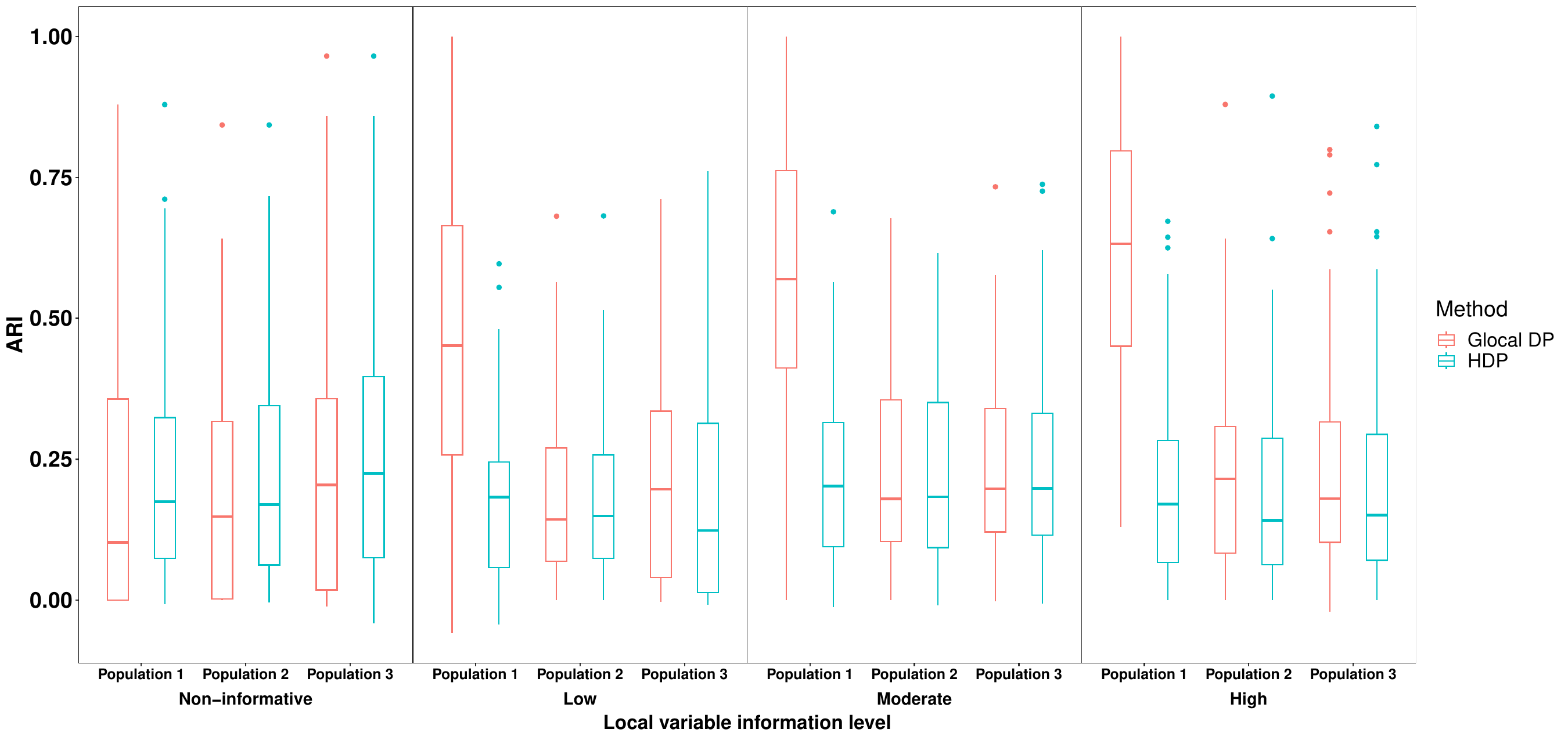}
    \caption{Boxplot of ARI, comparing the clustering performance of GLocal DP and HDP for varying level of information in the local variable of population 1. Boxplots were reported over 50 independent replications.}
    \label{fig:HDP_Comp_Noise}
\end{figure}

Furthermore, Figure \ref{fig:HDP_Comp_Noise} shows the boxplot of ARI, comparing the clustering performance of GLocal DP and HDP for varying separation of the local variable in population 1. We naively refer to the separation of the local variable by the level of information it contains e.g., ``Low'' level of information in local variable corresponds to low separation in the local variable etc. ``Non-informative'' local variable refers to the scenario where the local variable in population 1 is simply random noise.

\subsection{GLocal DP sampler for HDP sampling}
\begin{figure}[!htp]
        \centering
            \includegraphics[width= 0.95\linewidth]{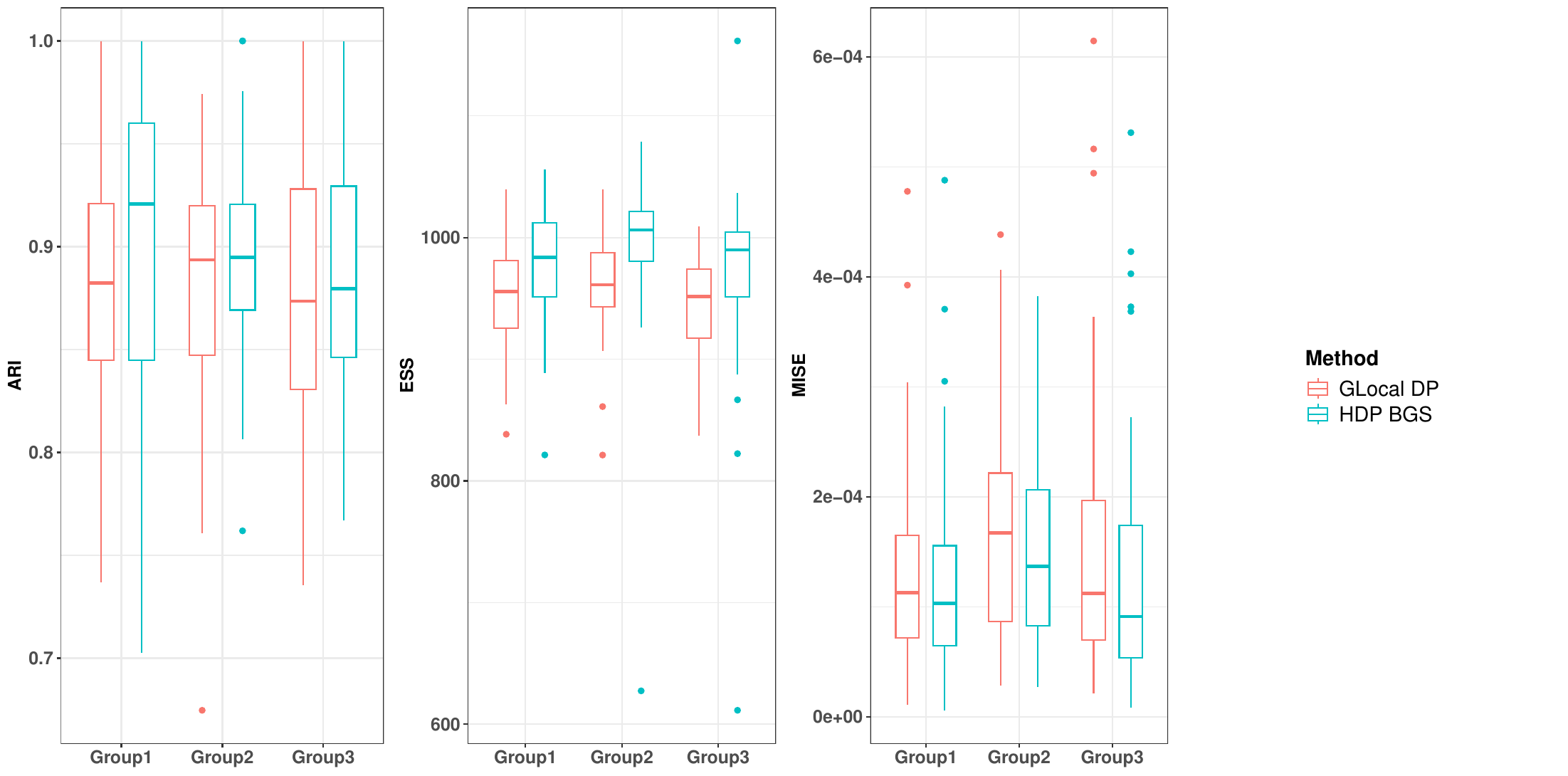}
        \caption{Clustering accuracy measured by adjusted Rand index (ARI), effective sample sizes (ESS) and mean integrated squared error (MISE) of the estimated densities of our proposed sampler and the blocked-Gibbs sampler for HDP. The means of the Gaussian mixture were taken to be $\phi^0 = (-6,- 2,2,6)$. Boxplots show variation across 50 independent replicates.}
        \label{fig:HDP_ARI_ESS_MISE_Replicated_Easy}
        \end{figure}

Recall that Algorithm \ref{alg:the_algorithm} in the Supplementary Section \ref{supp-posterior_inference} reduces to a blocked-Gibbs sampling algorithm for HDP. Furthermore, the blocked Gibbs algorithm arising as a special case of our proposed sampler is a novel contribution to the HDP sampling algorithms. We conducted a simulation study to compare the special case of our sampler with the blocked Gibbs sampler by \citealp{das_etal} for HDP sampling. Particularly, we consider 3 groups ($J = 3$) and consider a Gaussian mixture model having 4 true components, the means of which are taken to be $\phi^0 = (-6,- 2,2,6)$ with common precision $\tau = 1$. The mixture weights are chosen as  $\pi_1^0 = (0.5,0.5,0,0), \pi_2^0 = (0.25, 0.25,0.25,0.25)$ and $\pi_3^0 = (0,0.1,0.6,0.3)$. Considering equal sample sizes $n_j = 100$ for each group, we generate the true cluster labels $z_{ji}^0 \sim \pi_j^0$ and the observations $x_{ji} \sim N(\phi^0_{z_{ji}^0} ,\tau^{-1})$, for $i = 1,2,...,n_j$ and $j = 1,2,...,J$.  We assume a conjugate prior $N(0, 100)$ on each $\phi_k$. For both algorithms, we set the truncation level to 10, ran $20,000$ iterations, discarded the first $5,000$ iterations as burn-in, and retained every 15th posterior sample, resulting in 1,000 posterior samples. We estimated the clusters by the least-squares method and compared the clustering performance of the two samplers as indicated by the adjusted Rand Index (ARI) between the estimated and true clusters for each of the three groups. We also estimated densities for 100 equidistant grid points $\{y_h : h = 1,2,...,100\}$ in $[x_{\min} - 1, x_{\max} + 1]$, where $x_{\min} = \min\{x_{ji} : i,j\}$, $x_{\max} = \max\{x_{ji} : i,j\}$. We computed the effective sample sizes (ESS) and mean integrated squared error (MISE) of the estimated densities from the two samplers for each of the three groups. We performed 50 repeated simulations. Figure \ref{fig:HDP_ARI_ESS_MISE_Replicated_Easy} shows that our sampler had similar performance to BGS across all metrics. 
        
\begin{figure}[!htp]
        \centering
            \includegraphics[width= 0.95\linewidth]{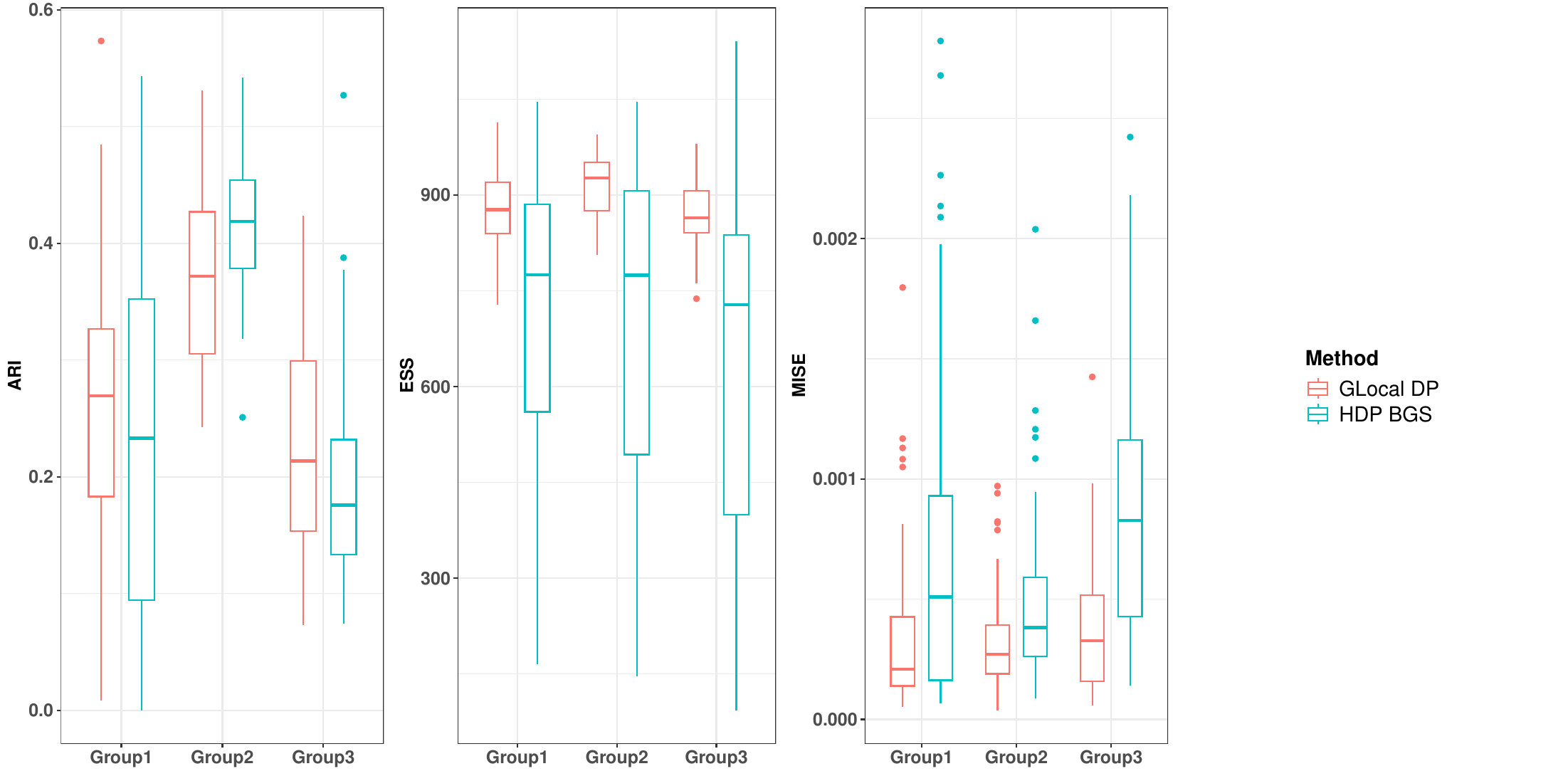}
        \caption{Clustering accuracy measured by adjusted Rand index (ARI), effective sample sizes (ESS) and mean integrated squared error (MISE) of the estimated densities of our proposed sampler and the blocked-Gibbs sampler for HDP.  The means of the Gaussian mixture were taken to be $\phi^0 = (-3,- 1,1,3)$. Boxplots show variation across 50 independent replicates.}
        \label{fig:HDP_ARI_ESS_MISE_Replicated_Hard}
        \end{figure}

        In addition, we also considered a slightly more difficult scenario with overlapping clusters across the three groups. In particular, the means of Gaussian mixtures were taken to be $\phi^0 = (-3,- 1,1,3)$. All other parameters were kept the same as before and we considered 50 independent replications. Figure \ref{fig:HDP_ARI_ESS_MISE_Replicated_Hard} shows that the clustering accuracy of our proposed sampler is comparable to that of BGS. However, our algorithm slightly outperformed BGS in ESS and MISE of the estimated densities. 

\end{document}